\documentclass[aps,prd,reprint,twocolumn,superscriptaddress,nofootinbib]{revtex4-1}

\usepackage{graphicx}  
\usepackage{subfigure}
\usepackage{latexsym}
\usepackage{slashed}
\usepackage{dcolumn}   
\usepackage{bm}        
\usepackage{amsmath}
\usepackage{amssymb}   
\usepackage{longtable}
\usepackage[section]{placeins}
\usepackage{float}
\usepackage{rotating}
\usepackage{multirow}
\usepackage{array}
\newcolumntype{L}[1]{>{\raggedright\let\newline\\\arraybackslash\hspace{0pt}}p{#1}}
\newcolumntype{C}[1]{>{\centering\let\newline\\\arraybackslash\hspace{0pt}}p{#1}}
\newcolumntype{R}[1]{>{\raggedleft\let\newline\\\arraybackslash\hspace{0pt}}p{#1}}

\def\askPmu{(p - p')^\mu}
\def\askGmu{g^{\mu\nu}}

\newcommand{\ewnur}{EW$\nu_R$ }
\newcommand{\uxu}{$U(1)_{\text{SM}} \times U(1)_{\text{MF}}$ }
\newcommand{\suxsu}{$SU(2)_L \times SU(2)_R$ }

\newcommand{\Htilde}{\widetilde{H}}

\newcommand{\Htil}{\widetilde{H} }
\newcommand{\Htill}{\widetilde{H}^\prime }
\newcommand{\Htilll}{\widetilde{H}^{\prime\prime} }
\newcommand{\be}{\begin{equation}}
\newcommand{\ee}{\end{equation}}
\newcommand{\bea}{\begin{eqnarray}}
\newcommand{\eea}{\end{eqnarray}}
\newcommand{\bfl}{\begin{flalign}}
\newcommand{\efl}{\end{flalign}}
\newcommand{\bma}{\begin{matrix}}
\newcommand{\ema}{\end{matrix}}

\newcommand{\bml}{\begin{mathletters}}
\newcommand{\eml}{\end{mathletters}}
\newcommand{\bes}{\begin{subequations}}
\newcommand{\ees}{\end{subequations}}

\newcommand{\bi}{\begin{itemize}}
\newcommand{\ei}{\end{itemize}}
\newcommand{\gev}{{\rm GeV }}
\newcommand{\tev}{{\rm TeV }}

\newcommand{\uva}{\affiliation{Department of Physics, University of Virginia, Charlottesville, VA 22904-4714, USA}}
\newcommand{\hue}{\affiliation{Center for Theoretical and Computational Physics, Hue University College of Education, Hue, Vietnam}}

\begin{document}
\title{Non-sterile electroweak-scale right-handed neutrinos and the dual nature of the 125-GeV scalar}
\author{Vinh Hoang}
\email{vvh9ux@virginia.edu}\uva
\author{Pham Q. Hung}
\email{pqh@virginia.edu}\uva\hue
\author{Ajinkya Shrish Kamat}
\email{ask4db@virginia.edu}\uva

\date{\today}

\begin{abstract}
Can, and under which conditions, the 125-\gev SM-like scalar with the signal strengths for its decays into $W^{+}W^{-}$, $ZZ$, $\gamma \gamma$, $b \bar{b}$ and $\tau \bar{\tau}$ being consistent with experiments be accommodated in models that go beyond the Standard Model? Is it truly what it appears to be, namely the SM Higgs boson, or could it be quite different?
A minimal extension of the original electroweak-scale right-handed neutrino model, in which right-handed neutrinos naturally obtain electroweak-scale masses, shows a scalar spectrum which includes either the 125-\gev SM-like scalar or a scalar which is quite {\em unlike} that of the Standard Model, both of which possessing signal strengths compatible with experiment. In other words, the 125-\gev scalar could be an {\em impostor}.
\end{abstract}

\maketitle

\section{Introduction}
The discovery of the 125-\gev SM-like scalar \cite{LHC} and the present absence of any new physics signals has opened up a whole host of questions as to the true nature of the electroweak symmetry breaking and to what may lie beyond the Standard Model. The sole existence of the 125-\gev particle would leave unanswered several deep questions such as the origin of neutrino masses, the hierarchy of quark and lepton masses among many others. It also implies that the electroweak vacuum is metastable with drastic consequences in the very far-distant future \cite{metastable}. It remains to be seen whether this most simple picture- albeit one with many question marks- will be the ultimate theory of nature or it is merely an effective theory at current accessible energies whose reality tests are incomplete and more non-SM phenomena will pop up in the not-too-distant future with Run II of the LHC.

Despite the present lack of new physics at the LHC, it does not imply that it is not there. On the contrary, new physics has already appeared in the neutrino sector through neutrino oscillation and its implication on neutrino masses. This evidence, although quite clear, is only indirect and does not show where the new physics that gives rise to the aforementioned phenomena may appear. This difficulty in finding a direct evidence for the new physics involved in generating neutrino masses is compounded by the fact that these masses are so tiny, more than seven orders of magnitude smaller than the lightest lepton: the electron. In the most generic scenario of the elegant seesaw mechanism for generating tiny masses, the right-handed neutrinos are sterile i.e. singlets under the electroweak gauge group. In a nutshell, the two mass eigenvalues are $m_D^2/M$ and $M_R$ where the Dirac mass $m_D$ is proportional to the electroweak scale while the Majorana mass $M_R$ is $\gg m_D$.
In addition to the fact that $\nu_R$'s are assumed to be electroweak singlets, the very large values for $M$ in a generic scenario makes it very very difficult to probe the crucial physics, namely that which gives rise to $M_R$ which is responsible for the lightness of the ``active" neutrinos. Another facet of this new physics is the Majorana nature of the ``active" neutrinos themselves which could manifest itself through neutrino less double beta decays which so far have not been observed. Through neutrino oscillations, we have a hint of new physics but what it might be  and where to look for it is still a big mystery at the present time. 

The aforementioned uncertainties rest in large part on the assumption that right-handed neutrinos are electroweak singlets. This usually comes from
a certain extension of the SM such as the Left-Right symmetric model $SU(2)_L \times SU(2)_R \times U(1)_{B-L}$ \cite{senjanovic} or the Grand Unified model $SO(10)$, among others. 
The L-R symmetric version has an advantage over the GUT version in that the right-handed neutrino Majorana mass is proportional to the breaking scale of $SU(2)_R$ which could be much lower than a typical GUT scale and which could open up the possibility for detecting right-handed neutrinos (and $W_R$ as well). A recent search for right-handed neutrinos and $W_R$ at the LHC \cite{CMSWR} has yielded a lower bound of around 3 TeV for $W_R$. Bounds on right-handed neutrino masses were also presented in \cite{CMSWR}. We will come back to the aforementioned remarks below.

If one is however willing to entertain the idea that right-handed neutrinos {\em are not} sterile, there is an entire panorama of accessible phenomena that can be searched for and studied. A non-sterile right-handed neutrino necessarily interacts with the electroweak gauge bosons and the Majorana mass term is expected to carry the electroweak quantum number and hence is proportional to the electroweak breaking scale. Right-handed neutrinos could then be searched for both from an interaction point of view and from an energetic one. A model of this kind was put forth by one of us (PQH) \cite{pqnur}  (the \ewnur model).

In the \ewnur model \cite{pqnur}, right-handed neutrinos are parts of $SU(2)$ doublets along with their charged partners (the mirror charged leptons). Anomaly freedom dictates the existence of doublets of right-handed mirror quarks. The particle content of the model is listed in the next section. The existence of extra doublets of chiral fermions, the mirror quarks and leptons, is potentially fatal for the model because of their contributions to the electroweak precision parameters, in particular the S-parameter. Those extra chiral doublets would make a ``large" contribution to the S-parameter, an undesirable outcome. Fortunately, the \ewnur model contains a Higgs triplet which makes opposite contributions to the S-parameter and thus offsetting those of the mirror fermions. An exhaustive study of the electroweak precision parameters within the framework of the \ewnur model has been carried out in \cite{hung2} with the main result being that there is a large parameter space which satisfies the precision constraints. 

The \ewnur model in its original inception \cite{pqnur} contains, beside one Higgs doublet which couples to both SM and mirror fermions, two scalar triplets, one (complex) with hypercharge $Y/2=1$ and another (real) with $Y/2=0$. Out of the thirteen degrees of freedom (4 for the doublet, 6 for the complex triplet and 3 for the real triplet), three are absorbed by W's and Z and the remaining {\em ten} become physical degrees of freedom. Can one of those ten physical scalars describe the observed 125-\gev SM-like scalar? If not, what minimal extension would be needed for that purpose? Where and how does one look for the more massive scalars which could be CP even or odd? 

The plan of the paper is as follows. Section~\ref{sec:summary} will be devoted to a summary of the \ewnur model with its particle content and, in particular for this paper, its scalar sector. For completeness, the electroweak precision parameter constraints will also be summarized. Section~\ref{sec:extewnur} presents some of the salient points concerning the scalar sector of the original \ewnur model. A particular attention is paid to what this sector has to say about the 125-\gev SM-like scalar. We show why the lightest spin-0 particle has to be CP-odd if one wishes to identify it with the 125-\gev object. This has to do with the fact that the production cross section for the scalar is very large compared with the equivalent SM quantity. This occurs when a single Higgs doublet couples to both SM and mirror fermions. The CP-odd option unfortunately is ruled out by the likelihood analysis which favors the CP-even case \cite{LHC2}. At the end of this section we present a simple extension of the original model by adding one extra Higgs doublet. In this extension, by imposing a global symmetry, one Higgs doublet is made to couple to SM fermions while the other one couples only to mirror fermions. The scalar mass eigenstates and eigenvalues are shown as well as their couplings to fermions and gauge bosons. Section~\ref{sec:data} discusses the implications of the extended model in light of the existence of the 125-\gev SM-like scalar. We will show in that section the dual nature of the 125-\gev SM-like scalar and only further measurements can tell whether or not it is an ``impostor".

\section{The \ewnur model: A summary}\label{sec:summary}

The main idea of the \ewnur model \cite{pqnur} was to search for a model in which right-handed neutrinos {\em naturally} acquire a mass proportional to the electroweak scale $\Lambda_{EW} = 246 ~\gev$. For this to occur, the {\em most natural} way to implement this idea is for right-handed neutrinos to be {\em non-sterile}. In particular, the simplest way is to put them in doublets along with right-handed mirror charged lepton partners. In this manner, a Majorana mass term of the type $M \nu_{R}^{T} \sigma_2 \nu_R$ necessarily carries an $SU(2) \times U(1)$ quantum number and transforms like an $SU(2)$ triplet. (Details are summarized below.) As shown in \cite{pqnur}, a new Higgs sector including triplets is needed and it obviously participates in the symmetry breaking of the electroweak gauge group. The \ewnur model of \cite{pqnur} is highly testable for the following reasons: 1)  $\nu_R$'s are sufficiently light; 2) $\nu_R$'s couple to W and Z and can be produced through these couplings; 3) The presence of an extended Higgs sector.

At this point, a comparison with the popular Left-Right symmetric model (L-R model) is in order here.  Basically, one would like to probe the physics that governs the right-handed neutrino Majorana mass scale $M_R$ since it is a cornerstone of the seesaw mechanism. 

As we will review below, the gauge structure of  the \ewnur model  \cite{pqnur} is the same as that of the SM, namely $SU(2) \times U(1)$. The gauge structure of the L-R model is $SU(2)_L \times SU(2)_R \times U(1)_{B-L}$. As a result, the highest mass scale of the \ewnur model is $\Lambda_{EW} \sim 246\; \gev$, while the L-R model is characterized by {\em two} scales:  $\Lambda_{L}$ and $\Lambda_{R}$ with $\Lambda_{R}$ being generally much larger than $\Lambda_{L}$.  The $\nu_R$ Majorana mass is proportional to $\Lambda_{EW}$ in the \ewnur model and is naturally {\em ``light"} while, for the L-R model, it is proportional to $\Lambda_{R}$ and could be ``heavy". 

The next difference lies in the production mechanism and detection of $\nu_R$'s. In the L-R model, $\nu_R$'s  are produced through the process $\bar{u} + d \rightarrow W_R \rightarrow N + l$ \cite{keung}. In contrast, the production of $\nu_R$'s in  \ewnur model proceeds through $q + \bar{q} \rightarrow Z \rightarrow \nu_R + \bar{\nu}_R$ or $\nu_R + \nu_R$ ($\nu_R$ is a Majorana particle). Since $M_Z $ is the mass scale that enters the production cross section for $\nu_R$ in the \ewnur model, one expects the number of events characteristic of that model to be {\em significantly} larger than that for the L-R model which is controlled by $M_{W_R}$, making it much easier to probe signals such as like-sign dileptons \cite{pqnur}. 

To summarize, the main difference between the \ewnur model and the L-R model is the question of energy scales. Right-handed neutrinos in the \ewnur model have masses proportional to $\Lambda_{EW}$ and hence are accessible experimentally through the direct coupling with the Z-boson at present (and future) colliders. Its physics is {\em bounded} by the {\em electroweak scale}. The scale $\Lambda_{R}$ of the L-R model on the other hand is unknown and is only bounded from below experimentally. It would be much harder to probe $SU(2)_R$ if that scale turns out to be much higher than the present experimental bound of 3 \tev's. The direct evidence for the seesaw mechanism through the production and decay of $\nu_R$ is certainly within the reach of the \ewnur model, both from the energetic and production points of view.

Next, concerning the scalar sector, the Higgs triplet of $SU(2)_R$ could be quite heavy, its mass being proportional to $\Lambda_{R}$. The physics of the Higgs triplet of the \ewnur model is controlled by the electroweak scale and its existence (or non-existence) can hopefully be verified in Run II of the LHC. This is the subject of the present manuscript.

Last but not least, the L-R model can be embedded in a grand unified group $SO(10)$. As mentioned in \cite{pqnur}, the \ewnur model can be embedded in the group $E_6$. However, we consider the testability of the \ewnur model to be more important than its embedding in a GUT group.

\subsection{Gauge structure and particle content of the \ewnur model }
Below is a summary of the gauge structure and particle content of the minimal \ewnur model of \cite{pqnur}.
The notations for the leptons and quarks are generic for any family.

\begin{itemize}

\item Gauge group: $SU(3)_C \times SU(2) \times U(1)_Y$

\item Lepton $SU(2)$ doublets (generic notation):

SM: 
\begin{equation}
l_{L} = \left( \begin{array}{c}
\nu_L \\
e_{L}
\end{array} \right)\;;
\end{equation}

Mirror: 
\begin{equation}
l^{M}_{R} = \left( \begin{array}{c}
\nu_R \\
e^{M}_{R}
\end{array} \right)\;.
\end{equation}

\item Lepton $SU(2)$ singlets (generic notation):

SM: $e_R$ ; Mirror: $e^{M}_L$

\item Quark $SU(2)$ doublets (generic notation):

SM: 
\begin{equation}
q_{L} = \left( \begin{array}{c}
u_{L} \\
d_{L}
\end{array} \right)\;;
\end{equation}

Mirror: 
\begin{equation}
q^{M}_{R} = \left( \begin{array}{c}
u^{M}_{R} \\
d^{M}_{R}
\end{array} \right)\;.
\end{equation}

\item Quark $SU(2)$ singlets (generic notation):

SM: $u_R$ ,  $d_R$  ;  Mirror: $u^{M}_L$ , $d^{M}_L$.

\item The Higgs sector:

a) One Higgs doublet: $\Phi$. This Higgs doublet couples to both SM and mirror fermions.

b) One complex Higgs triplet with $Y/2=1$ containing doubly-charged scalars:
\begin{equation}
\tilde{\chi} = \frac{1}{\sqrt{2}}\,\vec{\tau}.\vec{\chi}=
\left( \begin{array}{cc}
\frac{1}{\sqrt{2}}\,\chi^{+} & \chi^{++} \\
\chi^{0} & -\frac{1}{\sqrt{2}}\,\chi^{+}
\end{array} \right)\;.
\end{equation}

c) One real Higgs triplet with $Y/2=0$:
\begin{equation}
(\xi^+, \xi^0, \xi^-)\,.
\end{equation}

d) One SM singlet Higgs: $\phi_S$.

\end{itemize}

\subsection{Symmetry breaking in the \ewnur model}
$SU(2) \times U(1)$ is spontaneously broken by the vacuum expectation values (VEV) of the Higgs doublet and triplets. The Higgs potential \cite{ChanoGold,pqnur} has a global $SU(2)_L \times SU(2)_R$ symmetry. The triplets transform as $(3,3)$ and the doublet as $(2,2)$ under that global symmetry. Specifically,
\begin{equation}
\chi = \left( \begin{array}{ccc}
\chi^{0} &\xi^{+}& \chi^{++} \\
\chi^{-} &\xi^{0}&\chi^{+} \\
\chi^{--}&\xi^{-}& \chi^{0*}
\end{array} \right) \,,
\end{equation}
and
\begin{equation}
\Phi = \left( \begin{array}{cc}
\phi^{0} & \phi^{+} \\
\phi^{-} & \phi^{0} 
\end{array} \right)  \, .
\end{equation}
Proper vacuum alignment dictates $\langle \chi^{0} \rangle = \langle \xi^{0} \rangle = v_M$ i.e.
\begin{equation}
\langle \chi \rangle = \left( \begin{array}{ccc}
v_M &0&0 \\
0&v_M&0 \\
0&0&v_M
\end{array} \right) \,,
\end{equation}
\begin{equation}
\langle \Phi \rangle = \left( \begin{array}{cc}
v_2/\sqrt{2} &0 \\
0&v_2/\sqrt{2} 
\end{array} \right) \,.
\end{equation}
These VEVs leave an unbroken $SU(2)_D$ custodial symmetry i.e. $SU(2)_L \times SU(2)_R \rightarrow SU(2)_D$. This ensures that $\rho = M_{W}^2/M_{Z}^2 \cos^2 \theta_W=1$ at tree level and one now has
\begin{equation}
v= \sqrt{v_2^2+ 8\,v_M^2} \approx 246 \, \gev \,.
\end{equation}

As discussed in \cite{pqaranda,hung2,GeorgMach}, with respect to $SU(2)$, the two triplets (one real and one complex) and one doublet sum up to 13 degrees of freedom, 3 of which are Nambu-Goldstone bosons absorbed by W's and Z  leaving 10 physical degrees of freedom. Under the custodial symmetry $SU(2)_D$, these transform as
\begin{eqnarray}
		\text{five-plet (quintet)} &\rightarrow& H_5^{\pm\pm},\; H_5^\pm,\; H_5^0;\nonumber\\[0.5em]
		\text{triplet} &\rightarrow& H_3^\pm,\; H_3^0;\nonumber\\[0.5em]
		\text{two singlets} &\rightarrow& H_1^0,\; H_1^{0\prime}\,.  \nonumber
\end{eqnarray}  
The expressions for the scalar states can be explicitly found in Eq.~(\ref{eq:higgs}), by setting $s_{2M}\rightarrow 0$ and $\Phi_{2M} \rightarrow 0$.

\subsection{The seesaw mechanism in the \ewnur model}
The main purpose of the \ewnur model was to provide a scenario in which right-handed neutrinos are non-sterile and get their masses out of the symmetry breaking of $SU(2) \times U(1)$. A Majorana mass term of the form $M_{R} \, \nu_{R}^{T}\,  \sigma_{2} \, \nu_R$ in the \ewnur model comes from the following Yukawa interaction:
\begin{equation}
g_{M} l_R^{M,T} \,\sigma_2\, \tilde{\chi} \, l_R^{M} \, ,
\end{equation}
which gives 
\begin{equation}
g_M \, \nu_{R}^{T}\,  \sigma_{2} \, \nu_R \chi^{0} \,.
\end{equation}
The right-handed neutrino Majorana mass is now intrinsically linked to the breaking scale of $SU(2) \times U(1)$ through the VEV of $\tilde{\chi}$ as
\begin{equation}
M_R = g_M\,v_M  \, .
\end{equation}
As stressed in \cite{pqnur}, $M_R$ is bounded from below because $\nu_R$ are now members of an $SU(2)$ doublet and would contribute to the Z-boson width leading to the lower bound: 
\begin{equation}
M_R \geq M_Z/2 \approx 46 ~\gev\,.
\end{equation}

As discussed in \cite{pqnur}, a global symmetry $U(1)_{MF}$ (referred to as $U(1)_M$ in \cite{pqnur}) was imposed so as to forbid a term of the form $g_{L} l_L^{M,T} \,\sigma_2\, \tilde{\chi} \, l_L^{M}$ which would give a large Majorana mass $g_L v_M$ to the left-handed neutrino, unless $g_L$ is unnaturally fine-tuned to be tiny. As discussed in \cite{pqnur}, this is accomplished by the following transformation properties: $(l_R^M, ~e_L^M) \rightarrow e^{\imath\alpha_{MF}} (l_R^M, ~e_L^M)$, $\widetilde{\chi} \rightarrow e^{-2\imath \alpha_{MF}} \widetilde{\chi}$ and $\phi_S \rightarrow e^{-\imath\alpha_{MF}}\phi_S$, with all other SM particles being $U(1)_{MF}$ singlets. We will come back to this symmetry and its extended version below.

A Dirac mass term is of the form $m_D (\nu_{L}^{\dagger}\, \nu_R + h.c.)$. This is a product of two doublets and the simplest choice for the Higgs scalar is an $SU(2)$ singlet with zero hypercharge, namely $\phi_S$.
\be
{\cal L}_S = g_{Sl} \, \bar{l}_{L}\, \phi_S \, l^{M}_{R} + H.c. \,.
\ee
With $\langle \phi_S \rangle = v_S$, the Dirac mass is given by
\be
m_D = g_{Sl} v_S \,.
\ee

The magnitude of the light neutrino mass given by
\be
m_\nu = \frac{m_D^2}{M_R} <O(eV) \, ,
\ee
implying $v_S \sim O(10^5 \, eV)$ if we assume $g_{Sl} \sim O(1)$ or $v_S \sim O(\Lambda_{EW} \sim 246 \,\gev)$ for $g_{Sl} \sim O(10^{-7})$.

\subsection{Constraints from electroweak precision data}

The presence of extra $SU(2)$ doublets of chiral fermions in the form of mirror fermions would seriously affect the constraints from electroweak precision data. As first mentioned in \cite{pqnur}, the positive contribution of mirror fermions to the S-parameter could be compensated by the negative contribution to S from the Higgs triplets. A detailed analysis has been performed in \cite{hung2} which showed that there is a large parameter space in the model which satisfies the present constraints of the electroweak precision data. A sample of the plots summarizing the scatter plots of the model is given below. In those plots, we had 3500 points inside the $2\sigma$ constraint. Among them, 100 points fall inside the $1\sigma$ region. The number of these data points were taken for illustrative purpose and could be larger with a longer program run time. Moreover, it is possible to find any combination in the parameters space, $m_Z\leq m_H, m_{q^M}\leq 600\;\gev, \;150\leq m_{l^M} \leq 600 \;\gev, \; m_Z/2\leq m_{\nu_R} \leq 600\;\gev$, that can satisfy the constraints. As one can see,  the size of parameter space in the model is {\em not} fine-tuned. This is illustrated in the figures below. 

\begin{figure}[H]
\centering
 \includegraphics[scale=0.35]{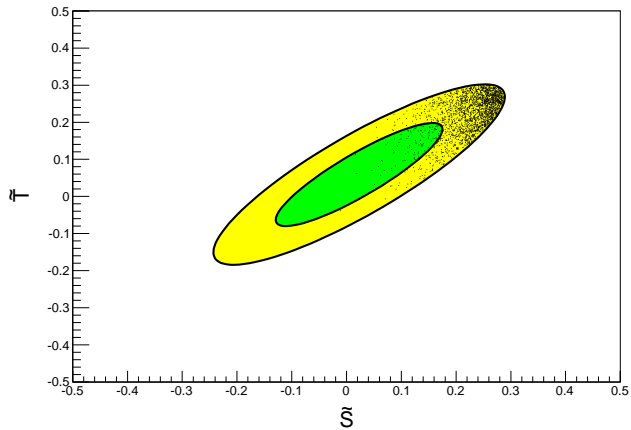} 
 \caption{\label{TStotal}{\small Total $\tilde{T}$ versus $\tilde{S}$ with the 1 and 2 $\sigma$ experimental contours}}
\label{TvsS}
\end{figure}

$\widetilde{S}$ and $\widetilde{T}$ are new Physics contributions to the $S$ and $T$ parameters respectively. One can see from Fig.~\ref{TvsS} that the \ewnur model satisfies very well the constraints from the electroweak precision data and has passed the first (indirect) test. 
Here and in \cite{hung2}, $\tilde{S}$ and $\tilde{T}$ are defined as the sum of the contributions from the scalar and mirror fermion sectors, namely $\tilde{S}=\tilde{S}_S + \tilde{S}_{MF}$ and $\tilde{T}=\tilde{T}_S + \tilde{T}_{MF}$. For clarity, we also show the plots of $\tilde{S}_S$ versus $\tilde{S}_{MF}$ and $\tilde{T}_S$ versus $\tilde{T}_{MF}$. 
\begin{figure}[H]
\centering
    \includegraphics[scale=0.35]{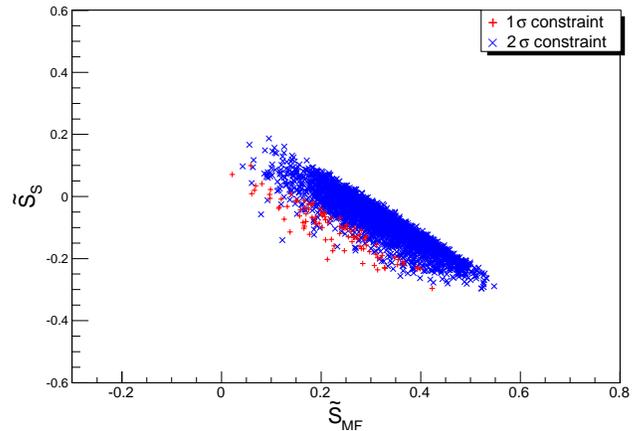} 
 \caption{{\small Constrained $\tilde{S}_S$ versus $\tilde{S}_{MF}$}}
\label{SsvsSmf}
\end{figure}
\begin{figure}[H]
\centering
    \includegraphics[scale=0.35]{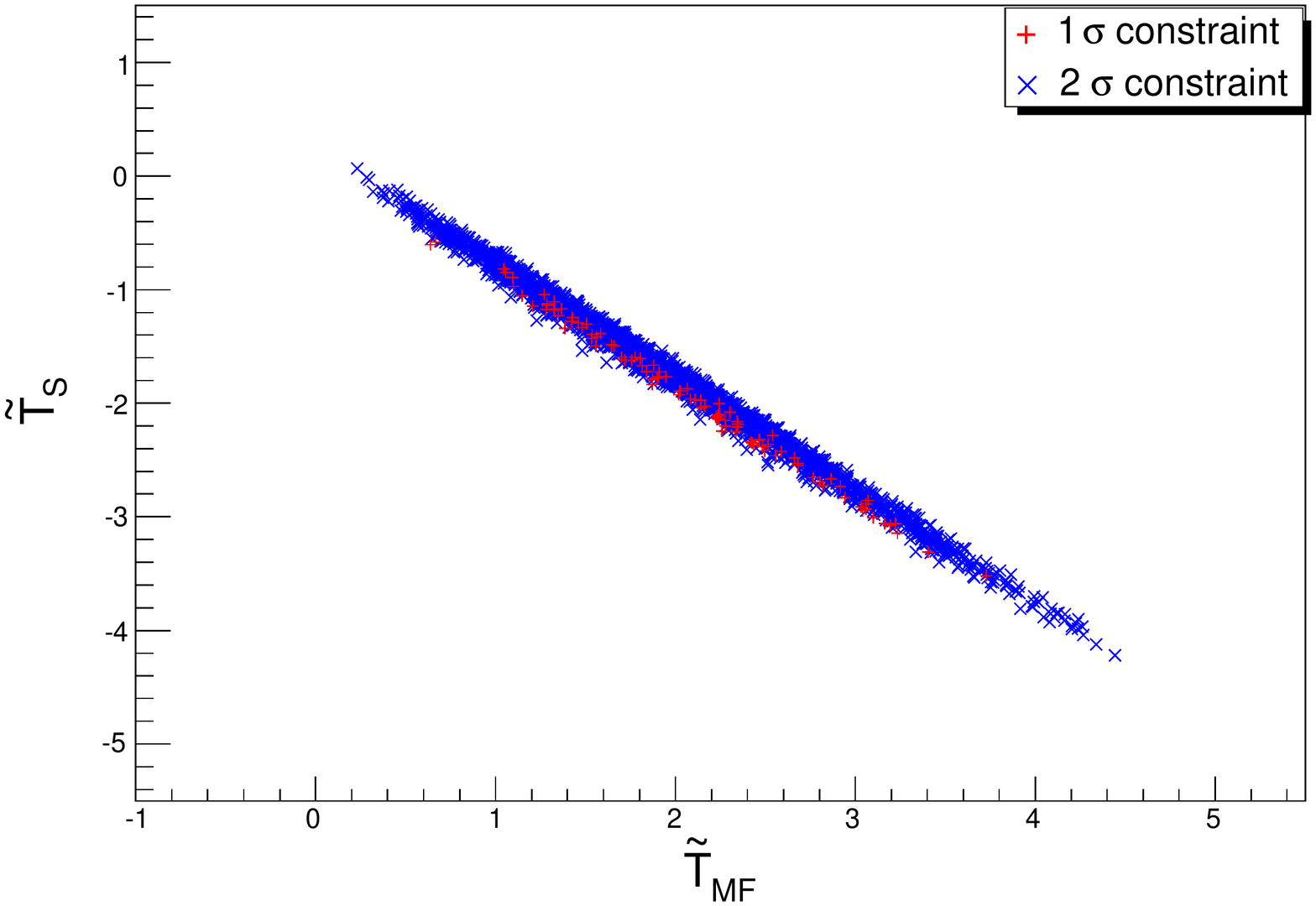} 
 \caption{{\small Constrained $\tilde{T}_S$ versus $\tilde{T}_{MF}$}}
\label{TsvsTmf}
\end{figure}
\begin{figure}[H]
	\centering
	\includegraphics[width=0.5\textwidth]{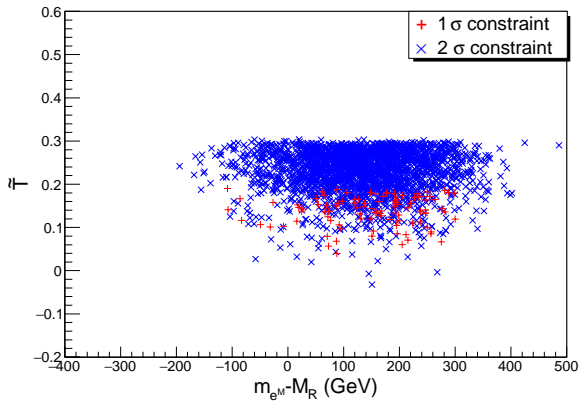}
	\caption{{\small Constrained $\tilde{T}$ versus mass splitting between $e^M$ and $\nu_R$: $(m_{e^M} - M_R)$}}
	\label{TvsmeM-MR}
\end{figure}
From these plots, one can see that the mirror fermions within a doublet generally tend to be non-degenerate although they can as well be degenerate (Fig.~\ref{TvsmeM-MR}). The mass-splittings within a mirror quark generation yield a similar-looking plot, not shown here.
(The statement made in \cite{hung2} which is based on a thorough analysis is more accurate than that made in the original paper \cite{pqnur} concerning the size of that mass splitting.) The interesting phenomenology of the mirror sector is beyond the scope of this manuscript and is under investigation.

Throughout the analysis, one can see the importance of the negative contributions coming from the scalar sector which provides a counterweigh to the positive ones coming from the mirror fermion sector in such a way that the sum of the two stays within the $2\sigma$ constraint. This is certainly not a fine-tuning procedure, since we do not require precise cancellations between the two sectors.
Further details can be found in \cite{hung2}.

An important point which is worth repeating here is the role played by the scalar triplet in regulating the new physics contribution to the $S$ and $T$-parameters. In fact, it has been pointed out in \cite{hung2} that the contribution to $S$ from the scalar triplet can be made increasingly negative by increasing the mass of the doubly-charged Higgs as an example. This can offset the positive contribution coming from the mirror fermions. One can see the importance of the scalar sector, in particular the Higgs triplets, in making the \ewnur model consistent with precision data.

The next step is to examine constraints coming from direct searches of the Higgs boson.

\subsection{Constraint on the ``minimal" \ewnur model from the 125-GeV SM-like Higgs boson}\label{sec:minimalH}

By ``minimal" we mean that the Higgs structure is as described above: one Higgs doublet and two Higgs triplets. Some phenomenology of these scalars has been investigated in \cite{pqaranda}. This topic will be revisited in a future publication. For the purpose of this manuscript, we shall focus on the four neutral states: $H_5^0 $, $H_3^0$, $H_1^0$ and $H_1^{0\prime}$  and in particular $H_3^0$ and $H_1^0$ since the other two do not couple to SM and mirror fermions \cite{hung2}. $H_3^0$ and $H_1^0$ are CP-odd and CP-even respectively. As shown in \cite{hung2}, because of the coupling $g_{H_{1}^0\,q\bar{q}}=-\imath \frac{m_{q} g}{2m_{W} c_{H}}$, the gluon fusion production cross section for $H_1^0$ was estimated to be $\sigma_{EW\nu_R} \agt 49\, \sigma_{SM}$ where the factor $49= (1+6)^2$ takes into account the contributions from the top and mirror quarks. This alone practically ruled out $H_1^0$ as the 125-\gev SM-like scalar. Also, since the coupling to fermions are very similar to that of the SM, modulo the factor $1/\cos \theta_H$, the various branching ratios (BR) are expected to be of the order of those of the SM and the signal strengths ($\mu = (\sigma \times BR)/(\sigma \times BR)_{SM}$) will largely exceed observations. 

It was shown in \cite{hung2} that the CP-odd (pseudoscalar) $H_3^0$ could, with the appropriate choice of parameters, can fit the bill for being the 125-\gev object both in terms of the production cross section and in terms of branching ratios. However, a likelihood analysis ruled this option out by more than 3 $\sigma$s \cite{LHC2}. Although a measurement of the spin and parity of the 125-\gev object is yet to be performed, it is fair to assume that it is more likely to be a $0^+$ state.

As one can see, the reason why the CP-even $H_1^0$ has such a large gluon fusion production cross section (at least 49 times larger than the SM one at at same mass) is because it comes from the Higgs doublet (the real part of the neutral component) which couples to SM fermions {\em as well as} mirror fermions. The loop controlling the gluon fusion production of $H_1^0$ is dominated by the top quark and the mirror quarks giving rise to the factor of 49 mentioned above while it is dominated only by the top quark contribution in the SM. An extension in the Higgs sector of the minimal \ewnur model is needed. This is shown in the next section.

\section{Extended \ewnur model}\label{sec:extewnur}
The simplest extension- and, in fact, the most natural one - of the minimal \ewnur model is to have {\em two} Higgs doublets with one coupled to SM fermions and the other one to mirror fermions. One extra Higgs doublet leads to 4 more degrees of freedom. To prevent cross coupling, a global symmetry will be imposed. Basically, we introduce the following Higgs doublets along with the corresponding global symmetries $U(1)_{SM} \times U(1)_{MF}$:
\begin{eqnarray}
\label{smtrans}
			U(1)_{SM}&:&\Phi_2 \;\rightarrow e^{\imath \alpha_{SM}} \; \Phi_2\nonumber\\
			&&(q_L^{SM}, l_L^{SM}) \;\rightarrow e^{\imath \alpha_{SM}} \; (q_L^{SM}, l_L^{SM})\,,
\end{eqnarray}
		
\begin{eqnarray}
\label{qmtrans}
			U(1)_{MF}&:&\;\Phi_{2M} \;\rightarrow e^{\imath \alpha_{MF}} \; \Phi_{2M}\nonumber\\
			&&\;(q_R^{M}, l_R^{M}) \;\rightarrow e^{\imath \alpha_{MF}} \; (q_R^{M}, l_R^{M})\,,
\end{eqnarray}
\be
		\phi_S \;\rightarrow e^{-\imath(\alpha_{MF} - \alpha_{SM})} \; \phi_S \,,
\ee
\be
		\label{chitrans}
		\tilde{\chi} \;\rightarrow e^{-2\imath \alpha_{MF}} \; \tilde{\chi} \,.
\ee
All other fields ($SU(2)$-singlet right-handed SM fermions, left-handed mirror fermions) are singlets under  $U(1)_{SM}\,\times U(1)_{MF}$. 	
	
These symmetries will forbid, at tree level, Yukawa couplings of the form $g_Y \bar{f}_L \Phi_{2}M f_R$ and $g_Y \bar{f}^{M}_R \Phi_2 f^{M}_L$ . Only Yukawa interactions of the type $g_Y \bar{f}_L \Phi_2 f_R$ and $g_Y \bar{f}^{M}_R \Phi_{2M} f^{M}_L$ are allowed. The Yukawa couplings of the physical states to SM and mirror fermions will involve mixing angles. This is detailed below.

%
\par When a $Y = 1$ complex scalar doublet is added to the minimal \ewnur model, under the global $SU(2)_L \times SU(2)_R$ we have the following scalar fields:
\begin{equation}
	\label{Phi1}
	\Phi_{2}=\left( \begin{array}{cc}
	\phi_{2}^{0,*} & \phi_{2}^{+} \\
	\phi_{2}^{-} & \phi_{2}^0
	\end{array} \right) \,,
\end{equation}
\begin{equation}
	\label{Phi2}
	\Phi_{2M}=\left( \begin{array}{cc}
	\phi_{2M}^{0,*} & \phi_{2M}^{+} \\
	\phi_{2M}^{-} & \phi_{2M}^0
	\end{array} \right) \,,
\end{equation}
and
\begin{equation}
	\label{chi}
	\chi = \left( \begin{array}{ccc}
	\chi^{0} &\xi^{+}& \chi^{++} \\
	\chi^{-} &\xi^{0}&\chi^{+} \\
	\chi^{--}&\xi^{-}& \chi^{0*}
	\end{array} \right) \,.
\end{equation}
Proper vacuum alignment for $SU(2)_L \times U(1)_Y \rightarrow U(1)_{em}$ gives
\begin{equation}
\label{phi1vev}
\langle \Phi_{2} \rangle = \left( \begin{array}{cc}
v_2/\sqrt{2} &0 \\
0&v_2/\sqrt{2} 
\end{array} \right) \,,
\end{equation}
\begin{equation}
\label{phi2vev}
\langle \Phi_{2M} \rangle = \left( \begin{array}{cc}
v_{2M}/\sqrt{2} &0 \\
0&v_{2M}/\sqrt{2} 
\end{array} \right) \,,
\end{equation}
and
\begin{equation}
\label{chivev}
\langle \chi \rangle = \left( \begin{array}{ccc}
v_M &0&0 \\
0&v_M&0 \\
0&0&v_M
\end{array} \right) \,.
\end{equation}
Thus, the VEVs of real parts of $\Phi_2$, $\Phi_{2M}$ and $\chi$ are $(v_2/\sqrt{2})$, $(v_{2M}/\sqrt{2})$ and $v_M$ respectively such that
\be
	v_2^2 + v_{2M}^2 + 8\,v_M^2 = v^2\,,
\ee
where $v \approx 246 ~\gev$. We define
\be\label{eq:sdefs}
	s_2 = \dfrac{v_2}{v}; ~~s_{2M} = \dfrac{v_{2M}}{v}; ~~s_M = \dfrac{2 \sqrt{2}\; v_M}{v}\,.
\ee
Notice that apart from $\xi$, this model contains the same number of triplet ($\widetilde{\chi}$) and the same number of doublets (2: $\Phi_2$, $\Phi_{2M}$) as the L-R model. The big difference comes from the fact that the VEV of the triplet $\widetilde{\chi}$ is bounded from above by the $\Lambda_{EW} \sim 246~\gev$, while the VEV of the L-R triplet is only bounded from below by experiment to be $\gtrsim 3~\tev$ \cite{CMSWR}.

\par A generic $SU(2)_L \times SU(2)_R$ preserving potential for these scalars can now be written as
\begin{widetext}
	\begin{flalign}
	\label{pot0}
		&V(\Phi_{2},\Phi_{2M},\chi) = \lambda_1 \Big[Tr\Phi_{2}^\dagger \Phi_{2} - v_2^2 \Big]^2 + \lambda_2 \Big[Tr\Phi_{2M}^\dagger \Phi_{2M} - v_{2M}^2 \Big]^2 + \lambda_3 \Big[Tr\chi^\dagger \chi - 3 v_M^2\Big]^2&\nonumber\\[0.5em]
		&\; +\lambda_4 \Big[Tr\Phi_{2}^\dagger \Phi_{2} - v_2^2 + Tr\Phi_{2M}^\dagger \Phi_{2M} - v_{2M}^2 + Tr\chi^\dagger \chi - 3 v_M^2 \Big]^2&\nonumber\\[0.5em]
		&\; + \lambda_5 \Big[(Tr\Phi_{2}^\dagger \Phi_{2})\; (Tr\chi^\dagger \chi) - 2\; (Tr\Phi_{2}^\dagger \frac{\tau^a}{2} \Phi_{2} \frac{\tau^b}{2})\; (Tr\chi^\dagger T^a \chi T^b) \Big]&\nonumber\\[0.5em]
		&\; + \lambda_6 \Big[(Tr\Phi_{2M}^\dagger \Phi_{2M})\; (Tr\chi^\dagger \chi) - 2\; (Tr\Phi_{2M}^\dagger \frac{\tau^a}{2} \Phi_{2M} \frac{\tau^b}{2})\; (Tr\chi^\dagger T^a \chi T^b) \Big]&\nonumber\\[0.5em]
		&\; + \lambda_7 \Big[(Tr\Phi_{2}^\dagger \Phi_{2})\; (Tr\Phi_{2M}^\dagger \Phi_{2M})\; - (Tr\Phi_{2}^\dagger \Phi_{2M})\; (Tr\Phi_{2M}^\dagger \Phi_{2}) \Big]
		 +\lambda_8 \Big[3\; Tr\chi^\dagger \chi \chi^\dagger \chi - (Tr\chi^\dagger \chi)^2 \Big]\,.&
	\end{flalign}
\end{widetext}
Note that this potential, like the one in the minimal \ewnur model is also invariant under $\chi \rightarrow -\chi$. Now it is also invariant under the global \uxu symmetry. The vacuum alignment given above breaks the global \suxsu down to the custodial $SU(2)_D$. One still has $M_W = g\,v/2$ and $M_Z = M_W/\cos \theta_W$, but now $v= \sqrt{v_2^2 + v_{2M}^2 + 8\,v_M^2} \approx 246 \, \gev$. It is found that three `massless' Nambu-Goldstone Bosons can be obtained after spontaneous breaking of $SU(2)_L \times U(1)_Y$ to $U(1)_{em}$, when a condition $\lambda_5 = \lambda_6 = \lambda_7$ imposed on the potential above. Thus, the potential that should be used to find the physical Higgs states is
\begin{widetext}
	\begin{flalign}
		\label{eq:pot}
		&V(\Phi_{2},\Phi_{2M},\chi) = \lambda_1 \Big[Tr\Phi_{2}^\dagger \Phi_{2} - v_2^2 \Big]^2 + \lambda_2 \Big[Tr\Phi_{2M}^\dagger \Phi_{2M} - v_{2M}^2 \Big]^2 + \lambda_3 \Big[Tr\chi^\dagger \chi - 3 v_M^2\Big]^2&\nonumber\\[0.5em]
		&\; +\lambda_4 \Big[Tr\Phi_{2}^\dagger \Phi_{2} - v_2^2 + Tr\Phi_{2M}^\dagger \Phi_{2M} - v_{2M}^2 + Tr\chi^\dagger \chi - 3 v_M^2 \Big]^2&\nonumber\\[0.5em]
		&\; + \lambda_5 \Big[(Tr\Phi_{2}^\dagger \Phi_{2})\; (Tr\chi^\dagger \chi) - 2\; (Tr\Phi_{2}^\dagger \frac{\tau^a}{2} \Phi_{2} \frac{\tau^b}{2})\; (Tr\chi^\dagger T^a \chi T^b) + (Tr\Phi_{2M}^\dagger \Phi_{2M})\; (Tr\chi^\dagger \chi)&\nonumber\\[0.5em]
		&\; - 2\; (Tr\Phi_{2M}^\dagger \frac{\tau^a}{2} \Phi_{2M} \frac{\tau^b}{2})\; (Tr\chi^\dagger T^a \chi T^b) + (Tr\Phi_{2}^\dagger \Phi_{2})\; (Tr\Phi_{2M}^\dagger \Phi_{2M})\; - (Tr\Phi_{2}^\dagger \Phi_{2M})\; (Tr\Phi_{2M}^\dagger \Phi_{2}) \Big]& \nonumber\\[0.5em]
		&\; +\lambda_8 \Big[3\; Tr\chi^\dagger \chi \chi^\dagger \chi - (Tr\chi^\dagger \chi)^2 \Big]\,.&
	\end{flalign}
\end{widetext}
After the spontaneous breaking of $SU(2)_L \times U(1)_Y \rightarrow U(1)_{em}$, besides the three Nambu-Goldstone Bosons, there are {\em twelve} physical scalars grouped into $5 ~+ ~3 ~+ ~3 ~+ ~1$ of the custodial $SU(2)_D$ with 3 custodial singlets.
%
%
%
To express the Nambu-Goldstone Bosons and the physical scalars let us adopt the following convenient notation:
\begin{eqnarray}
	v &=& \sqrt{v_2^2 + v_{2M}^2 + 8 v_M^2}\,,\nonumber\\[0.5em]
	s_2 &=& \frac{v_2}{v}\,,\; s_{2M} = \frac{v_{2M}}{v}\,,\; s_M = \frac{2\sqrt{2}\;v_M}{v}\,, \nonumber\\[0.5em]
	c_2 &=& \frac{\sqrt{v_{2M}^2 + 8 v_M^2}}{v}\,,\; c_{2M} = \frac{\sqrt{v_2^2 + 8 v_M^2}}{v}\,,\nonumber\\[0.5em]
	c_M &=& \frac{\sqrt{v_2^2 + v_{2M}^2}}{v}\,.
\end{eqnarray}
Thus,
\begin{equation}
	s_2^2 + c_2^2 = s_{2M}^2 + c_{2M}^2 = s_M^2 + c_M^2 = 1\,.
\end{equation}
[In the limit $s_{2M} \rightarrow 0$ i.e. the extended EW$\nu_R \rightarrow minimal~\text{EW}\nu_R$, $s_M \rightarrow s_H$ and $c_M \rightarrow c_H$]. Let us also define, like we did for the minimal \ewnur model,:
	\begin{eqnarray}\label{eq:subsid}
		\phi_{2}^0 &\equiv& \frac{1}{\sqrt{2}} \Big(v_2 + \phi_{2}^{0r} + \imath \phi_{2}^{0\imath}\Big),\nonumber\\[0.5em]
		\phi_{2M}^0 &\equiv& \frac{1}{\sqrt{2}} \Big(v_{2M} + \phi_{2M}^{0r} + \imath \phi_{2M}^{0\imath}\Big),\nonumber\\[0.5em]
		\chi^0 &\equiv& v_M + \frac{1}{\sqrt{2}} \Big(\chi^{0r} + \imath \chi^{0\imath}\Big);\\[0.5em]
		\psi^\pm &\equiv& \frac{1}{\sqrt{2}} \Big(\chi^\pm + \xi^\pm \Big),\hspace{1em}\zeta^\pm \equiv \frac{1}{\sqrt{2}} \Big(\chi^\pm - \xi^\pm\Big)\,,
	\end{eqnarray}
for the complex neutral and charged fields respectively. With these fields the Nambu-Goldstone Bosons are given by
\begin{eqnarray}
	G_3^\pm &=& s_2 \phi_{2}^\pm + s_{2M} \phi_{2M}^\pm + s_M \psi^\pm,\nonumber\\[0.5em]
	G_3^0 &=& \imath \Big(- s_2 \phi_{2}^{0\imath} - s_{2M} \phi_{2M}^{0\imath} + s_M \chi^{0\imath}\Big).
\end{eqnarray}
The physical scalars can be grouped, as stated in the previous section, based on their transformation properties under $SU(2)_D$ as follows:
	\begin{eqnarray}
		\text{five-plet (quintet)} &\rightarrow& H_5^{\pm\pm},\; H_5^\pm,\; H_5^0;\nonumber\\[0.5em]
		\text{triplet} &\rightarrow& H_{3}^\pm,\; H_{3}^0;\nonumber\\[0.5em]
		\text{triplet} &\rightarrow& H_{3M}^\pm,\; H_{3M}^0;\nonumber\\[0.5em]
		\text{three singlets} &\rightarrow& H_1^0,\; H_{1M}^0,\; H_1^{0\prime}\,,
	\end{eqnarray}
where
	\begin{eqnarray}\label{eq:higgs}
		H_5^{++} &=& \chi^{++},\; H_5^+ = \zeta^+,\; H_5^0 = \frac{1}{\sqrt{6}}\Big(2\xi^0 - \sqrt{2}\chi^{0r}\Big),\nonumber\\[0.5em]
		H_3^+ &=& - \frac{s_2 s_M}{c_M} \phi_{2}^+ - \frac{s_{2M} s_M}{c_M} \phi_{2M}^+ + c_M \psi^+,\nonumber\\[0.5em]
		H_3^0 &=& \imath \left(\frac{s_2 s_M}{c_M} \phi_{2}^{0i} + \frac{s_{2M} s_M}{c_M} \phi_{2M}^{0i} + c_M \chi^{0i}\right),\nonumber\\[0.5em]
		H_{3M}^+ &=& - \frac{s_{2M}}{c_M} \phi_{2}^+ + \frac{s_2}{c_M} \phi_{2M}^+,\nonumber\\[0.5em]
		H_{3M}^0 &=& \imath \left(- \frac{s_{2M}}{c_M} \phi_{2}^{0i} + \frac{s_2}{c_M} \phi_{2M}^{0i}\right),\nonumber\\[0.5em]
		H_1^0 &=& \phi_{2}^{0r},\; H_{1M}^0 = \phi_{2M}^{0r},\nonumber\\[0.5em]
		H_1^{0\prime} &=& \frac{1}{\sqrt{3}} \Big(\sqrt{2}\chi^{0r} + \xi^0\Big)\,,
	\end{eqnarray}
with phase conventions $H_5^{--} = (H_5^{++})^\ast$, $H_5^- = -(H_5^+)^\ast$, $H_{3}^- = -(H_{3}^+)^\ast$, $H_{3M}^- = -(H_{3M}^+)^\ast$, $H_{3}^0 = -(H_{3}^0)^\ast$ and $H_{3}^0 = -(H_{3}^0)^\ast$. The masses of these physical scalars can easily be obtained from Eq. (\ref{eq:pot}). Since, the potential preserves the $SU(2)_D$ custodial symmetry, members of the physical scalar multiplets have degenerate masses. These masses are
\begin{eqnarray}\label{eq:multiplet}
	m_5^2 &=& 3 ( \lambda_5 c_M^2 + \lambda_8 s_M^2 ) v^2\,,\nonumber\\[0.5em]
	m_{3}^2 &=& \lambda_5 v^2\,,\;\;\; m_{3M}^2 = \lambda_5 ( 1 + c_M^2 ) v^2\,.
\end{eqnarray}
In general, the $H_1^0$, $H_{1M}^0$ and $H_1^{0 \prime}$ can mix according to the mass-squared matrix
\bea\label{eq:m2sing}
	&&\mathcal{M}_{\text{singlets}}^2 = v^2 \times\nonumber\\[0.5em]
	&&\left( 
	\begin{array}{ccc}
		8 s_2^2 (\lambda_1 + \lambda_4)	&	8 s_2 s_{2M} \lambda_4				&	2 \sqrt{6} s_2 s_M \lambda_4\\[0.5mm]
		8 s_2 s_{2M} \lambda_4				&	8 s_{2M}^2 (\lambda_2 + \lambda_4)	&	2 \sqrt{6} s_{2M} s_M \lambda_4\\[0.5mm]
		2 \sqrt{6} s_2 s_M \lambda_4		&	2 \sqrt{6} s_{2M} s_M \lambda_4		&	3 s_M^2 (\lambda_3 + \lambda_4)
	\end{array}
	\right)\,.
\eea
Hence, after the electroweak symmetry breaking generic $SU(2)_D$ singlet mass eigenstates are given by:
\be\label{eq:Hmasseigen}
	\left( \begin{array}{l}		\Htil\\ \Htill\\ \Htilll\end{array}\right) = 
	\left( \begin{array}{ccc}  	a_{1,1} 		& a_{1,1M} 		& a_{1,1'} \\
								a_{1M,1} 		& a_{1M,1M} 		& a_{1M,1'} \\
								a_{1',1} 		& a_{1',1M} 		& a_{1',1'}
	\end{array} \right) ~
	\left( \begin{array}{l} 		H_1^0\\ H_{1M}^0\\ H_1^{0\prime}\end{array}\right)\,.
\ee
We denote the mass eigenstates by $\widetilde{H}$, $\widetilde{H}^\prime$, and $\widetilde{H}^{\prime\prime}$. We adopt a convention of denoting the lightest of the three by $\Htil$, the next heavier one by $\Htill$ and the heaviest state by $\Htilll$. Thus, the $125$-$\gev$ candidate in the \ewnur model will be denoted by $\Htil$. Here the $3\times 3$ matrix on the right-hand side lists mixings between $H_1^0$, $H_{1M}^0$ and $H_1^{0\prime}$. 

To obtain the mass eigenstates, including the 125-\gev candidate in this model, we need to diagonalize the mass matrix in Eq. (\ref{eq:m2sing}). Due to the complicated dependence of $\mathcal{M}_{\text{singlets}}$ on many parameters, it is necessary to diagonalize Eq.~(\ref{eq:m2sing}) numerically. As a result, the elements of the (numerical) eigenvector matrix $\{a_{ij}\}$ depend on:
\bi
	\item
	Ratios of VEVs: $s_2$, $s_{2M}$, $s_M$, and
	\item
	Scalar couplings: $\lambda_1$, $\lambda_2$, $\lambda_3$, $\lambda_4$.
\ei
It should be noted that in the limit $\lambda_4 \rightarrow 0$ the off-diagonal elements in $\mathcal{M}_{\text{singlets}}$ vanish. 

Such mixed physical scalars have interesting implications for decay properties of the 125-\gev as well as the two heavier mass eigenstates. This is where we will focus on in the next two sections.

\begin{flushleft}
{\bf A note on the pseudo Nambu-Goldstone bosons in the \ewnur model:}
\end{flushleft}

Last but not least is the important question concerning whether or not Nambu-Goldstone (NG) bosons arise in the spontaneous breakdown of the aforementioned global symmetry $U(1)_{SM} \times U(1)_{MF}$. This issue has been discussed in \cite{ChanoGold} as well as in the original \ewnur minimal model \cite{pqnur}. The conclusion was the fact that there is a term in the potential which is required for proper vacuum alignment and which breaks {\em explicitly} the global symmetry ($U(1)_{MF}$ for the minimal model) and the would-be NG bosons acquire a mass proportional to the corresponding couplings. For the sake of clarity and also because we now also have $U(1)_{SM}$ in addition to $U(1)_{MF}$, we will briefly review the essential points discussed in \cite{ChanoGold} and \cite{pqnur}.

Let us first look at the minimal model of \cite{pqnur} where $\Phi_{2M}$ is absent. Eq.~({\ref{pot0}) without $\Phi_{2M}$ is of the same form as that used in \cite{ChanoGold} where it has been shown that the term containing $\lambda_5$ is needed for proper vacuum alignment of $\Phi_{2}$ and $\chi$ so that $SU(2) \times U(1)_Y \rightarrow U(1)_{em}$.  In this minimal model, only the mirror fermions, $\tilde{\chi}$ and $\phi_S$ transform non-trivially under $U(1)_{MF}$ as shown in Eq.~(\ref{qmtrans}) and Eq.~(\ref{chitrans}) with all other fields being singlets under $U(1)_{MF}$ including the SM fermions, $\Phi_{2}$ and $\xi^{\pm,0}$. The term proportional to $\lambda_5$ contains mixings such as $\xi^{0} \chi^{0}$, $\xi^{+} \chi^{-}$ and breaks explicitly the $U(1)_{MF}$ symmetry. The would-be triplet of NG bosons, $H_{3}^{\pm,0}$, acquire a mass proportional to $\lambda_5$ as shown in Eq.~(\ref{eq:multiplet}). The proper vacuum alignment ($\lambda_5 \neq 0$) prevents the appearance of NG bosons in the minimal model! 

The above argument can be generalized to the extended model with $U(1)_{SM}\,\times U(1)_{MF}$. The various fields transform under $U(1)_{SM}\,\times U(1)_{MF}$ as shown in Eq.~(\ref{smtrans}), Eq.~(\ref{qmtrans}) and Eq.~(\ref{chitrans}). Again, one notices that proper vacuum alignment requires the term proportional to $\lambda_5$ to be present as discussed above. From Eq.~(\ref{eq:pot}), one can again see that this term contains a mixing between $\xi$ and $\tilde{\chi}$ and now also a mixing between $\Phi_{2M}$ and $\Phi_{2}$. As a result, $U(1)_{SM}\,\times U(1)_{MF}$ is broken explicitly by the term proportional to $\lambda_5$. Consequently, the would-be NG bosons, $H_{3}^{\pm,0}$ and $H_{3M}^{\pm,0}$, now acquire a mass proportional to $\lambda_5$ as shown in Eq.~(\ref{eq:multiplet}). The proper vacuum alignment ($\lambda_5 \neq 0$) prevents the appearance of NG bosons in the extended model as well!

\section{125-GeV candidate in \ewnur Model}


Measured properties of the 125-\gev scalar particle that was discovered at the LHC so far tend to be close to the properties of SM Higgs boson. Hence, in every model of BSM Physics it is imperative to (i) have at least one Higgs particle with mass of about 125 \gev having SM-like decay signal strengths, and (ii) study the implications of these properties in the `allowed' parameter space of the model (e.g. allowed masses of any BSM particles in the model, etc.). To check the viability of a model or to search for the model experimentally, decay properties of the 125-\gev Higgs boson candidate in the model must be studied.

We denote the $125$-$\gev$ candidate in the \ewnur model by $\Htil$. From Eq.~(\ref{eq:Hmasseigen}) we see that it is a mixture of $H_1^0$, $H_{1M}^0$ and $H_1^{0\prime}$. Recall that $H_1^0$ comes from the SM-like scalar doublet $\Phi_2$, $H_{1M}^0$ comes from doublet $\Phi_{2M}$ and $H_1^{0\prime}$ from triplet $\chi$. 

Because the measured decay signal strengths of the 125-\gev Higgs boson are close to SM predictions, intuitively one might expect that $H_1^0$ has to be the dominant component of $\Htil$. But our investigation shows that the 125-\gev $\Htil$ can have SM-like decay signal strengths, even if $H_1^0$ is a sub-dominant component in it. Hence the dual-like nature of 125-\gev Higgs boson from perspective of the \ewnur model. In this section we will discuss this dual-like nature and its implications.

In the first subsection we will explain the methodology used in the analysis; the next subsection presents the analysis and results for $\Htil\sim H_1^0$ case. In the third subsection, we present a more interesting case, where, although $H_1^0$ is a subdominant component of $125$-$\gev$ $\Htil$, it still satisfies the experimental constraints on the signal strengths. 

\begin{flushleft}
{\bf A note on CP of the 125-GeV Higgs in the \ewnur model:}
\end{flushleft}
As seen in Sec.~\ref{sec:extewnur}, the \ewnur model has 6 neutral physical scalars, of which 3 are CP-even states ($H^0_1$, $H^0_{1M}$, $H^\prime_1$) and 3 are CP-odd states ($H^0_3$, $H^0_{3M}$, $H^0_5$). Their couplings to fermions and gauge bosons are listed in Tables~\ref{table:h_ferm} -~\ref{table:feyn22n} in Appendix~\ref{sec:extewnurcoup}. Among these $H^0_5$ does not couple to charged fermions. 

It can be seen from Table~\ref{table:h_ferm} that decay widths of $H^0_3/H^0_{3M} ~\rightarrow f ~\bar{f}$ can be close to the SM predictions for some combinations of the BSM parameters in the couplings. But as mentioned in Section~\ref{sec:minimalH}, $H^0_3, ~H^0_{3M}$ are disfavored as $125$-$\gev$ candidates as compared to the CP-even hypothesis \cite{LHC2}. 

Hence, in this paper while considering 125-\gev candidate in the \ewnur model, we proceed with the hypothesis that this candidate is a CP-even eigenstate \footnote{The possibility that the 125-\gev Higgs boson is a linear combination of CP-even and CP-odd state has not been thoroughly checked experimentally yet. The spin and parity of the 125-\gev scalar are yet to be measured at CMS and ATLAS. Thus, in this paper, we will stick to CP-eigenstate hypothesis based on the likelihood analysis}. 

Out of the 3 CP-even Higgs bosons, only $H^0_1$ can have decay widths to SM fermions similar to the SM predictions. Therefore, one might expect that in the \ewnur model $H^0_1$ is the candidate for 125-\gev Higgs boson. However, in the absence of explicit decay widths and based on current available signal strengths for various decay channels, we will show that the 125-\gev Higgs boson can be very different from the standard expectation. It is in the spirit of our analysis that we may coin the term ``Dr.~Jekyll" to the Standard Model expectation (a mild impostor) and the term ``Mr.~Hyde" to the definite ``impostor" (which mainly comes from the scalar triplet) scenario presented in the paper. 

\subsection{Methodology for comparing the \ewnur model predictions with data}\label{sec:data}

For any given decay channel of a Higgs, CMS and ATLAS experiments at the LHC measure the total cross section of the decay process. 
The cross section of any decay channel of the Higgs boson that is measured at the LHC is given by
\be
	\sigma(H \text{-decay}) = \sigma(H \text{-production}) \times BR(H \text{-decay})\,,
\ee
where $\sigma(H \text{-production})$ is the production cross section of $H$ and $BR(H \text{-decay})$ is the Branching Ratio of the decay channel of $H$ that is under consideration.
\be
	BR(H \text{-decay}) = \frac{\Gamma(H \text{-decay})}{\Gamma_H}\,,
\ee
where $\Gamma(H \text{-decay})$ is the partial width of the $H$-decay channel, and $\Gamma_H$ is the total width of $H$. To compare the data with the Standard Model predictions, the ratio of the measured signal strength to its SM-predicted value is presented, denoted by $\mu$. 
\be\label{eq:mudef}
	\mu(H \text{-decay}) = \frac{\sigma(H \text{-decay})}{\sigma_{SM}(H \text{-decay})}\,,
\ee
$\sigma(H \text{-decay})$ being measured experimentally or predicted by a model.

Therefore, to compare the EW$\nu_R$-predicted decay signal strengths with the data, we investigate the agreement between the ratio of EW$\nu_R$ prediction with the SM prediction $\mu_{\text{\ewnur}} = \sigma_{\text{\ewnur}}(H \text{-decay}) / \sigma_{SM}(H \text{-decay})$, to the ratio of measured decay cross section with the SM-prediction $\mu_{data} = \sigma_{data}(H \text{-decay}) / \sigma_{SM}(H \text{-decay})$. $\sigma_{H \text{-decay}}$ in the \ewnur model with the predictions SM for that decay channel. 

Hence, we need to calculate 
\bi
	\item
	partial decay widths for these channels, 
	\item
	the total width of $\Htil$ and
	\item
	the production cross-section of $gg\rightarrow\Htil$.
\ei

The analysis is done in the following steps:
\begin{enumerate}
	\item
	Identify all the decay channels that contribute significantly to the total width of the 125-\gev $\Htil$
	Identify the variables on which the aforementioned three quantities depend.
	\item
	Identify the limits on the variables.
	\item
	Select a set of values for the variables within their respective limits.
	\item
	Calculate the signal strengths $\mu$ in various channels for the $125$-$\gev$ $\Htil$, and compare them with the measured values from CMS.
\end{enumerate}
%

\subsubsection{Decay channels under consideration}

For this analysis we calculate signal strengths $\mu$ for decay channels $\Htil\rightarrow ZZ,~W^+W^-,~\gamma\gamma,~b\bar{b},~\tau\bar{\tau}$. We calculate the production cross-section and partial widths of various decay channels as explained in Appendix~\ref{sec:hdecay}. The total width of the 125-\gev $\Htil$ is calculated by adding individual partial decay widths:
\bea\label{eq:Hwidth}
\Gamma_{\Htil} = &&~\Gamma_{\Htil\rightarrow b\bar{b}} ~+ ~\Gamma_{\Htil\rightarrow \tau\bar{\tau}} ~+ ~\Gamma_{\Htil\rightarrow c\bar{c}} ~+ ~\Gamma_{\Htil\rightarrow W^+ W^-}\nonumber\\[0.5em]
&& + ~\Gamma_{\Htil\rightarrow ZZ} ~+ ~\Gamma_{\Htil\rightarrow gg} ~+ ~\Gamma_{\Htil\rightarrow \gamma \gamma}\,.
\eea
Among all the partial widths considered above, $\Gamma_{\Htil\rightarrow b\bar{b}}$ and $\Gamma_{\Htil\rightarrow W^+ W^-}$ are the most dominant for the SM-Higgs. Because of the constraint $m_{f^M} > 100~\gev$, the decay channel $\Htilde \rightarrow f^M \bar{f}^M$ does not occur at the leading order, when $f^M$ is on-shell.

In what follows we identify the relevant variables in the analysis and estimate their allowed ranges.

\subsubsection{Lower limit on the masses of charged mirror fermions}

The lower limit of $102 ~\gev$ on the masses of charged mirror leptons and mirror quarks is imposed based on the results of search for sequential heavy charged leptons and quarks at LEP3 (refer `Heavy Charged Leptons' and `Heavy Quarks' sections in \cite{pdg} and references therein). Strictly speaking these constraints apply only to sequential heavy fermions, such as $L' \rightarrow \tau Z \rightarrow \tau l\bar{l}, \tau q \bar{q}$ or $Q^\prime\rightarrow bZ\rightarrow b~ l\bar{l}, b~ q \bar{q}$ or $Q^\prime\rightarrow bW^+\rightarrow b~ l\bar{\nu}, ~b~q\bar{q^\prime}$ etc.

However, charged mirror fermions in the \ewnur model couple to the SM fermions in an altogether different way, through the scalar singlet $\phi_S$ \cite{pqnur,pqnur2}: $q^M\rightarrow q\phi_S$, $l^M\rightarrow l\phi_S$. This $\phi_S$ would appear as missing energy in the detector. Thus, the signature of final states   in charged mirror fermion decay would involve a lepton + missing $E_T$ or a jet + missing $E_T$. Moreover, at CMS or ATLAS these decays could occur outside the beam-pipe and inside the silicon vertex detector \cite{pqnur,pqnur2}. Therefore, the constraints from the aforementioned searches do not directly apply to charged mirror fermions. We still impose these constraints on charged mirror fermions, arguing that if these mirror fermions were lighter than $\sim 100 ~\gev$, they would have been discovered at 200 \gev LEP3 \cite{pdg}.

\subsubsection{Limits on VEVs, scalar and Yukawa couplings}

We consider only the cases where the scalar couplings and Yukawa couplings of mirror fermions are perturbative. The perturbative constraint on scalar and Yukawa couplings are $\lambda_i/4\pi \lesssim \mathcal{O}(1)$ and $\alpha_{f^M} = g_{MF}^2/4\pi \lesssim \mathcal{O}(1)$ respectively. For numerical analysis we limit ourselves to cases, where $\lambda_i /4\pi\leq 1.3$ and $\alpha_{f^M} \leq 1.5$.

As discussed towards the end of Sec.~\ref{sec:extewnur}, the $SU(2)_D$ singlet mass eigenstates depend on $s_2$, $s_{2M}$ and $s_M$. Therefore, they also depend on the vacuum expectation values (VEVs) of the real parts of $\Phi_2$, $\Phi_{2M}$ and $\chi$. While investigating different numerical forms of $\{a_{ij}\}$, one needs to vary the VEV's. Hence, it is necessary to estimate the limits on these VEVs before analyzing the 125-\gev candidate in detail. 

Recall that the charged SM fermions, the charged mirror fermions and the right handed neutrinos get their masses due to $v_2$, $v_{2M}$, and $v_M$ respectively. Various constraints on these masses constrain the ranges of the VEVs. 

If the pole mass of top quark ($173.5~\gev$), the heaviest SM fermion, is {\em perturbative} and comes from $v_2$, then $v_2 \gtrsim 69 ~\gev$ (because $g^2_{top} \leq 4 \pi$). We set the lower bound on the masses of all the charged mirror fermions at $102~\gev$, which is the LEP3 \cite{pdg} bound on the heavy BSM quarks and BSM charged leptons. Hence, considering a constraint of $g^2_{MF}/4\pi \leq 1.5$ on the Yukawa couplings of all the charged mirror fermions, $v_{2M} \gtrsim 27 ~\gev$, implying $v_M \lesssim 80 ~\gev$. Thus, for perturbative determination of $M_R$ requires $M_R \lesssim 283 ~\gev$. We also know that $M_R \geq M_Z / 2\; \approx\; 45.5 ~\gev$ \cite{pqnur}, and, hence, $v_M \gtrsim 13 ~\gev$. This implies that $v_2 \lesssim 241 ~\gev$ and $v_{2M} \lesssim 233 ~\gev$. This limit on $v_{2M}$ along with the perturbative limit on $g_{MF}$ sets an upper limit on the masses of the mirror fermions: $m_{MF} \lesssim 715 ~\gev$. The allowed ranges for VEVs and for parameters defined in Eq~(\ref{eq:sdefs}) are summarized in the table below.
\begin{table}[!htb]
	\renewcommand{\arraystretch}{2}
	\centering
	\caption{\label{tab:v_ranges}Allowed ranges of VEVs and parameters defined in Eq. (\ref{eq:sdefs}). All values are given in \gev.}
	\begin{tabular}{l|l}
	\hline
	$69 ~\lesssim ~v_2 ~~\lesssim ~241$ & $~0.28 ~\lesssim ~s_2 ~~\lesssim ~0.98$\\
	$33 ~\lesssim ~v_{2M} \lesssim ~233~~$ & $~0.13 ~\lesssim ~s_{2M} \lesssim ~0.95$\\
	$13 ~\lesssim ~v_M ~\lesssim ~~83$ & $~0.15 ~\lesssim ~s_M ~\lesssim ~0.95$\\
	\hline
	\end{tabular}
\end{table}
%

\subsubsection{Common predictions for multiple decay channels}

In the \ewnur model, predictions for the signal strengths of $\Htil\rightarrow W^+W^-$ and $\Htil\rightarrow ZZ$ are equal. Similarly, predictions for the signal strengths of $\Htil\rightarrow b\bar{b}$ are equal to those for $\Htil\rightarrow\tau\bar{\tau}$. This is expected, since as seen in Appendix~\ref{sec:hdecay},
\bea\label{eq:ww_zz}
	\frac{\Gamma^{EW\nu_R}(\Htil\rightarrow W^+W^-)}{\Gamma^{SM}(H_{SM}^0\rightarrow W^+W^-)} &=& \frac{\Gamma^{EW\nu_R}(\Htil\rightarrow ZZ)}{\Gamma^{SM}(H_{SM}^0\rightarrow ZZ)}\,,\nonumber\\[0.5em]
	\frac{\Gamma^{EW\nu_R}(\Htil\rightarrow b\bar{b})}{\Gamma^{SM}(H_{SM}^0\rightarrow b\bar{b})} &=& \frac{\Gamma^{EW\nu_R}(\Htil\rightarrow \tau\bar{\tau})}{\Gamma^{SM}(H_{SM}^0\rightarrow \tau\bar{\tau})}\,.
\eea

Keeping all this in mind, in the next two subsections we analyze in detail the decay properties of the 125-\gev candidate in the \ewnur model.
%

\subsubsection{Numerical Analysis}\label{sec:numerical}

For this analysis a C++ code was written, also using some functionality of ROOT \cite{ROOT}. We investigated this case in following steps:
\begin{itemize}
	\item
	We generated random combinations of $s_2,~s_{2M},~s_M,~\lambda_1,~\lambda_2,~\lambda_3$ and $\lambda_4$, using TRandom3 random number generator in ROOT. These parameters were varied over the following ranges:
\bea\label{eq:hyde125para}
	-4 \pi ~~\leq ~~\lambda_1, ~&\lambda_2, ~\lambda_3&, ~\lambda_4 ~~\leq ~~4 \pi\,,\nonumber\\[0.5em]
	0.28 ~~\leq ~~&s_2& ~~\leq ~~0.98\,,\nonumber\\[0.5em]
	0.13 ~~\leq ~~&s_{2M}& ~~\leq ~~0.95\,,\nonumber\\[0.5em]
	0.15 ~~\leq ~~&s_M& ~~\leq ~~0.95\,.
\eea
The limits $|\lambda|/4\pi \lesssim 1$ are set so that $\lambda$'s are perturbative. Limits on $s_2$, $s_{2M}$, $s_M$ are based on Table~\ref{tab:v_ranges}.
	\item
	We numerically diagonalized the singlet mass matrix in Eq.~(\ref{eq:m2sing}) formed by every combination of the parameters to find the mass eigenvalues and corresponding eigenvector matrix (mixing matrix) in Eq.~(\ref{eq:Hmasseigen}). Only those combinations of parameters, which yielded the lightest mass eigenvalue in the range $125.7\pm 1.0~\gev$, were saved. 4 million such parameter combinations were found. 
	\item
	For all the saved combinations we calculated various signal strengths for each of these combinations. The gluon-gluon fusion channel was considered to calculate the predicted production cross section of the $\Htil$. The partial decay widths were calculated according to Appendix~\ref{sec:hdecay}, and the total width was calculated using Eq.~(\ref{eq:Hwidth}).
	\item
	In addition to the parameters in Eq~(\ref{eq:hyde125para}), following parameters are required to calculate the partial widths of $\Htilde\rightarrow \gamma\gamma$ and $\Htilde\rightarrow g g$, and the cross section of $g g \rightarrow \Htilde$:
\bea\label{eq:hydemupara}
	0 ~~&\leq& ~~\lambda_5 ~~\leq ~~15\,, \text{ varied with }\Delta\lambda_5 \sim 1.07\,,\nonumber\\[0.5em]
	\lambda_8 &=& -1\,, ~m_{H_3^+} = ~m_{H_{3M}^+} = 500 ~\gev\,,\nonumber\\[0.5em]
	m_{H_5^+} &=& 200 ~\gev\,, ~m_{H_5^{++}} = 320 ~\gev\,, ~m_{q_3^M} = 120 ~\gev\,,\nonumber\\[0.5em]
	m_{q_1^M} &=& m_{q_2^M} = m_{l^M} = 102 ~\gev\,.
\eea
	\item
	We checked if the signal strengths $\mu$'s of the 125-\gev $\Htil$ in various decay channels are within the $1\sigma$ constraints on the signal strengths, as measured by CMS experiment. We did not impose constraints from both the CMS and ATLAS, because for some of the decay channels considered here, the signal strength measurements from CMS and ATLAS do not agree with each other within the $1\sigma$ constraints. Also, CMS and ATLAS have not published their combined measurements from the recent analyses. We therefore chose to check agreement with the CMS measurements. 
	
	Depending on their $1\sigma$ constraints, certain combinations out of the 4 million would agree with either only with CMS or with ATLAS results. Thus, imposing the constraints from ATLAS would discard some of the combinations that the CMS constraints would allow and vice versa. However, this would not change any of the conclusions of the paper.
	\item
	We found 1501 out of 4 million combinations of the parameters that satisfy $1\sigma$ constraints from CMS on the 125-\gev Higgs signal strengths in $WW$, $ZZ$, $b\bar{b}$, $\tau\bar{\tau}$ and $\gamma\gamma$ decay channels. Table~\ref{tab:hyde} lists 16 examples out of 1501 cases, with the masses of $\Htil$, $\Htill$, $\Htilll$, their mixing-matrix elements, and the signal strengths of the 125-\gev $\Htil$ for various decay channels. 
	\item
	In the code, there was no constraint imposed as to what is to be the dominant component in $\Htil$. Interestingly, hardly any combinations among the 4 million had $H_1^0$ as a dominant component in the 125-\gev $\Htil$. This means that either
	\begin{enumerate}
		\item
		at the mass of about 125-\gev, 4 million combinations do not yield enough resolution in the parameter space so as to find the $\Htil\sim H_1^0$ case, OR
		\item
		the $\Htil\sim H_1^0$ case cannot be found with the imposed limits on the parameters, and it requires at least some of these parameters to have values outside of these limits. 
	\end{enumerate}
	\item
	Thus, this scan of the parameter space only yielded {\em Mr.~Hyde} cases, where the SM-like $H_1^0$ is a subdominant component in the 125-\gev $\Htil$. Implications of these cases will be further discussed in section \ref{sec:hyde}.
	\item
	On the other hand, to find the combinations of the parameters for which the 125-\gev $\Htil$ has a dominant SM-like $H_1^0$ component, and which also satisfy the CMS constraints on the signal strengths, we had to choose some of the scalar couplings to have values outside $[-4\pi,~4\pi]$. These {\em Dr.~Jekyll} cases thus require some interactions within the scalar sector to be in the strong-coupling regime. In the next subsection we discuss this scenario in detail.

\ei


\subsection{$\Htil$ as 125-GeV Higgs candidate with a dominant SM-like component}\label{sec:jekyll}


We illustrate the step-by-step process which we followed to analyze this case. 
\bi
	\item
	A Mathematica code was written to numerically diagonalize the custodial-singlet mass matrix in Eq.~(\ref{eq:m2sing}) and obtain its mass eigenvalues and eigenvector matrix i.e. the mixing matrix in Eq.~(\ref{eq:Hmasseigen}). 
	\item
	In this code, the values of $s_2 = 0.92, ~s_{2M} = 0.16$ (and thus, $s_M \approx 0.36$) were fixed. The analysis was performed for different $s_2$ values, but, for $\Htil \sim H_1^0$, only the cases with $s_2 \gtrsim 0.9$ were found to satisfy the experimental constraints on the signal strengths of the 125-\gev Higgs at LHC.
	\item
	After fixing $s_2$ and $s_{2M}$, the scalar couplings $\lambda_1$, $\lambda_2$, $\lambda_3$ and $\lambda_4$ were manually varied so that $|\lambda|/4\pi \leq 1.3$, in order to find the combinations of $\lambda$'s that yield the lowest eigenvalue of the mass matrix to be $125.7\pm 1.0 ~\gev$ and the corresponding eigenstate to have dominant $H_1^0$ component.
	\item
	Recall (refer to Eq.~(\ref{eq:pot})) that $\lambda_1$, $\lambda_2$ and $\lambda_3$ are the self-couplings of $\Phi_2$, $\Phi_{2M}$ and $\chi$ respectively. $\lambda_5$ is the measure of cross couplings of $\Phi_2$, $\Phi_{2M}$ and $\chi$.
	\item
	As stated in section \ref{sec:numerical}, we found combinations of the parameters which satisfy the CMS constraints on the signal strengths, when $\lambda_2, ~\lambda_5 > 4\pi$. $|\lambda_1|, ~|\lambda_4|, ~|\lambda_8|$ are still $\leq 4\pi$, while $\lambda_3 \approx 15$. For illustrative purpose, we show below two of many cases which satisfy the CMS constraints.
	\item
The calculation of the partial width of the $\Htil~\rightarrow~\gamma\gamma$ channel necessitates fixing the values or ranges for the remaining parameters. In the example cases shown below we fix other parameters as follows:
\bi
	\item
	$m_{H_3^+} = 600 ~\gev, ~m_{H_{3M}^+} = 700 ~\gev$,
	\item
	masses of all three charged mirror leptons $m_{l^M} = 102 ~\gev$,
	\item
	mass of lightest two generations of mirror quarks $m_{q_1^M} = m_{q_2^M} = 102 ~\gev$,
	\item
	for the purpose of partial widths of $\Htil$-decays in scenarios above, we also fix mass of the third mirror quark generation at $m_{q^M} = 120 ~\gev$. This mass will be varied to analyze constraints on $\Htil \sim H_{1M}^0$.
\ei
	\item
The values of $m_{H_3^+}$ and $m_{H_{3M}^+}$ are chosen so as to have largest allowed ranges for $m_{H_5^+}$ and $m_{H_5^{++}}$. We vary the latter two over the range $\sim 400 ~- ~730 ~\gev$ for Example 1 and 2. This variation does not affect much the predictions for the signal strengths of the $\Htil$ decays to $W^+W^-$, $ZZ$ and $f\bar{f}$, but only changes that for $\Htil \rightarrow \gamma\gamma$. $m_{H_5^+}$ and $m_{H_5^{++}}$ vary in correlation when the CMS constraints on the signal strength of the diphoton decay channel are imposed. For the numerical calculation of other signal strengths in the following two examples we chose on of these correlated pairs of the two masses.

	\item
Example 1: $\lambda_1 = -0.077, ~\lambda_2 = 14.06, ~\lambda_3 = 15.4, ~\lambda_4 = 0.1175, ~\lambda_5 = 15, ~\lambda_8 = -1$ and $m_{H_5^+} = 500~\gev, ~m_{H_5^{++}} = 540~\gev$. 
Fixing these along with $s_2 = 0.92, ~s_{2M} = 0.16,~s_M \approx 0.36$, fully determines the singlet mass matrix, and hence the mixing matrix, given by:
\be\label{eq:jekyll1}
	\left( \begin{array}{l}		\Htil\\ \Htill\\ \Htilll\end{array}\right) = 
	\left( \begin{array}{ccc}  	0.998 		& -0.0518 		& -0.0329 \\
								0.0514 		& 0.999 		& -0.0140 \\
								0.0336 		& 0.0123 		& 0.999
	\end{array} \right) 
	\left( \begin{array}{l} 		H_1^0\\ H_{1M}^0\\ H_1^{0\prime}\end{array}\right)\,,
\ee
with $\Htil \sim H_1^0$, $\Htill \sim H_{1M}^0$, $\Htilll \sim H_1^{0\prime}$ and $m_{\Htil} = 125.7 ~\gev$, $m_{\Htill} = 420 ~\gev$, $m_{\Htilll} = 601 ~\gev$. $a_{1,1M}$ - the (1,2) element of the $3\times 3$ matrix can actually vary between (-0.0515, -0.05295) and still satisfy CMS constraints. \\
Another example is Example 2: $\lambda_1 = 0.0329, ~\lambda_2 = 14.2, ~\lambda_3 = 15.4, ~\lambda_4 = 0.0056, ~\lambda_5 = 15, ~\lambda_8 = -1$, and $m_{H_5^+} = 590~\gev, ~m_{H_5^{++}} = 600~\gev$,
\begin{flalign}\label{eq:jekyll2}
	&\left( \begin{array}{l}		\Htil\\ \Htill\\ \Htilll\end{array}\right) =&\nonumber\\[0.5em]
	&\left( \begin{array}{ccc}  	0.99999...	& -2.49\times10^{-3}		& -1.60\times10^{-3} \\
								2.49\times10^{-3}	& 0.99999...			& -5.30\times10^{-4} \\
								1.60\times10^{-3}	& 5.26\times10^{-4}	& 0.99999..
	\end{array} \right) 
	\left( \begin{array}{l} 		H_1^0\\ H_{1M}^0\\ H_1^{0\prime}\end{array}\right),&
\end{flalign}
with $\Htil \sim H_1^0$, $\Htill \sim H_{1M}^0$, $\Htilll \sim H_1^{0\prime}$ and $m_{\Htil} = 125.7 ~\gev$, $m_{\Htill} = 420 ~\gev$, $m_{\Htilll} = 599 ~\gev$. The allowed range for $a_{1,1M}$ - the (1,2) element of the $3\times 3$ matrix is $(-1.20, -3.40)\times10^{-3}$.

\begin{table*}[!htb]
	\renewcommand{\arraystretch}{1.5}
	\centering
	\caption{\label{tab:sigXBRjekyll}Partial width of $H\rightarrow gg$ as the measure of the production cross section, partial widths and branching ratios for various channels in SM (for $m_{H_{SM}} = 125.7~\gev$ with total width = $4.17\times 10^{-3}$ \gev, and the \ewnur model for {\em Dr.~Jekyll} example 2 scenario: $a_{1,1M} = -0.0025$, where $m_{\Htil} = 125.7~\gev$, total width = $4.45\times 10^{-3}$ \gev and $\Htil \sim H_1^0$. All the partial widths are given in \gev.}
	\begin{tabular}{|c|c|c|c|c|c|c|c|}
	\hline
    ~ & \multicolumn{3}{c|}{SM}   & \multicolumn{3}{c|}{\ewnur}  &  \multirow{2}{*}{$\mu = \dfrac{(\sigma_{Hgg} \times BR)_{\text{SM}}}{(\sigma_{\Htil gg} \times BR)_{\text{\ewnur}}}$}      \\
     \cline{2-7}
    ~ & $\Gamma_{H\rightarrow gg}$      & Partial Width   & BR       & $\Gamma_{\Htil\rightarrow gg}$      & Partial Width   & BR       & ~         \\
    ~ & $\propto \sigma_{gg\rightarrow H}$ & ~ & ~ & $\propto \sigma_{gg\rightarrow H}$ & ~ & ~ & ~ \\
     \hline
	$\Htil\rightarrow W^+W^-$   	& 3.55E-04 		& 9.42E-04 & 2.26E-01 & 3.46E-04 & 7.63E-04 & 1.72E-01 & 0.74 \\
	$\Htil\rightarrow ZZ$   	& 3.55E-04 		& 1.17E-04 & 2.81E-02 & 3.46E-04 & 9.49E-05 & 2.13E-02 & 0.74 \\
	$\Htil\rightarrow b\bar{b}$   	& 3.55E-04 		& 2.36E-03 & 5.66E-01 & 3.46E-04 & 2.79E-03 & 6.26E-01 & 1.07  \\
	$\Htil\rightarrow \tau\bar{\tau}$	& 3.55E-04	& 2.59E-04 & 6.21E-02 & 3.46E-04 & 3.06E-04 & 6.87E-02 & 1.07  \\
	$\Htil\rightarrow \gamma\gamma$		& 3.55E-04	& 9.51E-06 & 2.28E-03 & 3.46E-04 & 1.26E-05 & 2.82E-03 & 1.21  \\
	\hline
	\end{tabular}
\end{table*}
%
%

	\item
Notice that, although Examples 1 and 2 have very different values for the off-diagonal elements in $\{a_{ij}\}$, they yield comparable numerical signal strength predictions, the reason being principally that in both the cases $\Htil\sim H_1^0$. We can also find other cases having intermediate values for the off-diagonal elements yielding comparable signal strengths.

	\item
	Table \ref{tab:sigXBRjekyll} shows cross section of 125-\gev $H \rightarrow gg$ (as a measure of production cross section), partial widths and branching ratios in the SM and the \ewnur model, for example 2. We see that these partial widths are not very different from those in SM. This is expected as, in this case, the couplings of $H_1^0$ with the SM gauge bosons and fermions are also close to those of the SM Higgs.
	\item
The partial widths and the signal strengths for $W^+W^-$ and $ZZ$ decay channels are smaller, whereas those for $b\bar{b}$, $\tau\bar{\tau}$ and $\gamma\gamma$ decay channels are larger, than the corresponding values in SM. 

It is because, for the example in Table \ref{tab:sigXBRjekyll}, $s_2 < |a_{1,1}| < 1$, and as per Eq.~(\ref{eq:WWZZwid}) the partial width $\Gamma^{EW\nu_R}(\Htil\rightarrow W^+W^-, ~ZZ) \sim |s_2 ~a_{1,1}|^2 \times \Gamma^{SM}(H^0_{SM}\rightarrow W^+W^-, ~ZZ)$ . 

On the other hand, as seen in Eq.~(\ref{eq:ffwid}), $\Gamma^{EW\nu_R}(\Htil\rightarrow f\bar{f}) \sim |a_{1,1}/s_2|^2 ~\Gamma^{SM}(H^0\rightarrow f ~\bar{f})>\Gamma^{SM}(H^0\rightarrow f ~\bar{f})$. 

$\Gamma^{EW\nu_R}(\Htil\rightarrow \gamma\gamma)$ is larger than the corresponding SM value, because in the \ewnur model, charged scalars and mirror fermions also contribute to this decay through triangle loops (refer to Appendix \ref{sec:diphoton}). Recall that in SM this decay is dominated only by the $W$ loop.

	\item
Fig.~\ref{fig:hdecaylim} shows the comparison between the CMS data for signal strengths $\mu(H\text{-decay})$ of the 125-\gev Higgs boson, and the corresponding predictions for the 125-\gev $\Htil$ in the \ewnur model, for examples 1 and 2 in {\em Dr.~Jekyll} scenario and examples 1, 2 and 3 in {\em Mr.~Hyde} scenario, discussed in the next subsection. 

For calculating the \ewnur predictions, we have considered the gluon-gluon fusion production channel $(gg\rightarrow \Htil)$, which is the most dominant Higgs-production channel at the LHC. Calculations of the predictions in the \ewnur model are explained in Appendix~\ref{sec:hdecay}.
	\item
	Notice that the predicted ranges for $\mu(\Htil\rightarrow W^+W^-, ~ZZ)$ and $\mu(\Htil\rightarrow b\bar{b}, \tau\bar{\tau})$ are much narrower than the allowed ranges by the CMS constraints. 
	
	A wider range of $a_{1,1M}$ than shown in Eqs.~(\ref{eq:jekyll1}), (\ref{eq:jekyll2}) is allowed if we impose the constraints on only, say, $\Htil \rightarrow W^+W^-$ decay. However, for a part of the $a_{1,1M}$ range that satisfies the constraints on $\mu(\Htil\rightarrow W^+W^-)$, the constraints on one or more of the other decay channels are not satisfied, and vice versa. So is true for all the other decay channels. Hence, when we seek the range of $a_{1,1M}$ that satisfies the constraints on {\em all 4} of the $\Htil\rightarrow W^+W^-, ~ZZ, ~b\bar{b}, ~\tau\bar{\tau})$ decay channels, the predicted ranges for the signal strengths of these different channels are correlated. This shortens the range of $a_{1,1M}$ and of the signal strength predictions. These correlated predictions are shown in Fig.~\ref{fig:hdecaylim}.

	\item
The predicted range for $\mu(\Htil\rightarrow \gamma\gamma)$ spans over $0-2.5$, because over the ranges of $m_{H_5^+}$ and $m_{H_5^{++}}$, $\mu(\Htil\rightarrow \gamma\gamma)$ can easily vary without significantly affecting the predictions for the signal strengths of other decay channels.

	\item
	{\bf Conclusions from Fig.~\ref{fig:hdecaylim}:}
	We see that in the $\Htil \sim H_1^0$ scenario predictions of the \ewnur model for various signal strengths agree with those of the 125-\gev Higgs boson, as measured by CMS. A slightly, but not very, different mixing matrices can also agree with the ATLAS measurements. Future measurements of partial widths would therefore be required to disentangle this scenario from that the SM.

\ei
\begin{figure}
	\centering
	\includegraphics[width=0.5\textwidth]{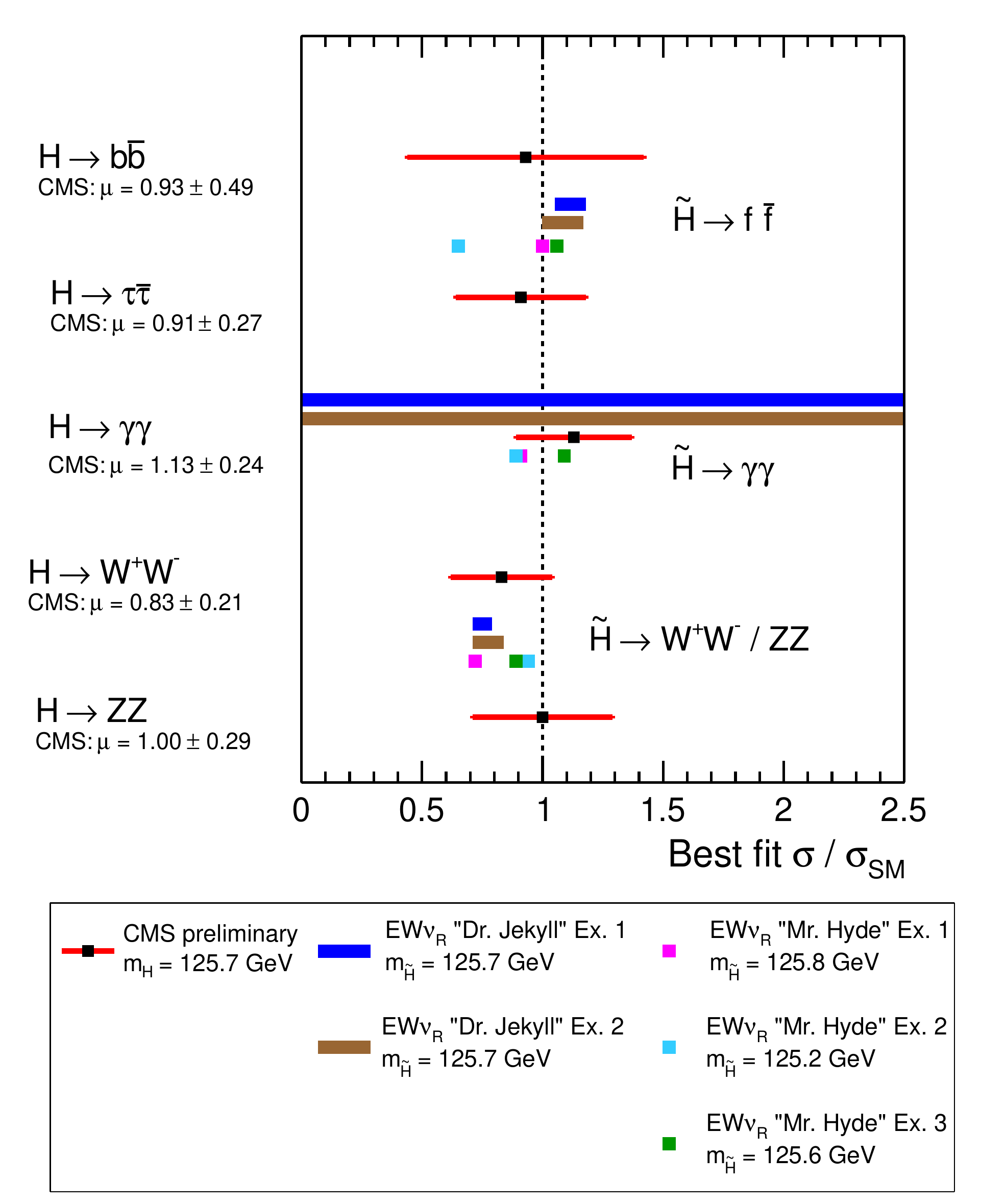}
	\caption{\label{fig:hdecaylim}Figure shows the predictions of $\mu(\Htil \rightarrow ~b\bar{b}, ~\tau\bar{\tau}, ~\gamma\gamma, ~W^+W^-, ~ZZ)$ in the \ewnur model for examples 1 and 2 in {\em Dr.~Jekyll} and example 1, 2 and 3 in {\em Mr.~Hyde} scenarios, in comparison with corresponding best fit values by CMS \cite{h_ww_122013, h_zz_4l_122013, h_bb_102013, h_tautau_012014}.}
\end{figure}

We now come to the most interesting part of our analysis, the one in which the 125-\gev Higgs boson is very unlike the SM Higgs.


\subsection{$\Htil$ as the 125-GeV Higgs candidate with a sub-dominant SM-like component}\label{sec:hyde}


Can the 125-\gev $\Htil$ in the \ewnur model have $H_1^0$ as a subdominant component and still satisfy the experimental constraints on its signal strengths? There are only two CP-even, neutral scalar states other than $H_1^0$, and they are $H_{1M}^0$ and $H_1^{0\prime}$. The analysis explained in section \ref{sec:numerical} revealed 1501 out of 4 million parameter combinations, for which $H_1^0$ can, indeed, be a subdominant component in 125-\gev $\Htil$ while agreeing with the measured signal strengths of the 125-\gev Higgs at the LHC - the scenario we earlier referred to as {\em Mr.~Hyde} scenario.

\subsubsection{Results of the analysis}

\bi
	\item
	Table~\ref{tab:hyde} shows 16 out of the 1501 combinations of the parameters. 
		\item
	It can be seen from Table~\ref{tab:hyde} that in {\em Mr.~Hyde} scenario, the CMS constraints on the signal strength can be satisfied, even when the scalar couplings satisfy $|\lambda/4\pi|<1$. This means that the scalar particles heavier than the 125-\gev Higgs, need not be strongly coupled, and could be potentially detected as {\em narrow resonances} at the LHC.
	
	\item
	Similarly, $s_{2M}$ can be larger than in {\em Dr.~Jekyll} scenario. The mirror fermion masses are given in terms of $s_{2M}$ by
\be
	m_{f^M} = \dfrac{g_{MF} s_{2M} v}{\sqrt{2}}\,.
\ee

Consequently, larger (than in {\em Dr.~Jekyll} scenario) masses of the mirror fermions are allowed by the perturbative limit on their Yukawa couplings. In other words, for a given mass of the mirror fermions, their Yukawa couplings in {\em Mr.~Hyde} scenario can be smaller than those in {\em Dr.~Jekyll} scenario.

	\item
	To highlight interesting features of this scenario, we consider three examples listed in Table~\ref{tab:hyde}.\\
		
	\item
Example 1 (row 1 of Table~\ref{tab:hyde}): $s_2 = 0.900$, $s_{2M} = 0.270$, $s_M = 0.341$, $\lambda_1 = -0.481, ~\lambda_2 = 6.00, ~\lambda_3 = 1.46, ~\lambda_4 = 2.99, ~\lambda_5 = 2, ~\lambda_8 = -1$,
\be\label{eq:hyde1}
	\left( \begin{array}{l}		\Htil\\ \Htill\\ \Htilll\end{array}\right) = 
	\left( \begin{array}{ccc}  	0.300 		& -0.094 		& -0.949 \\
								0.334 		& -0.921	 	& -0.197 \\
								0.893	 	& 0.376 		& 0.246
	\end{array} \right) 
	\left( \begin{array}{l} 		H_1^0\\ H_{1M}^0\\ H_1^{0\prime}\end{array}\right),
\ee
with $\Htil \sim H_1^{0\prime}$, $\Htill \sim H_{1M}^0$, $\Htilll \sim H_1^0$; $m_{\Htil} = 125.8 ~\gev$, $m_{\Htill} = 416 ~\gev$, $m_{\Htilll} = 1100 ~\gev$, $M_R \lesssim 105 ~\gev$, and $\mu(\Htil\rightarrow W^+W^-/ZZ) = 0.72$, $\mu(\Htil\rightarrow \gamma\gamma) = 0.91$, $\mu(\Htil\rightarrow b\bar{b}/\tau\bar{\tau}) = 1.00$.\\ 


	\item
Example 2 (row 2 of Table~\ref{tab:hyde}): $s_2 = 0.514$, $s_{2M} = 0.841$, $s_M = 0.168$, $\lambda_1 = 6.15, ~\lambda_2 = 7.68, ~\lambda_3 = 8.84, ~\lambda_4 = -2.131502, ~\lambda_5 = 5, ~\lambda_8 = -1$,
\be\label{eq:hyde2}
	\left( \begin{array}{l}		\Htil\\ \Htill\\ \Htilll\end{array}\right) = 
	\left( \begin{array}{ccc}  	0.188 		& 0.091 		& 0.978 \\
								-0.941 		& -0.268	 	& 0.207 \\
								-0.281	 	& 0.959 		& -0.035
	\end{array} \right) 
	\left( \begin{array}{l} 	H_1^0\\ H_{1M}^0\\ H_1^{0\prime}\end{array}\right),
\ee
with $\Htil \sim H_1^{0\prime}$, $\Htill \sim H_1^0$, $\Htilll \sim H_{1M}^{0}$; $m_{\Htil} = 125.2 ~\gev$, $m_{\Htill} = 633 ~\gev$, $m_{\Htilll} = 1427 ~\gev$, $M_R \lesssim 52.0 ~\gev$, and $\mu(\Htil\rightarrow W^+W^-/ZZ) = 0.94$, $\mu(\Htil\rightarrow \gamma\gamma) = 0.89$, $\mu(\Htil\rightarrow b\bar{b}/\tau\bar{\tau}) = 0.65$.\\ 


	\item
Example 3 (row 3 of Table~\ref{tab:hyde}): $s_2 = 0.401$, $s_{2M} = 0.900$, $s_M = 0.151$, $\lambda_1 = 4.76, ~\lambda_2 = 3.41, ~\lambda_3 = 7.71, ~\lambda_4 = -1.29, ~\lambda_5 = 4, ~\lambda_8 = -1$,
\be\label{eq:hyde3}
	\left( \begin{array}{l}		\Htil\\ \Htill\\ \Htilll\end{array}\right) = 
	\left( \begin{array}{ccc}  	0.187		& 0.115 		& 0.976 \\
								0.922 		& 0.321	 		& -0.215 \\
								0.338	 	& -0.940 		& 0.046
	\end{array} \right) 
	\left( \begin{array}{l} 		H_1^0\\ H_{1M}^0\\ H_1^{0\prime}\end{array}\right),
\ee
with $\Htil \sim H_1^{0\prime}$, $\Htill \sim H_1^0$, $\Htilll \sim H_{1M}^{0}$; $m_{\Htil} = 125.6 ~\gev$, $m_{\Htill} = 454 ~\gev$, $m_{\Htilll} = 959 ~\gev$, $M_R \lesssim 46.4 ~\gev$, and $\mu(\Htil\rightarrow W^+W^-/ZZ) = 0.89$, $\mu(\Htil\rightarrow \gamma\gamma) = 1.09$, $\mu(\Htil\rightarrow b\bar{b}/\tau\bar{\tau}) = 1.06$.

	\item
	In example 1, $H_{1M}^0$ is the dominant component in $\Htill$, whereas $H_1^0$ is the dominant in $\Htill$ in examples 2 and 3. Although the mixing matrices in examples 2 and 3 are not very different, the ratio of VEV's $s_2$, $s_{2M}$ are different enough to result in the signal strengths that are not very similar (especially for $\Htil\rightarrow f\bar{f}$). As the partial width of $\Htil\rightarrow f\bar{f}$ is proportional to $|a_{1,1}/s_2|^2$, it changes rapidly with $s_2$. Also, because we have 6 mirror quarks which contribute to the cross section of gluon-gluon fusion, the production cross section dominantly changes as $\sim |a_{1,1}/s_2+6 ~a_{1,1M}/s_{2M}|^2$. Thus, any change in $a_{1,1M}/s_{2M}$ is amplified while calculating the signal strengths. 

	\item
Comparison of the signal strengths for the three examples with the CMS constraints on them can be seen in Fig.~\ref{fig:hdecaylim}. Notice the agreement between the predictions for the signal strengths with the CMS constraints in the figure. This agreement demonstrates that the {\em SM-like signal strengths of 125-\gev Higgs at the LHC are not sufficient to conclude that it is a SM-like Higgs, or even if it has a dominant SM-like component}.

	\item
Table~\ref{tab:sigXBRhyde} shows the partial widths, branching ratios and the signal strengths for {\em Mr.~Hyde} scenario in the \ewnur model and SM. It can be seen that the partial widths in this scenario are very different from the SM (smaller by a factor of $\sim 5$ for the example in the table), but it results in similar signal strengths. Measurements of the partial widths are therefore necessary to be able to experimentally distinguish between {\em Mr.~Hyde} scenario and SM.
\ei
\begin{table*}[!htb]
	\renewcommand{\arraystretch}{1.5}
	\centering
	\caption{\label{tab:sigXBRhyde}Partial width of $H\rightarrow gg$ as the measure of the production cross section, partial widths and branching ratios for various channels in SM (for $m_{H_{SM}} = 125.6~\gev$ and total width $4.15\times 10^{-3}$ \gev), and the \ewnur model for row 3 in Table \ref{tab:hyde}, also in Eq.~(\ref{eq:hyde3}) where $\Htil \sim H_1^{0\prime}$ (with $m_{\Htil} = 125.6~\gev$ and total width $1.34\times 10^{-3}$ \gev). All the partial widths are given in \gev.}
	\begin{tabular}{|c|c|c|c|c|c|c|c|}
	\hline
    ~ & \multicolumn{3}{c|}{SM}   & \multicolumn{3}{c|}{\ewnur}  &  \multirow{2}{*}{$\mu = \dfrac{(\sigma_{Hgg} \times BR)_{\text{SM}}}{(\sigma_{\Htil gg} \times BR)_{\text{\ewnur}}}$}      \\
     \cline{2-7}
    ~ & $\Gamma_{H\rightarrow gg}$      & Partial Width   & BR       & $\Gamma_{\Htil\rightarrow gg}$      & Partial Width   & BR       & ~         \\
    ~ & $\propto \sigma_{gg\rightarrow H}$ & ~ & ~ & $\propto \sigma_{gg\rightarrow H}$ & ~ & ~ & ~ \\
     \hline
	$\Htil\rightarrow W^+W^-$   	& 3.54E-04 		& 9.30E-04 & 2.24E-01 & 5.75E-04 & 1.64E-04 & 1.23E-01 & 0.89 \\
	$\Htil\rightarrow ZZ$   		& 3.54E-04 		& 1.16E-04 & 2.79E-02 & 5.75E-04 & 2.04E-05 & 1.53E-02 & 0.89 \\
	$\Htil\rightarrow b\bar{b}$   	& 3.54E-04 		& 2.35E-03 & 5.67E-01 & 5.75E-04 & 5.07E-04 & 3.79E-01 & 1.06  \\
	$\Htil\rightarrow \tau\bar{\tau}$	& 3.54E-04	& 2.58E-04 & 6.22E-02 & 5.75E-04 & 5.42E-05 & 4.06E-02 & 1.06  \\
	$\Htil\rightarrow \gamma\gamma$		& 3.54E-04	& 9.46E-06 & 2.28E-03 & 5.75E-04 & 2.04E-06 & 1.53E-03 & 1.09  \\
	\hline
	\end{tabular}
\end{table*}
%

\subsubsection{Remarks on the $H_{1M}^0$ component in $\Htil$}

A few remarks are in order here:
\bi
	\item
	Notice that for all the cases listed in Table~\ref{tab:hyde} $H_1^{0\prime}$ is the dominant component in the 125-\gev $\Htil$. In all 1501 cases we found, the modulus of the coefficient of $H_{1M}^0$ in the 125-\gev $\Htil$ was $\leq 0.32$.
	\item
	In the gluon-gluon fusion channel $H_{1M}^0$ is produced through triangle loops of 6 mirror quarks. Therefore, if $H_{1M}^0$ is the dominant component in $\Htil$, then the production cross section of $\Htil$ could become too high to be compensated by small branching ratios. Thus, it makes sense that $H_{1M}^0$ is disfavored to be the dominant component in $\Htil$, by the constraints on the signal strengths.
	\item
	Even if $H_{1M}^0$ is a sub-dominant component in $\Htil$, one should not think that it has decoupled from the other two singlets. In other words, the scalar doublet $\Phi_{2M}$ does not really decouple from $\Phi_2$ and $\chi$. This is because:
	\bi
		\item
		Even if $H_{1M}^0$ has a small coefficient in $\Htil$, its production amplitude through 6 mirror quarks has a significant contribution to the production cross section of $\Htil$.
		\item
		The real degree of freedom of $\Phi_{2M}$ leads to $H_{1M}^0$. But its other degrees of freedom also contribute to other physical particles such as $H_3^{0,\pm}$, $H_{3M}^{0,\pm}$. These particles contribute to $\Htil \rightarrow \gamma\gamma$ and the total width. Hence, they play a role in ensuring that the branching ratios are in the appropriate range to achieve an agreement with the signal strength constraints.
	\ei
	\item
	Thus, although $H_{1M}^0$ is a sub-dominant component in $\Htil$, the scalar doublet $\Phi_{2M}$, newly added to the minimal \ewnur model, plays a crucial role in accommodating the 125-\gev Higgs boson in the \ewnur model, in {\em Mr.~Hyde} as well as {\em Dr.~Jekyll} scenario.
\ei

\begin{turnpage}
\begin{table*}[!]
	\renewcommand{\arraystretch}{1.2}
	\centering
	\caption{\label{tab:hyde}All the masses and the total width of $\Htil$ are given in \gev. Fixed parameters as given in Eq. (\ref{eq:hydemupara}).}
\begin{tabular}{|c|c|c|c|c|c|c|c|c|c|c|c|c|c|c|c|c|c|c|c|c|c|c|c|c|c|}
\hline
~ & $s_2$   & $s_{2M}$  & $s_M$   & $\lambda_1$   & $\lambda_2$   & $\lambda_3$  & $\lambda_4$   & $\lambda_5$   & $m_{\Htil}$   & $m_{\Htill}$ & $m_{\Htilll}$ & $a_{1,1}$ & $a_{1,1M}$ & $a_{1,1'}$ & $a_{1M,1}$ & $a_{1M,1M}$ & $a_{1M,1'}$ & $a_{1',1}$ & $a_{1',1M}$ & $a_{1',1'}$ & $\Gamma_{\Htil}$  & $\mu_{WW/ZZ}$ & $\mu_{\gamma\gamma}$ & $\mu_{bb/\tau\tau}$ & $M_R\lesssim$ \\
\hline
1  & 0.90 & 0.27 & 0.34 & -0.48   & 6.00    & 1.46    & 2.99    & 2       & 125.8 & 416  & 1100   & 0.301 & -0.094 & -0.949  & 0.334  & -0.922  & 0.197    & 0.893   & 0.376    & 0.246     & 1.66E-03 & 0.719  & 0.914    & 1.002  & 105.0 \\
2  & 0.51 & 0.84 & 0.17 & 6.15    & 7.68    & 8.84    & -2.13   & 5       & 125.2 & 633  & 1427   & 0.189 & 0.091  & 0.978   & -0.941 & -0.268  & 0.207    & -0.281  & 0.959    & -0.035    & 9.61E-04 & 0.941  & 0.895    & 0.647  & 52.0  \\
3  & 0.41 & 0.90 & 0.15 & 4.76    & 3.41    & 7.71    & -1.29   & 4       & 125.6 & 454  & 959    & 0.187 & 0.115  & 0.976   & 0.922  & 0.321   & -0.215   & 0.338   & -0.940   & 0.046     & 1.34E-03 & 0.891  & 1.089    & 1.062  & 46.4  \\
4  & 0.87 & 0.32 & 0.36 & -0.39   & 4.40    & 1.21    & 4.48    & 1       & 126.2 & 420  & 1382   & 0.303 & -0.087 & -0.949  & 0.357  & -0.913  & 0.197    & 0.884   & 0.398    & 0.246     & 1.12E-03 & 0.753  & 1.108    & 0.849  & 111.9 \\
5  & 0.35 & 0.91 & 0.22 & 8.73    & 5.26    & 4.88    & -1.59   & 9       & 126.1 & 617  & 1237   & 0.143 & 0.090  & 0.986   & 0.966  & 0.204   & -0.159   & 0.216   & -0.975   & 0.057     & 1.12E-03 & 0.994  & 0.995    & 0.682  & 69.2  \\
6  & 0.31 & 0.87 & 0.38 & 4.99    & 9.67    & 2.10    & -1.02   & 1       & 126.4 & 435  & 1786   & 0.238 & 0.041  & 0.970   & 0.970  & 0.035   & -0.239   & 0.044   & -0.999   & 0.031     & 2.62E-03 & 0.953  & 1.128    & 1.092  & 118.1 \\
7  & 0.31 & 0.87 & 0.38 & 4.99    & 9.67    & 2.10    & -1.02   & 2       & 126.4 & 435  & 1786   & 0.238 & 0.041  & 0.970   & 0.970  & 0.035   & -0.239   & 0.044   & -0.999   & 0.031     & 2.62E-03 & 0.954  & 0.894    & 1.093  & 118.1 \\
8  & 0.92 & 0.21 & 0.31 & -0.73   & 9.34    & 1.83    & 9.25    & 5       & 125.6 & 412  & 1988   & 0.239 & -0.096 & -0.966  & 0.210  & -0.966  & 0.148    & 0.948   & 0.239    & 0.211     & 2.91E-03 & 0.882  & 1.341    & 0.708  & 95.7  \\
9  & 0.36 & 0.92 & 0.16 & 6.78    & 3.10    & 7.16    & -1.34   & 4       & 126.1 & 501  & 905    & 0.156 & 0.129  & 0.979   & -0.901 & -0.388  & 0.195    & -0.405  & 0.912    & -0.056    & 1.37E-03 & 1.031  & 1.001    & 1.043  & 49.6  \\
10 & 0.92 & 0.19 & 0.35 & -0.34   & 11.69   & 1.17    & 2.79    & 1       & 126.7 & 428  & 1067   & 0.272 & -0.047 & -0.961  & 0.237  & -0.965  & 0.114    & 0.932   & 0.259    & 0.251     & 9.77E-04 & 0.729  & 1.107    & 0.663  & 108.4 \\
11 & 0.95 & 0.12 & 0.29 & -0.46   & 9.71    & 1.68    & 3.37    & 9       & 126.0 & 248  & 1167   & 0.230 & -0.126 & -0.965  & 0.113  & -0.981  & 0.155    & 0.967   & 0.145    & 0.211     & 1.57E-02 & 0.742  & 1.083    & 0.678  & 89.2  \\
12 & 0.38 & 0.89 & 0.26 & 3.47    & 5.24    & 3.25    & -0.99   & 10      & 125.2 & 409  & 1281   & 0.241 & 0.065  & 0.968   & 0.964  & 0.097   & -0.247   & 0.110   & -0.993   & 0.039     & 1.82E-03 & 0.849  & 0.967    & 1.080  & 79.6  \\
13 & 0.30 & 0.93 & 0.22 & 5.01    & 2.67    & 3.06    & -0.75   & 12      & 125.1 & 415  & 906    & 0.131 & 0.075  & 0.989   & 0.979  & 0.146   & -0.141   & 0.155   & -0.986   & 0.054     & 1.06E-03 & 0.770  & 1.266    & 0.666  & 69.0  \\
14 & 0.36 & 0.89 & 0.28 & 2.34    & 2.56    & 2.04    & -0.52   & 12      & 126.0 & 333  & 890    & 0.211 & 0.071  & 0.975   & 0.971  & 0.104   & -0.217   & 0.117   & -0.992   & 0.047     & 1.74E-03 & 1.006  & 1.267    & 0.984  & 87.0  \\
15 & 0.48 & 0.86 & 0.20 & 2.26    & 2.90    & 4.11    & -0.74   & 6       & 126.0 & 376  & 896    & 0.217 & 0.090  & 0.972   & 0.950  & 0.208   & -0.231   & 0.223   & -0.974   & 0.040     & 1.29E-03 & 1.012  & 0.917    & 0.857  & 60.9  \\
16 & 0.32 & 0.91 & 0.26 & 2.76    & 2.07    & 2.18    & -0.49   & 11      & 126.3 & 323  & 804    & 0.182 & 0.076  & 0.980   & 0.975  & 0.115   & -0.190   & 0.127   & -0.991   & 0.053     & 1.69E-03 & 0.919  & 0.930    & 1.019  & 80.6  \\
   \hline
\end{tabular}
\end{table*}
\end{turnpage}
Before concluding this section, we will briefly discuss some indirect constraints on the next heavier scalar $\Htill$, in both these scenarios.
%
\subsection{The next heavier neutral scalar $\Htill$}

In {\em Dr.~Jekyll} scenario, examples 1 and 2 that we considered have $H_{1M}^0$ as the dominant component in $\Htill$, which is the next heavier physical scalar after the 125-\gev $\Htil$. Here the total width of $\Htill$ is also greater than its mass, with the scalar coupling $\lambda_2>4\pi$. Thus, it is a strongly coupled scalar, which is difficult to detect as a narrow resonance.

In example 1 of {\em Mr.~Hyde} scenario, $H_{1M}^0$ is the dominant component in $\Htill$, while in examples 2 and 3 $H_1^0$ is the dominant component in $\Htill$. In all 3 examples, $\Htill$ has a total width $<10\%$ of its mass.

This subsection compares the signal strength of $\Htill\rightarrow W^+W^-$ and the $\sigma \times BR(\Htill\rightarrow \gamma\gamma)$ with the CMS constraints on SM-like heavy Higgs, for examples having $m_{\Htill} \lesssim 600 ~\gev$. These CMS constraints \cite{CMS_heavy_Higgs_WW, HWWsearch, ATLAS_diphoton_july2014, CMS_diphoton_july2014} assume the Standard model background, whereas, in the \ewnur model, extra processes involving mirror fermions and extra scalars also contribute to the background in addition to the SM processes. The background in this model is therefore expected to be larger than that in the SM. A detailed study of this background is out of the scope of this paper.

Although the SM background does not strictly apply to $\Htill$ in the \ewnur model, we show how the \ewnur predictions compare with the experimental constraints.

For our calculations we computed the total width of $\Htill$ using
\bea\label{eq:widthH1M}
\Gamma_{\Htill} = &&~\sum_{i=1}^3\Gamma_{\Htill\rightarrow q_i^M\bar{q}_i^M} ~+ \sum^3_{j=1} ~\times ~\Gamma_{\Htill\rightarrow l_j^M\bar{l}_j^M}\nonumber\\[0.5em]
 &&+ ~\Gamma_{\Htill\rightarrow t\bar{t}} ~+ ~\Gamma_{\Htill\rightarrow b\bar{b}} \nonumber\\[0.5em]
&& + ~\Gamma_{\Htil\rightarrow \tau\bar{\tau}} ~+ ~\Gamma_{\Htil\rightarrow c\bar{c}} ~+\Gamma_{\Htill\rightarrow W^+ W^-}\nonumber\\[0.5em]
&& + ~\Gamma_{\Htill\rightarrow ZZ} ~+ ~\Gamma_{\Htill\rightarrow gg} ~+ ~\Gamma_{\Htill\rightarrow \gamma \gamma}\,.
\eea
The partial decay widths were calculated using the method illustrated in Appendix~\ref{sec:hdecay}.

\subsubsection{Constraints on the signal strength of $\Htill\rightarrow W^+W^-$}
%
\begin{figure}
	\centering
	\includegraphics[width=0.5\textwidth]{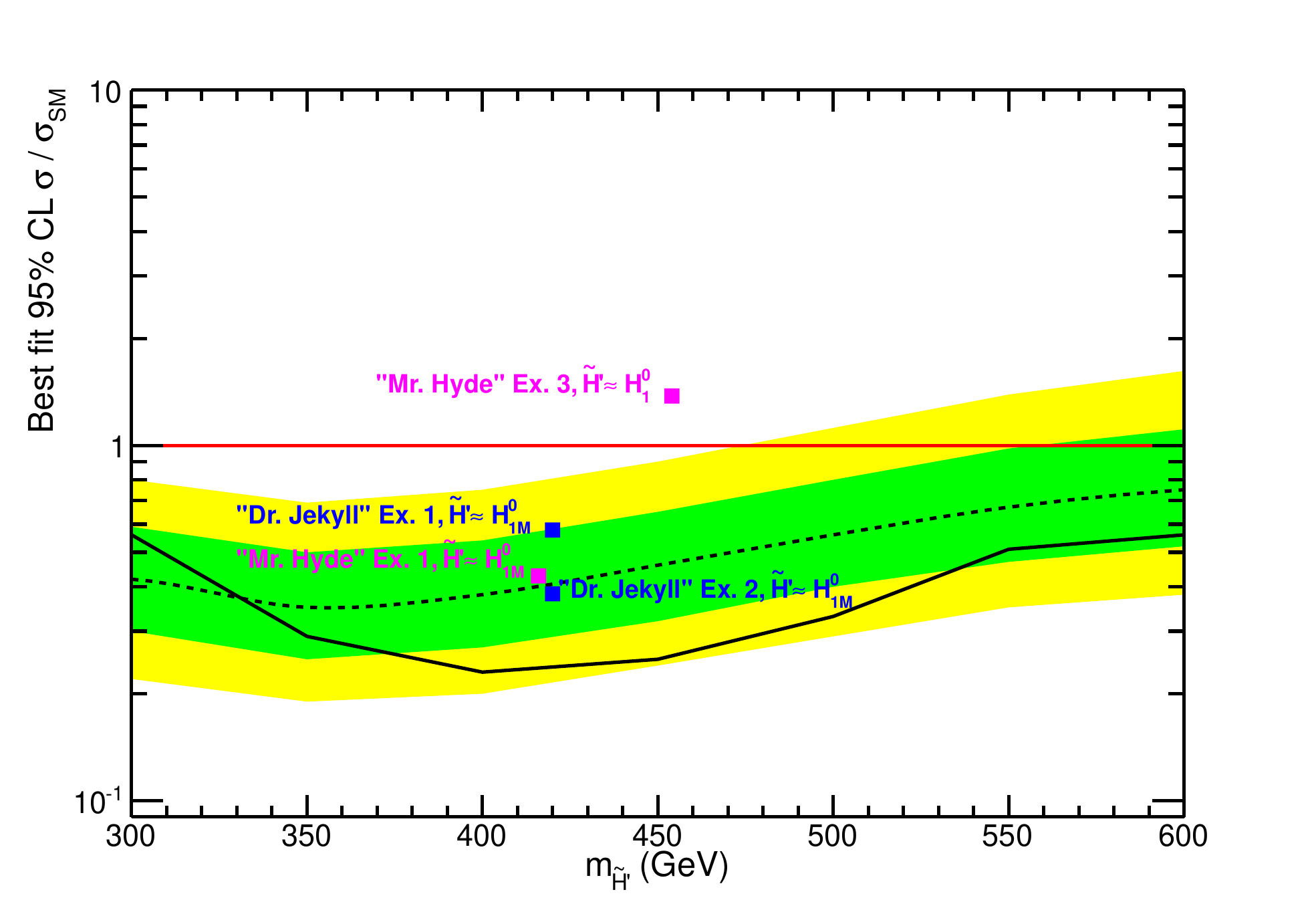}
	\caption{\label{fig:h1mww}Predicted signal strength of $\Htill\rightarrow W^+W^-$ in 4 example scenarios (blue and purple squares). The results of the search for SM-like Higgs boson up to 600 \gev with the $1\sigma$ (green band) and $2\sigma$ (yellow band) limits on the SM background (dotted curve) and CMS data (solid black curve) are also displayed.}
\end{figure}
\bi
	\item
	Fig.~\ref{fig:h1mww} shows the signal strength of $\Htill\rightarrow W^+W^-$ for examples 1 and 2 in {\em Dr.~Jekyll} scenario (blue squares) and exmaples 1 and 3 in {\em Mr.~Hyde} scenario (purple squares). The $1-$ and $2-\sigma$ SM background bands and the CMS data \cite{CMS_heavy_Higgs_WW, HWWsearch} are also displayed. The signal strength for example 2 in {\em Mr.~Hyde} scenario is not displayed as $m_{\Htill} = 633~\gev$ for this example, but the CMS data for this decay channel are only available up to 600~\gev. 
	
	\item
	In the figure, notice that the predicted signal strengths for examples 1 and 2 in {\em Dr.~Jekyll} case and example 1 in {\em Mr.~Hyde} case are within the $\pm 1\sigma$ SM-background bands. Therefore, the CMS data are surely not conclusive for confirming or ruling out these examples.
	
	\item
	Example 1 in {\em Mr.~Hyde} scenario predicts a signal strength $\mu(\Htill\rightarrow W^+W^-) \approx 1.3$, which is certainly larger the SM-background band and the data. $H_1^0$ - the SM-like Higgs is the dominant component in $\Htill$ in this example. However, the SM-background still does not strictly apply here, since additional background processes can contribute to it. For example, production of $W^+W^-$ from two gluons through a box loop of mirror quarks.
\ei
\subsubsection{Constraints on $\Htill\sim H_{1M}^0$ from $\gamma\gamma$-decay channel}

The constraints on $\sigma(gg\rightarrow \Htill)\times BR(\Htill\rightarrow \gamma\gamma)$ from CMS \cite{CMS_diphoton_july2014} and ATLAS \cite{ATLAS_diphoton_july2014} are accompanied by assumptions that the total width of the SM-like heavy Higgs is $0.1 ~\gev$ or $10\%$ of its mass. The total width of $\Htill$ in our scenarios does not follow either of these patterns. We observed that $\sigma(gg\rightarrow\Htill)\times BR(\Htill\rightarrow \gamma\gamma)$ predictions for all the examples in both the scenarios is consistently lower than the CMS and ATLAS constraints.

\subsubsection{A comment on $\Htilll$}

In the examples of {\em Dr.~Jekyll} scenario considered in this paper, $\Htilll \sim H_1^{0\prime}$, and in the examples of {\em Mr.~Hyde} scenario that we have considered, $\Htilll \sim H_1^0$ or $H_{1M}^0$. For all these examples, $m_{\Htilll} \gtrsim 600~\gev$. So far, the CMS data in Fig~\ref{fig:h1mww} are not sensitive to the signal strengths of the order of SM predictions in this mass range. A detailed analysis of its signal strengths will be of interest, when more data for this higher mass range are available and are analyzed with full \ewnur background.
%
\subsection{Conclusions about the 125-GeV Higgs candidate in the \ewnur model}\label{sec:125conclude}

\bi
	\item
	In this section we investigated two very different scenarios of the 125-\gev Higgs boson at the LHC from the perspective of the \ewnur model:
	\begin{enumerate}
		\item
		{\em Dr.~Jekyll}: $H_1^0$, which is the real part of the SM-like scalar doublet $\Phi_2$, is the dominant component in the 125-\gev $\Htil$, and
		\item
		{\em Mr.~Hyde}: $H_1^0$ is a sub-dominant component in the 125-\gev $\Htil$ - a more interesting scenario.
	\end{enumerate}
	
	\item
	We demonstrated that in both these scenarios the signal strengths of the 125-\gev $\Htil$ in $W^+W^-, ~ZZ, ~\gamma\gamma, ~b\bar{b}$ and $\tau\bar{\tau}$ decay channels agree with the constraints from CMS (and also ATLAS) data. Thus, from the perspective of the \ewnur model, the present data at the LHC are inconclusive about whether SM-like $H_1^0$ is {\em the dominant} or {\em a sub-dominant component} in the 125-\gev particle. Hence ``the dual nature" of the 125-\gev Higgs in the \ewnur model.
	
	More data, measurements of individual partial widths and study of the heavier physical scalars in the \ewnur model are necessary to distinguish between either of these scenarios and the SM-Higgs.

	\item
	As expected, the individual partial widths of the 125-\gev $\Htil$ in {\em Dr.~Jekyll} scenario are not very different from those in SM. Here, the scalar couplings $|\lambda_2|, ~|\lambda_3|, ~|\lambda_5|$ need to be greater than  $4\pi$ to satisfy the constraints on the signal strengths. This means that the heavier scalars in this scenario tend to be strongly coupled and have large widths.

	Dominant SM-like component in the 125-\gev Higgs also leads to $v_2$ (the VEV of $\Phi_2$) being the dominant part in $v$, and  smaller $v_{2M}$, which gives masses to the mirror fermions. Consequently, the masses of the mirror fermions, allowed by the perturbative limit on their Yukawa couplings, cannot be much greater than $\sim 120 ~\gev$. We adopt a lower limit of 102~\gev set by the constraints from LEP3 \cite{pdg}.

	\item
	Hence, if future measurements of the individual decay widths of 125-\gev Higgs result in SM-like widths, then it is more likely to be consistent with {\em Dr.~Jekyll} scenario. In this case, the heavier scalars would appear not as narrow resonances, but as broad resonances or enhancement in the background in this model.

	Since the SM-like $H_1^0$ is the dominant component in the 125-\gev $\Htil$, the effective theory around this energy looks like SM, in which the heavier scalars are {\em integrated out}.
	
	\item
	In contrast, the individual partial widths of the 125-\gev $\Htil$ are very different from those in SM, in {\em Mr.~Hyde} scenario. In all 1501 combinations of the parameters that we found to agree with the experimental $1\sigma$ constraints on the signal strengths contain $H_1^{0\prime}$ as the dominant component in the 125-\gev $\Htil$. The predicted signal strengths of this $\Htil$ agree with the experimental $1\sigma$ constraints on the signal strengths even when the scalar couplings $|\lambda|$'s are smaller than $4\pi$. The heavier scalars in this case are not strongly coupled, as a result.
	
	The $H_{1M}^0$ as a dominant component in $\Htil$ is disfavored to agree with the constraints on the signal strengths, due to its large contribution to the cross section of $gg \rightarrow \Htil$. 
	
	Because $v_{2M}$ is not constrained to be small in this case, the perturbative upper limit on the mirror fermions is about 700~\gev.
	
	\item
	Therefore, if the partial widths of the 125-\gev $\Htil$ are measured to be very different from those in SM, it would point towards {\em Mr.~Hyde} scenario. The heavier scalars in this case have narrow widths and can be detected as resonances.
	
	The SM-like $H_1^0$ is the dominant component in one of the heavier scalars, $\Htill$ or $\Htilll$. Thus, the effective theory around 125~\gev is very different from SM, while the SM-like $H_1^0$ is {\em integrated out} with the heavier scalars.

	\item
	As can be seen from Eq~(\ref{eq:hydemupara}) we scanned only a part of the entire parameter space in the \ewnur model, by fixing values or ranges of the parameters. A more thorough scan of the parameter space could be of interest, especially if more data from the LHC Run II show any signs of physics Beyond the Standard Model. 

\ei

In the next section we will briefly discuss some of the decay properties of the CP-odd neutral spin-0 states - $H_3^0$ and $H_{3M}^0$.

\section{Signals of CP-odd spin zero states }

In addition to the 125-\gev $\tilde{H}$, the \ewnur~model also contains CP-odd spin zero states - $H_{3}^0, H_{3M}^0 $ - and two other heavy CP-even spin zero states - $\Htill$, and $\Htilll$. In this section, we show possibilities to probe the signal of the neutral pseudo-scalars in various decay channels at LHC. To do so, we will investigate the product of production cross section and the branching ratio, a.k.a the absolute signal strength, in $H_{3,3M}^0\rightarrow \gamma\gamma,~\tau\tau$ channels. We will also calculate the signal strengths ($\mu$) for $H_{3}^0,~H_{3M}^0$ and the SM Higgs boson $H_{SM}$ in other channels.\\
\be\label{eq:muy}
\mu \; = \; \dfrac{\sigma(gg \rightarrow H_{3,3M}^0) Br( H_{3,3M}^0 \rightarrow XX )}{\sigma(gg \rightarrow H_{SM}) Br( H_{SM} \rightarrow XX)}
\\
\ee

In this extension of the \ewnur model, the degenerate masses of two $SU(2)_D$ custodial triplets are related by
\be\label{eq:mH3H3M}
 \dfrac{m_{H_3}^2}{m_{H_{3M}}^2} = \dfrac{1}{1+c_M^2}
\ee
We assume that the neutral states $H_3^0$ and $H_{3M}^0$ obey this relationship, and use two cases of  $s_M = 0.168;\;0.36$. The lighter one, $m_{H_3^0}$, is scanned over ranges $130 - 440\;\gev$, whereas the mass $m_{H_{3M}^0}$ of the heavier one over $182 - 618\; \gev$ and $177 - 601\; \gev$, respectively.

\subsection{Ratio of production cross sections }

At the LHC, $H_3^0,~H_{3M}^0$ are expected to be produced mainly through gluon-gluon fusion, similar to $H_{SM}$. By using effective coupling approximation, we have
\be\label{eq:csection}
  R=\dfrac{\sigma(gg \rightarrow H_{3,3M}^0)}{\sigma(gg \rightarrow H_{SM})} \approx \dfrac{\Gamma(H_{3,3M}^0 \rightarrow gg)}{\Gamma(H_{SM} \rightarrow gg)}\,.
\ee
$H_{3,3M}^0$ do not couple directly to the gauge bosons $W, ~Z, ~\gamma$. And triplet couplings, such as $ H_{3,3M}^0 H_{3,3M}^+ H_{3,3M}^-,\; H_{3,3M}^0 H_5^+ H_5^-,\; H_{3,3M}^0 H_5^{++} H_5^{--}$, are forbidden by CP conservation. Therefore, only fermionic loops involving the top quark and the mirror quarks contribute to the gluonic decay of $H_3^0$ and $H_{3M}^0$ \cite{qcd_corr3}:
 \bea\label{eq:H3ogg}
  \Gamma \left( H_{3,3M}^0 \rightarrow gg \right) =&& \dfrac{G_F \alpha_s^2}{16\sqrt{2} \pi^3} m_{H_{3,3M}^0}^3\nonumber\\[1em]
  &\times& \left| \sum\limits_{Q} g_Q^{H_{3,3M}^0} F_Q^{H_{3,3M}^0}(\tau_f)\right|^2\,,
 \eea
 \be\label{eq:F_H3o} 
 F_Q^{H_{3,3M}^0}(\tau_f) = \tau_f f(\tau_f)\,,
 \ee
 \be
 \tau_t = 4 m_f^2/m_{H_{3,3M}^0}^2\,,
 \ee

where $g_Q^{H_{3,3M}^0}$ are couplings of $H_3^0$ and $H_{3M}^0$ to the top quark and mirror quarks, listed in Table \ref{table:h_ferm}.

Here, $\sum\limits_{Q}$ is summed over the top quark and mirror quarks. However, the contributions from mirror quarks can be suppressed due to the fact that the mirror-up quarks and the mirror-down quarks couple to $H_{3,3M}^0$ with opposite signs. In this work, we particularly consider degenerate mirror fermion doublets, meaning $m_{u^M} = m_{d^M}$, for simplicity. As a result, the contributions from mirror quarks cancel out. Thus, only the top-quark loop appears in the production cross section of $H_3^0, H_{3M}^0$. Then, the ratios of production cross section are given by
 \be\label{eq:csectionapprox}
 R_{H_3^0}= \tau_t^2  \dfrac{\left|\tan\theta_M f(\tau_t)\right|^2}{\left|\tau_t+(\tau_t-1)f(\tau_t)\right|^2}
 \ee
for $H_3^0$, and
 \be\label{eq:csection3oM}
 R_{H_{3M}^0}= \tau_t^2 \dfrac{\left|s_{2M}f(\tau_t)\right|^2}{\left|s_2(\tau_t+(\tau_t-1)f(\tau_t))\right|^2}
 \ee
for $H_{3M}^0$.

\subsection{ In \;$ \gamma\gamma$ channel}

ATLAS \cite{ATLAS_diphoton_july2014} and CMS \cite{CMS_diphoton_july2014} have recently reported their results of the search for narrow resonances in spin-0 state to diphoton decay channel, up to $600~\gev$ for CMS and $840~\gev$ for ATLAS. Both the reports make certain assumptions about the total width of the decaying spin-0 state. They present the upper limit on the production cross section times branching ratio for this channel at $ 95\% $ confidence level. So far, no significant excess has been found, except for the two $2\sigma$ resonances above the background at $m = 201\;\gev$ and $m = 530\;\gev$ in the ATLAS analysis. We compare our predictions in $\gamma\gamma$ channel with those results, even though the assumptions about the total width of resonances are not generally applicable to the \ewnur~model.

Similar to the gluonic decay, only fermionic loops contribute to the partial width of $H_{3,3M}^0 \rightarrow \gamma \gamma$, given by \cite{qcd_corr3}
\be\label{eq:H3ogamma}
\Gamma\left( H_{3,3M}^0\rightarrow \gamma\gamma\right) = \dfrac{g^2\;\alpha^2\; m^3_{H_{3,3M}^0}}{256\;\pi\; m_W^2}\left|\sum\limits_{i} ~ N_{ci}\; e_i^2 \; g_i \; F_i^{H_{3,3M}^0} \right|^2\,.
\ee
Here, $i=$ top quark, six mirror quarks, and three charged mirror leptons. The total widths of $H_{3,3M}^0$ are calculated by summing all the partial widths.
\begin{eqnarray}\label{eq:width33M}
\Gamma_{H_{3,3M}}=&&~\Gamma(H_{3,3M}^0\rightarrow\gamma\gamma)+\Gamma(H_{3,3M}^0\rightarrow gg)\nonumber\\
    &&+\Gamma(H_{3,3M}^0\rightarrow W^+W^-)+\Gamma(H_{3,3M}^0\rightarrow ZZ)\nonumber\\
    &&+\Gamma(H_{3,3M}^0\rightarrow\tau\bar{\tau})+\Gamma(H_{3,3M}^0\rightarrow t\bar{t})\nonumber\\
    &&+\Gamma(H_{3,3M}^0\rightarrow c\bar{c})+\Gamma(H_{3,3M}^0\rightarrow b\bar{b})\nonumber\\
    &&+ \sum\limits_{i=1}^{6}\Gamma(H_{3,3M}^0\rightarrow q^M_i\bar{q}^M_i)\nonumber\\
    && +\sum\limits_{j=1}^{3}\Gamma(H_{3,3M}^0\rightarrow l^M_j\bar{l}^M_j)
\end{eqnarray}
The branching ratio of $H_{3,3M}^0\rightarrow\gamma\gamma$ is
\be
 \label{eq:Branching33M}
  Br(H_{3,3M}^0\rightarrow\gamma\gamma)=\dfrac{\Gamma(H_{3,3M}^0\rightarrow \gamma\gamma)}{\Gamma_{H_{3,3M}^0}}
\ee
The absolute signal strength of $H_{3,3M}^0\rightarrow\gamma\gamma$ is defined as\\
\begin{flalign*}
\label{eq:Xbranching33M}
 \sigma \times BR(H_{3,3M}^0\rightarrow\gamma\gamma) = R\times\sigma(gg \rightarrow H_{SM})\\
  \times Br(H_{3,3M}^0 \rightarrow \gamma \gamma)\\
\end{flalign*}
At any particular mass, $R_{H_{3,3M}^0}$ and $Br(H_{3,3M}^0 \rightarrow \gamma\gamma)$ are calculated directly, while  $\sigma(gg \rightarrow H_{SM})$ is taken from the handbook of Higgs cross section \cite{HXreport3}. To be consistent with the previous analysis, we also provide two scenarios which correspond to the dual nature of the 125-\gev scalar. For illustrative purposes, we consider up- and down- members of mirror quark doublets to have degenerate masses. The first two generations of mirror quarks and all charged mirror leptons have the same mass, $m_{q^M_1} = m_{q^M_2} = m_{l^M} = 102 \;\gev$. The heaviest mirror quark generation has a mass $m_{q^M_3} = 120 \;\gev$. Masses of all right-handed neutrinos are at $M_R = 50 \;\gev$. 
  \begin{itemize}
  \item
In the case of {\em Dr.~Jekyll}, as $\Htil\sim H_1^0$, the mixing angles are $s_2 = 0.92,\;s_{2M} = 0.16,\;s_M = 0.36$, which corresponds to the first example in the section \ref{sec:jekyll}. In the plot below, we present the dependence of the production cross section times branching ratio of $H_{3,3M}$ on mass. Moreover, the next heavy CP-even state is $\tilde{H}^{\prime}$ with the mass of $m_{\Htill}=420\;\gev$. So we incorporate the production cross section times branching ration of $\Htill\rightarrow\gamma\gamma$ also. 

  \begin{figure}[H]
    \includegraphics[scale=0.45]{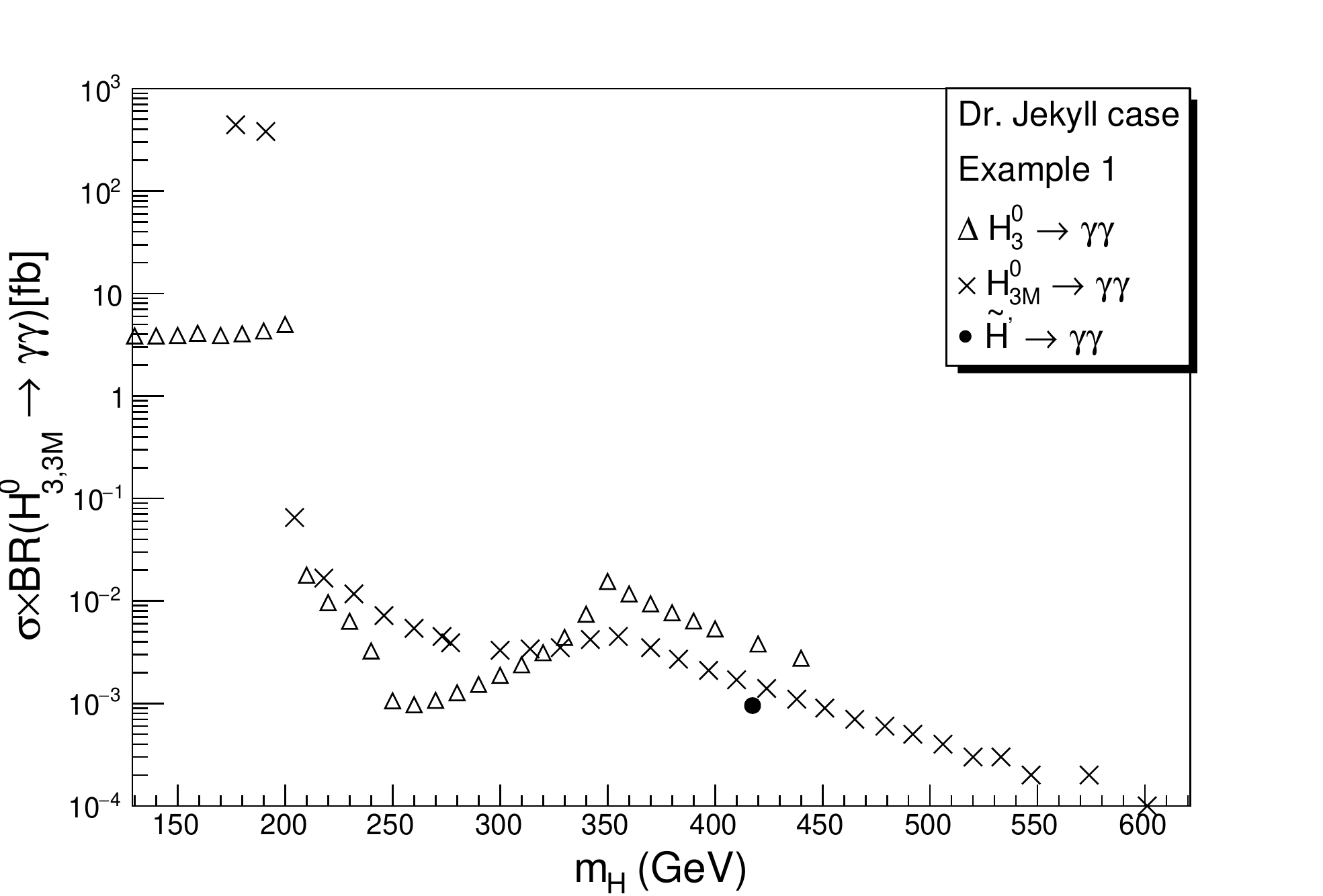}
   \caption{\label{fig:gamma_H1oDominant} {\small The production cross section times branching ratio in $\gamma\gamma$ channel of $H_3^0$ and $H_{3M}^0$, in {\em Dr.~Jekyll} scenario. $m_{H_3^0}=130 - 440 \;\gev$, $m_{H_{3M}^0}=177 - 601\;\gev$ } }
   \end{figure}
   \item
   In the other case when $H_1^0$ is sub-dominant in $\Htil$ or {\em Mr.~Hyde}, a set of parameters is chosen as $s_2=0.514,\;s_{2M}=0.841,\;s_M=0.168$ corresponding to the example 2 in the section \ref{sec:hyde}. In this scenario, all the heavy CP-even states are above $600 \gev$. So that we just present the dependence of the production cross section times branching ratio of $H_{3,3M}\rightarrow\gamma\gamma$. 
   \begin{figure}[H]
    \includegraphics[scale=0.45]{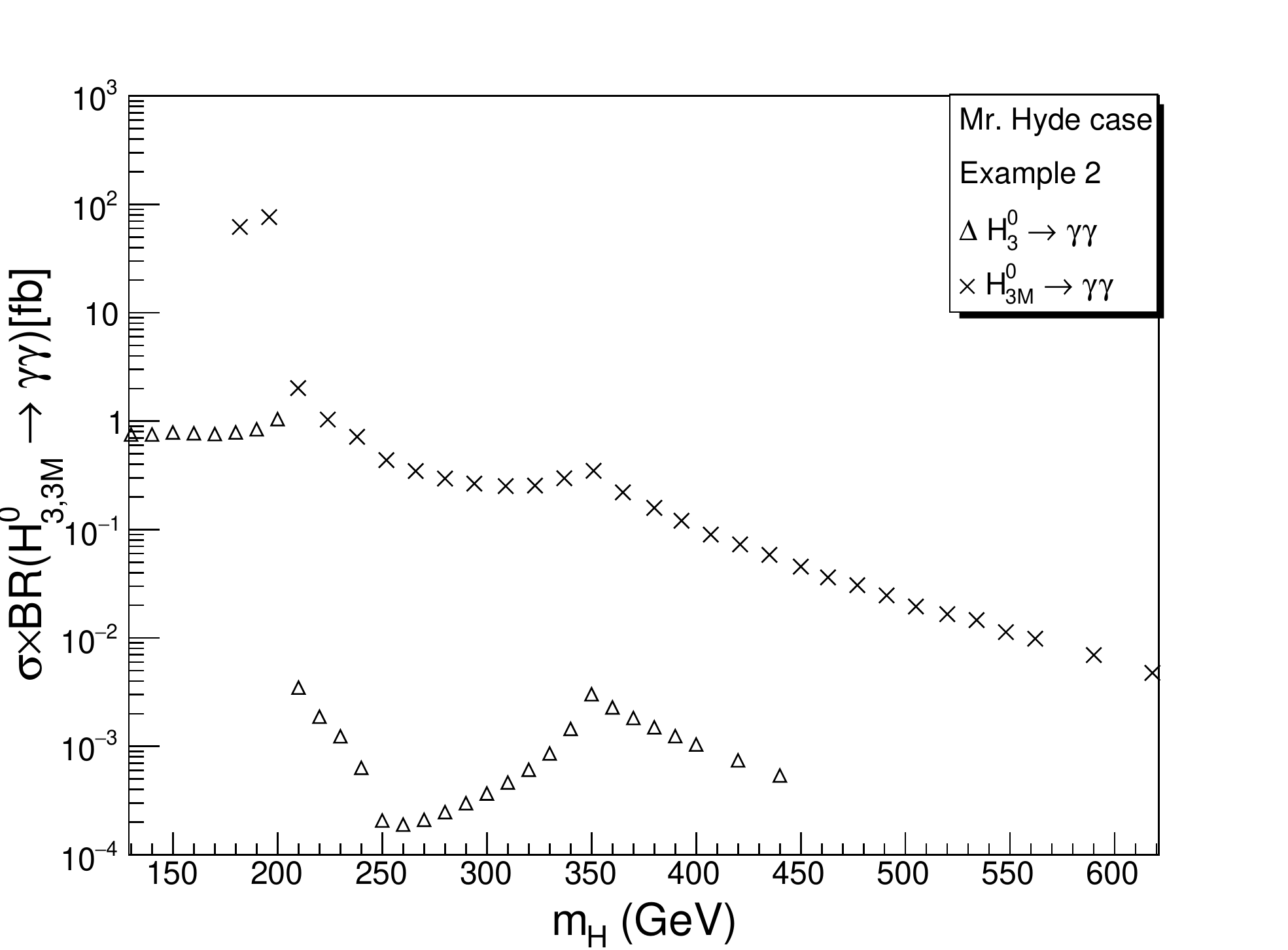}
   \caption{\label{fig:gamma_H1oSubdominant} {\small The production cross section times branching ratio in $\gamma\gamma$ channel of $H_3^0$ and $H_{3M}^0$, in {\em Mr.~Hyde} scenario. $m_{H_3^0}=130 - 440 \;\gev$, $m_{H_{3M}^0}=182 - 618\;\gev$ } }
   \end{figure}
\end{itemize}
Remarks:
\begin{enumerate}
  \item
   Before fermionic thresholds, $2m_{q^M_{1,2}}$ and $2m_{l^M}$, the signal strength for $H_{3,3M}^0$ can be larger than what is measured by ATLAS and CMS for heavy SM-like scalar. To be conservative, we can exclude $m_{H_{3}^0} \lesssim 150~\gev$ and $m_{H_{3M}^0} \lesssim 205~\gev$ (FIG.\ref{fig:gamma_H1oDominant}) and $\lesssim 210~\gev$ (FIG.\ref{fig:gamma_H1oSubdominant}). However, for some other set of parameters, the signal strengths could be well below the upper limit set by ATLAS and CMS.
  \item
    As $m_{H_3^0}$ increases, more mirror fermionic decay channels are kinematically allowed. On the other hand, the production cross section decreases. The branching ratios of $H_{3,3M}^0\rightarrow\gamma\gamma$ therefore decrease rapidly beyond the thresholds, $2m_{q^M_{1,2}}, \;2m_{q^M_3},\;2m_{l^M},\;2m_t$. As a result, the signal strengths in both the cases are below the experimental upper limits.
  \item
   At the same mass, the signal strengths of CP-odd spin-0 states are generally larger than those of CP-even scalars. Consequently, it is easier to detect CP-odd spin-0 states than the CP-even ones. 
\end{enumerate}

\subsection{In $\tau\bar{\tau}$ channel}

Recently, ATLAS \cite{ATLAS_tau} and CMS \cite{CMS_tau} also reported their new results in $\tau\bar{\tau}$ channel. Although, the main aim of their reports is to look for MSSM neutral boson, they provide a model independent limit on the production cross section times branching ratio of a general spin-zero state. Therefore, in this part we investigate the signal strength of our $H_{3,3M}^0\rightarrow\tau\bar{\tau}$ with two sets of parameters considered in the previous subsection.

  \begin{itemize}
   \item In the {\em Dr.~Jekyll} case. 
   \begin{figure}[H]
    \includegraphics[scale=0.45]{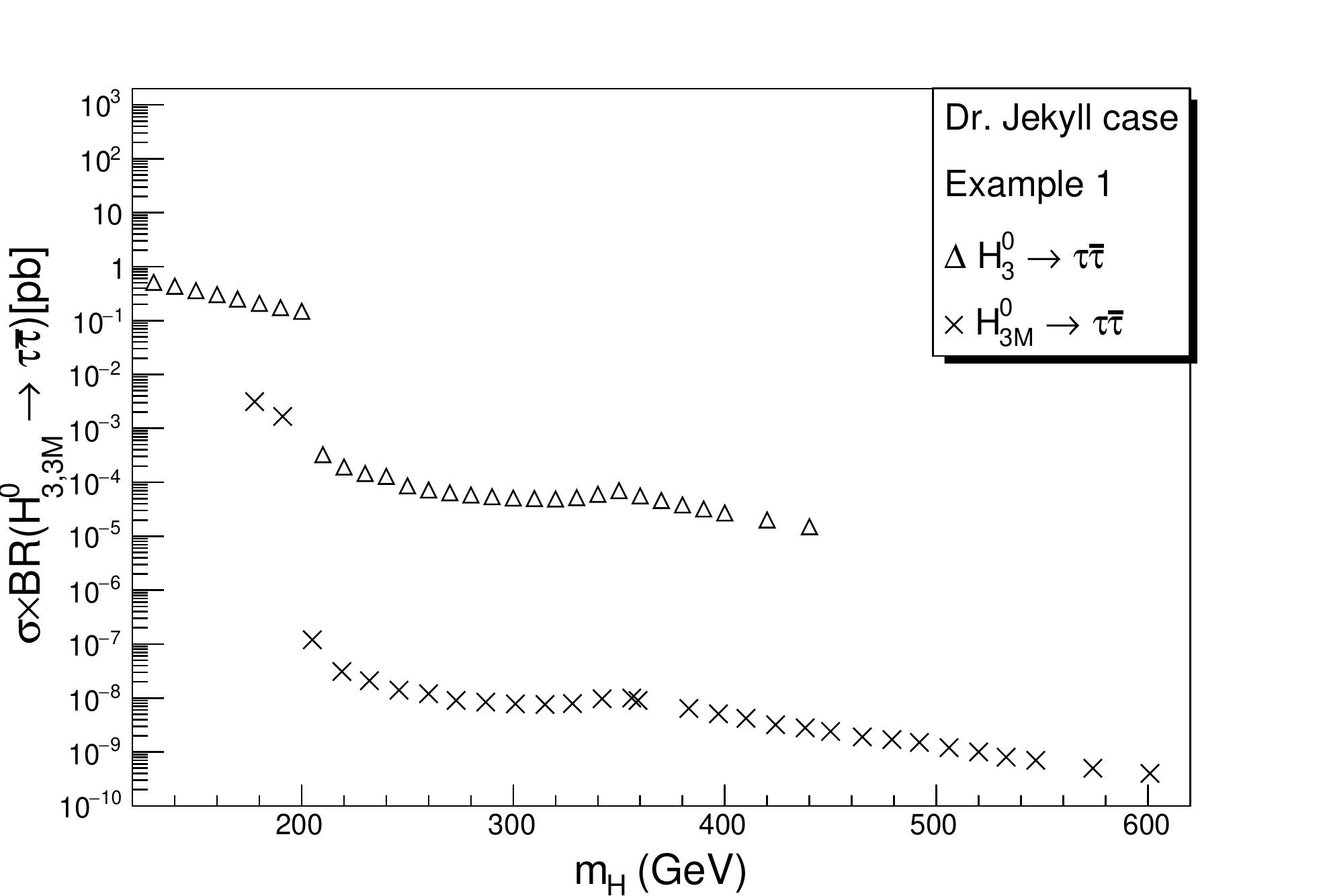}
    \caption{\label{fig:tau_H1oDominant} {\small The production cross section times branching ratio in $\tau\bar{\tau}$ channel of $H_3^0$ and $H_{3M}^0$, in {\em Dr.~Jekyll} scenario. $m_{H_3^0}=130 - 440 \;\gev$, $m_{H_{3M}^0}=177 - 601\;\gev$ } }
    \end{figure}
    \item In the {\em Mr.~Hyde} case.
     \begin{figure}[H]
       \includegraphics[scale=0.45]{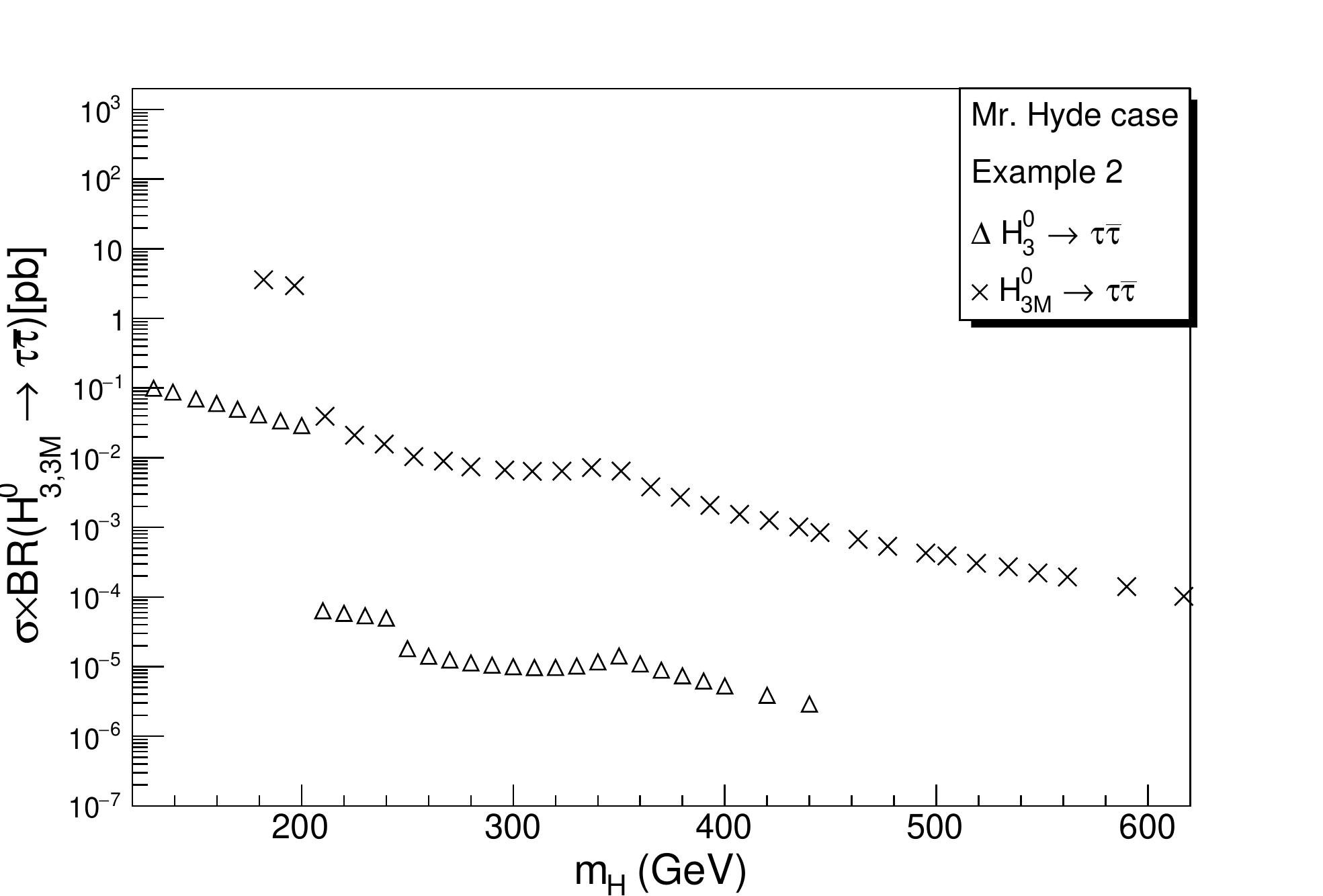}
    \caption{\label{fig:tau_H1oSubdominant} {\small The production cross section times branching ratio in $\tau\bar{\tau}$ channel of $H_3^0$ and $H_{3M}^0$, in {\em Mr.~Hyde} scenario. $m_{H_3^0}=130 - 440 \;\gev$, $m_{H_{3M}^0}=182 - 618\;\gev$ }} 
      \end{figure}
\end{itemize}
Remarks:
\begin{enumerate}
  \item
   In both the cases, the signal strengths can exceed the upper limit from ATLAS and CMS before the thresholds of 2 times mirror fermion masses, which is $204\;\gev$. It happens because, unlike SM Higgs, the decay processes such as $H_{3,3M}^0\rightarrow WW/ZZ$ occur only at the loop level, and their partial widths are relatively small. Consequently, the branching ratios of $H_{3,3M}^0\rightarrow\tau\bar{\tau}$ are not as small as in SM. Hence, the signal strength for this channel is one order above the upper limits set by ATLAS and CMS. However, we believe that in a wide range of parameter space of the $EW\nu_R$ model, it is still possible to agree with the limits in this mass region.    
	\item
	After passing the first threshold, the signal strengths of both $H_{3,3M}^0\rightarrow\tau\bar{\tau}$ decrease rapidly, because the total widths $\Gamma_{H_{3,3M}^0}$ are dominated by the fermionic decays. Then they reach another peak at $2m_t$. Over the entire region after the first threshold, the signal strengths for both $H_{3,3M}^0$ are below the limits.
  \end{enumerate}

\subsection{In $WW/ZZ$  channels}

In this model, pseudo-scalars $H_{3,3M}^0$ do not couple directly to $W^\pm$ and $Z$. Decay processes $H_{3,3M}^0 \rightarrow WW/ZZ$ occur only at loop levels. It is expected that these processes will be highly suppressed. To prove that, we calculate the signal strengths ($\mu$) for $H_{3}^0\rightarrow WW/ZZ$ and $H_{SM}\rightarrow WW/ZZ$. $\mu$ is defined in Eq.~(\ref{eq:muy}). 
\bea\label{eq:muyVV}
\mu_{VV} &=& \dfrac{\sigma(gg \rightarrow H_3^0) Br(H_3^0 \rightarrow VV)}{\sigma(gg \rightarrow H_{SM}) Br(H_{SM} \rightarrow VV)}\nonumber\\[1em]
&=& R_{H_3^0}\dfrac {Br(H_3^0 \rightarrow VV)}{ Br(H_{SM} \rightarrow VV)}\,,
\eea
where $V=W,Z$. Once again, $Br(H_{SM} \rightarrow VV)$ is taken from \cite{HXreport3}, while the ratio of production cross sections $R_{H_3^0}$ in Eq.~(\ref{eq:csectionapprox}) and $Br(H_3^0 \rightarrow VV)$ are calculated directly. At one loop order, the partial decay width for these processes are \cite{GunionKao}
\begin{itemize}
\item $H_3^0 \rightarrow WW$
 \be\label{eq:H3oWW}
 \Gamma( H_3^0 \rightarrow WW ) = \dfrac{3^2g^6(m_{H_3^0}^2 - 4m_{W}^2)^{3/2}}{2^{14}\pi^5m_W^2}|\textit{A}_{WW}|^2
 \ee
 \be\label{eq:Aww}
\textit{A}_{WW} = m_t^2t_MA_t^W - m_b^2t_M^2A_b^W + \dfrac{m_{l^M}^2}{\sqrt{2}}A_{l^M}^W + \dfrac{M_R^2}{\sqrt{2}c_M}A_{\nu_R}^W\,;
 \ee
\item $H_3^0\rightarrow ZZ$
 \be\label{eq:H3ozz}
 \Gamma( H_3^0 \rightarrow ZZ ) = \dfrac{3^2g^6(m_{H_3^0}^2 - 4m_{Z}^2)^{3/2}}{2^{15}\pi^5m_W^2}|\textit{A}_{ZZ}|^2
 \ee
 \be\label{eq:Azz}
\textit{A}_{ZZ} = m_t^2t_MA_t^Z - m_b^2t_MA_b^Z + \dfrac{m_{l^M}^2}{\sqrt{2}}A_{l^M}^Z\,.
 \ee
\end{itemize}
$A_f^{W/Z}$ are amplitudes with top and bottom quarks, mirror charged leptons, and right-handed neutrinos in the loops. They have specific forms given in the Appendix~\ref{sec:AH3}.

 \begin{figure}[H] 
   \includegraphics[scale=0.45]{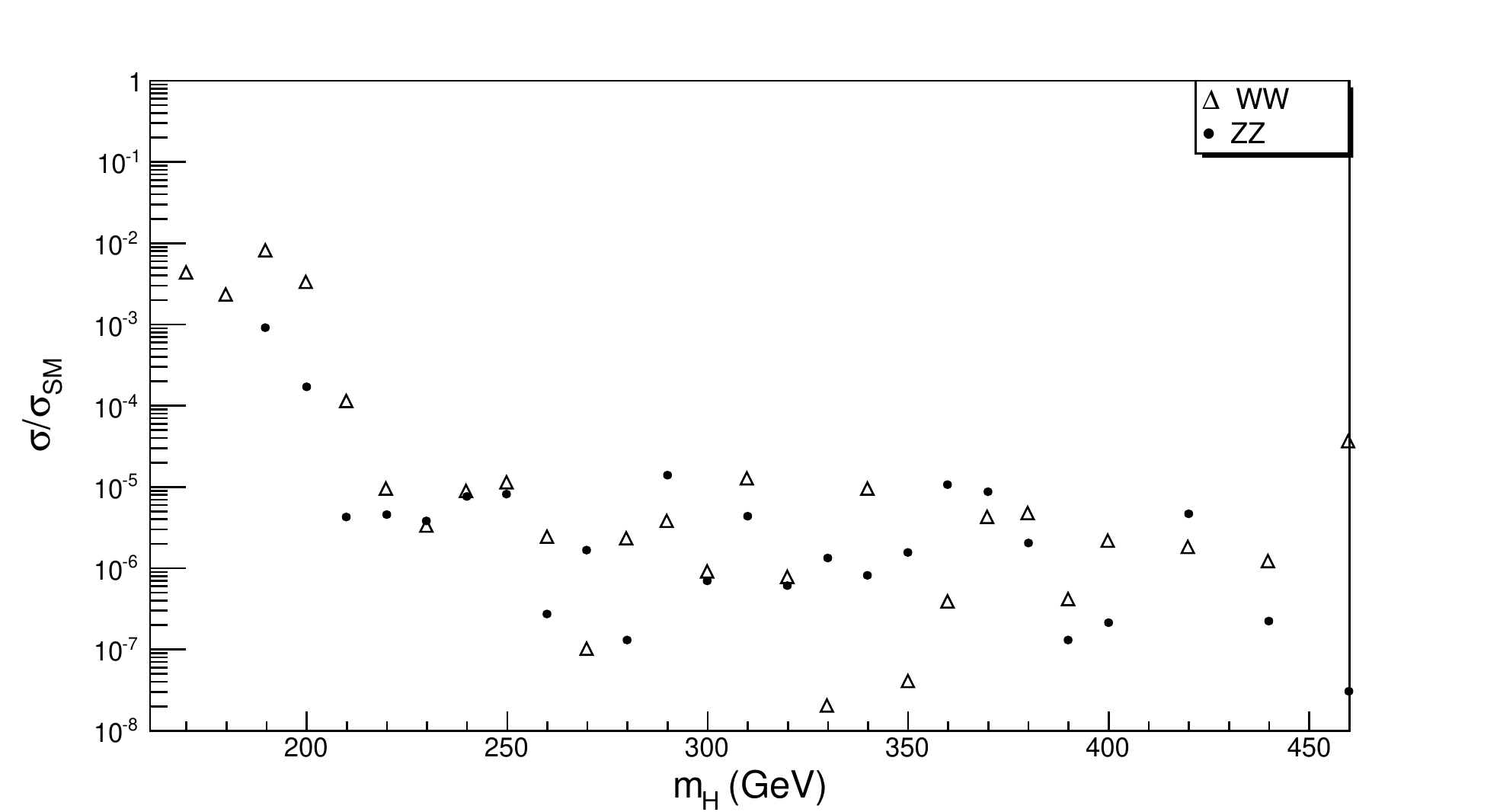}
   \caption{\label{fig:muyWW}{\small Ratio of strength signal in $WW/ZZ$ channel of $H_3^0$ comparing to $H_{SM}$}}
  \end{figure}
As expected, the signal strengths of $H_3^0$ in vector boson channels are highly suppressed. 

\subsection{$(H_{3}^0 \rightarrow \bar{l}^M l^M )/ (H_{SM}\rightarrow WW)$}

At the LHC, $H\rightarrow WW$ is an important channel to search for new scalars in the high-mass region. From Fig.~\ref{fig:muyWW}, we see that the ratio of the signal strengths is suppressed very much for pseudoscalar $H_3^0$ in this model. However, it can also decay through $H_{3}^0 \rightarrow \bar{l}^M l^M \rightarrow \bar{l} \phi_S l \phi_S^* $, where $\phi_S$ is invisible and considered as missing transverse energy ($E_T^M$). Thus, the signal of this process is identified with 2 leptons and missing $E_T^{M}$, which mimics the signal for $WW$ decay of a scalar such as the SM Higgs boson: $H_{SM} \rightarrow W^+ W^- \rightarrow \bar{l} \nu l \bar{\nu}$. For the case $m_{l^M} = 102~\gev$, $m_H = 210 - 500 \gev$, both the mirror lepton intermediate states are on-shell.
\be\label{eq:H3olmlm}
\Gamma \left(H_{3}^0 \rightarrow \bar{l}^M l^M\right) =  \frac{g^2m_{H_3^0}^3t_M^2}{256\pi ~m_W^2} \tau_{l^M}\sqrt{1 - \tau_{l^M}}\,, 
\ee
where $\tau_{l^M}= 4m_{l^M}^2/m_{H_3^0}^2$, and
\be\label{eq:lMlphiS}
 \Gamma \left(l^M\rightarrow l\phi_S^*\right) = g_{sl}^2\dfrac{m_{l^M}}{32\pi}\,.
\ee 
Similar to the previous comparisons, we define the signal strength
\be\label{eq:RH3oHsm} 
  \mu_{l\bar{l}}=R_{H_3^0} \frac{Br\left(H_{3}^0 \rightarrow \bar{l}^M l^M\right) Br\left(l^M \rightarrow l\phi_S^* \right)}{Br \left( H_{SM} \rightarrow W^+ W^- \right) Br\left(  W\rightarrow l\bar{\nu}\right)}\,,
\ee
where $Br\left(  W^-\rightarrow l\bar{\nu}\right) = 0.108 $ \cite{pdg}. With $M_R=50\;\gev$, $l^M\rightarrow \nu_R\bar{\nu} l$ is kinamatically possible.
\be\label{eq:BrlMPhiS}
 Br\left(l^M \rightarrow l\phi_S^* \right) = \dfrac{\Gamma \left(l^M\rightarrow l\phi_S^*\right)}{\Gamma \left(l^M\rightarrow l\phi_S^*\right)+\Gamma \left(l^M\rightarrow \nu_R\bar{\nu} l\right)}\,.
\ee
It is clear that the ratio $\mu_{l\bar{l}}$ depends on the value of $g_{sl}$. The search for a high-mass Higgs boson in $H\rightarrow W^+W^-\rightarrow l\bar{\nu} \bar{l}\nu$ was carried out at both ATLAS (in the range $260 - 1000 \;\gev$ \cite{ATLAS-CONF-2013-067}) and CMS (in the range $145 - 1000\;\gev$) \cite{CMS_heavy_Higgs_WW}. No excess over the background was detected in the entire mass region that was scanned. 

The observed $95\%$ CL upper limit on the ratio of the signal strengths is below $\mu=1$ all the way up to $m_H\approx 600\;\gev$ \cite{CMS_heavy_Higgs_WW}. Therefore, we can set upper limits as $\mu_{l\bar{l}}\leq 1$ and hence, $g_{sl}\leq 10^{-3}$. 


\section{Conclusions}

The 125-\gev object has presented us with a challenge to understand its nature: Is it really the SM Higgs boson as it appears to be or is it simply an {\em impostor}? So far, the only data available to us are given in terms of the so-called signal strengths, $\mu$, as defined in Eq.~(\ref{eq:mudef}). The signal strengths for the various decay modes of the SM Higgs boson are consistent with data. However,  It turns out that it might be possible for various BSM models to be consistent with experiment also based solely on such signal strengths. This is what we have shown in this paper in the context of the $EW\nu_R$ in its extended version.

As we have described in the beginning of our paper, the $EW\nu_R$ \cite{pqnur} was invented with the purpose of realizing the seesaw mechanism at the electroweak scale instead of some GUT scale. As such one can {\em directly} test the seesaw mechanism at the LHC and at the proposed ILC through the physics associated with the model such as lepton-number violating production of electroweak-scale Majorana right-handed neutrinos and -this is the subject of the present paper- Higgs physics beyond that of the SM. 

The extended $EW\nu_R$ model discussed in this paper contains three neutral CP-even mass eigenstates, $\Htil$, $\Htill$ and $\Htilll$, which are linear combinations of $H_{1}^{0}$, $H_{1M}^0$ which couple to SM fermions and mirror fermions respectively and $H_1^{0\prime}$ which couples only to $\nu_R$'s. The notation for the mass eigenstates $\Htil$, $\Htill$ and $\Htilll$ refers to states with increasing masses. We scanned the parameter space with the following requirements in mind: 1) The mass of the lightest state should be $\sim 125 \, \gev$; 2) The mixing angles should be such that the signal strengths fit the data from CMS and ATLAS. We found many combinations of $H_{1}^{0}$, $H_{1M}^0$ and $H_1^{0\prime}$ which satisfy those requirements. What is interesting here is the dual nature of the 125-\gev scalar that we uncovered in our scanning of the parameter space: 1) There are states with the SM-like scalar $H_{1}^{0}$ as a dominant component; 2) There are states with $H_1^{0\prime}$ as a dominant component and is thus {\em very unlike} that of the SM model. In other words, these states are impostors. All of these states - and we are far from exhausting the parameter space - yield the signal strengths compatible with the CMS and ATLAS data.

It goes without saying that detailed studies of various properties of the 125-\gev  scalar such as the total width, partial widths, ..., are needed to determine if it were indeed the SM Higgs boson or just simply an {\em impostor}. Of course, a discovery of one or several extra scalars would definitely point toward physics beyond the SM. In the extended $EW\nu_R$ model, although the aforementioned 125-\gev-like scalars all yield comparable signal strengths, details such as production cross sections, branching ratios, total widths and partial widths can differ quite a bit from one another. States with $H_{1}^{0}$ as a dominant component ({\em Dr.~Jekyll}) tend to behave more like the SM Higgs boson while the scenario in which $H_{1}^{0}$ as a sub-dominant component ({\em Mr.~Hyde}) is very different. In other words, we may have discovered a scalar which is involved in the electroweak symmetry breaking but which {\em may not be} the SM Higgs boson. 

If, in the absence of direct measurements of decay widths, how could one tell {\em Dr.~Jekyll} from {\em Mr.~Hyde}? First, it goes without saying that a discovery of one or more extra scalars will definitely point to BSM physics. In the context of the \ewnur model, if the extra states are {\em broad} and strongly interacting, we would be dealing with {\em Dr.~Jekyll}'s scenario which is more SM-like in terms of the 125-\gev scalar. On the other hand, if the extra states are narrow resonances, we would be facing a truly interesting scenario, that of {\em Mr.~Hyde} in which the 125-\gev scalar is truly an {\em impostor}. Direct measurements of decay widths in this case would confirm whether or not this is the case.


\section{Acknowledgements}

We would like to thank Giuseppe Cerati for providing results of the search for SM-like heavy Higgs boson at CMS. This work was supported by US DOE grant DE-FG02-97ER41027. ASK was supported by the Graduate Fellowship of the Department of Physics, University of Virginia.

\appendix
\section{Some couplings in the Extended \ewnur Model}\label{sec:extewnurcoup}
Here we tabulate all the couplings of the custodial scalars, which are relevant for calculating the signal strengths in various decay channels, in different scenarios considered in this paper.
\begin{table*}[!htb]
	\renewcommand{\arraystretch}{3}
	\centering
	\caption{\label{table:h_ferm}Yukawa couplings with SM quarks and mirror-quarks in the \ewnur model.}
  \begin{tabular}{|C{0.1\textwidth}C{0.38\textwidth}||C{0.1\textwidth}C{0.38\textwidth}|}
	\hline
	~	&	SM Quarks	&	~	&	Mirror Quarks\\
	\hline
	$\;g_{H_1^0 q \overline{q}}$		&	$-\imath \dfrac{m_q\; g}{2\; M_W\; s_2}\;....(q=t,b)$
	&	$\;g_{H_{1M}^0 q^M \overline{q}^M}$		&	$-\imath \dfrac{m_q^M\; g}{2\; M_W\; s_{2M}}$\\
	$\;g_{H_{3}^0 t \overline{t}}$		& $\imath \dfrac{m_t\; g\; s_M}{2\; M_W\; c_M}\gamma_5$
	&	$\;g_{H_{3}^0 u_i^M \overline{u}_i^M}$	&	$-\imath \dfrac{m_{u_i^M}\; g\; s_M}{2\; M_W\; c_M}\gamma_5$\\
	$\;g_{H_{3}^0 b \overline{b}}$		& $-\imath \dfrac{m_b\; g\; s_M}{2\; M_W\; c_M}\gamma_5$
	&	$\;g_{H_{3}^0 d_i^M \overline{d}_i^M}$	&	$\imath \dfrac{m_{d_i^M}\; g\; s_M}{2\; M_W\; c_M}\gamma_5$\\
	$\;g_{H_{3}^- t \overline{b}}$		& $\imath \dfrac{g\; s_M}{2\sqrt{2}\; M_W\; c_M} \big[m_t (1 + \gamma_5) - m_b (1 - \gamma_5)\big]$
	&	$\;g_{H_{3}^- u_i^M \overline{b}_i^M}$	&	$\imath \dfrac{g\; s_M}{2\sqrt{2}\; M_W\; c_M} \big[m_{u_i^M} (1 - \gamma_5) - m_{d_i^M} (1 + \gamma_5)\big]$\\
	$\;g_{H_{3M}^0 t \overline{t}}$		& $-\imath \dfrac{m_t\; g\; s_{2M}}{2\; M_W\; s_2}\gamma_5$
	&	$\;g_{H_{3M}^0 u_i^M \overline{u}_i^M}$	&	$-\imath \dfrac{m_{u_i^M}\; g\; s_2}{2\; M_W\; s_{2M}}\gamma_5$\\
	$\;g_{H_{3M}^0 b \overline{b}}$		& $\imath \dfrac{m_b\; g\; s_{2M}}{2\; M_W\; s_2}\gamma_5$
	&	$\;g_{H_{3M}^0 d_i^M \overline{d}_i^M}$	&	$\imath \dfrac{m_{d_i^M}\; g\; s_2}{2\; M_W\; s_{2M}}\gamma_5$\\
	$\;g_{H_{3M}^- t \overline{b}}$		& $\imath \dfrac{g\; s_{2M}}{2\sqrt{2}\; M_W\; s_2\; c_M} \big[m_t (1 + \gamma_5) - m_b (1 - \gamma_5)\big]$
	&	$\;g_{H_{3M}^- u_i^M \overline{d}_i^M}$	&	$\imath \dfrac{g\; s_2}{2\sqrt{2}\; M_W\; s_{2M}\; c_M} \big[m_{u_i^M} (1 - \gamma_5) - m_{d_i^M} (1 + \gamma_5)\big]$\\[0.5em]
\hline
	$\;g_{H_1^0 l \overline{l}}$		&	$-\imath \dfrac{m_l\; g}{2\; M_W\; s_2}\;....(l=e, ~\mu, ~\tau)$
	&	$\;g_{H_{1M}^0 l^M \overline{l}^M}$		&	$-\imath \dfrac{m_l^M\; g}{2\; M_W\; s_{2M}}$\\
%
	$\;g_{H_{3}^0 l \overline{l}}$		& $-\imath \dfrac{m_l\; g\; s_M}{2\; M_W\; c_M}\gamma_5$
	&	$\;g_{H_{3}^0 l_i^M \overline{l}_i^M}$	&	$\imath \dfrac{m_{l_i^M}\; g\; s_M}{2\; M_W\; c_M}\gamma_5$\\
	$\;g_{H_{3}^- \nu_L \overline{l}}$		& $-\imath \dfrac{g\; m_l\; s_M}{2\sqrt{2}\; M_W\; c_M} (1 - \gamma_5)$
	&	$\;g_{H_{3}^- \nu_{Ri} \overline{l}_i^M}$	&	$-\imath \dfrac{g\; m_{l_i^M}\; s_M}{2\sqrt{2}\; M_W\; c_M} (1 + \gamma_5)$\\
%
	$\;g_{H_{3M}^0 l \overline{l}}$		& $\imath \dfrac{m_l\; g\; s_{2M}}{2\; M_W\; s_2}\gamma_5$
	&	$\;g_{H_{3M}^0 l_i^M \overline{l}_i^M}$	&	$\imath \dfrac{m_{l_i^M}\; g\; s_2}{2\; M_W\; s_{2M}}\gamma_5$\\
	$\;g_{H_{3M}^- \nu_L \overline{l}}$		& $-\imath \dfrac{g\; m_l\; s_{2M}}{2\sqrt{2}\; M_W\; s_2\; c_M} (1 - \gamma_5)$
	&	$\;g_{H_{3M}^- \nu_{Ri} \overline{l}_i^M}$	&	$-\imath \dfrac{g\; m_{l_i^M}\; s_2}{2\sqrt{2}\; M_W\; s_{2M}\; c_M} (1 + \gamma_5)$\\[0.5em]
	\hline
	\end{tabular}
\end{table*}
  \begin{table*}[!htbp]
  \renewcommand{\arraystretch}{2}
  \centering
  \caption{\label{table:feyn21}$S_1S_2V$ type couplings: $V$ is a vector gauge boson and $S_1$, $S_2$ are Higgs/Nambu-Goldstone bosons. Common factor: $\imath g \askPmu$, where $p (p')$ is the incoming momentum of the $S_1 (S_2)$.}
  \bigskip
  \begin{tabular}{|C{0.15\textwidth}C{0.33\textwidth}||C{0.15\textwidth}C{0.33\textwidth}|}
  	\hline
  	\hspace{2em}$g_{H_5^0H_5^-W^+}$		& $-\dfrac{\sqrt{3}}{2}$ 	& \hspace{2em}$g_{H_5^{++}H_5^{--}Z}$	& $-\dfrac{(1 - 2 s_W^2)}{c_W}$\\
  	\hspace{2em}$g_{H_5^+H_5^{--}W^+}$	& $-\dfrac{1}{\sqrt{2}}$ 	& \hspace{2em}$g_{H_5^+H_5^-Z}$		& $\dfrac{(1 - 2 s_W^2)}{2 c_W}$\\
  	\hspace{2em}$g_{H_3^0H_3^-W^+}$		& $-\dfrac{1}{2} s_M^2$ 		& \hspace{2em}$g_{H_3^+H_3^-Z}$		& $\dfrac{(1 - 2 s_W^2)}{2 c_W}$\\
  	\hspace{2em}$g_{H_{3M}^0 H_{3M}^-W^+}$		& $\dfrac{1}{2}$ 	& \hspace{2em}$g_{H_{3M}^+H_{3M}^-Z}$		& $\dfrac{(1 - 2 s_W^2)}{2 c_W}$ \\
  	\hspace{2em}$g_{H_3^+H_5^{--}W^+}$	& $-\dfrac{1}{\sqrt{2}} c_M$ & \hspace{2em}$g_{H_3^+H_5^-Z}$		& $-\dfrac{1}{2 c_W} c_M$\\
  	\hspace{2em}$g_{H_3^0H_5^{-}W^+}$	& $-\dfrac{1}{2} c_M$ 		& \hspace{2em}$g_{H_3^0H_5^0Z}$		& $\dfrac{1}{\sqrt{3}} \dfrac{c_M}{c_W}$\\
  	\hspace{2em}$g_{H_5^0H_3^{-}W^+}$	& $-\dfrac{1}{2\sqrt{3}} c_M$ & \hspace{2em}$g_{G_3^+G_3^-Z}$		& $\dfrac{(1 - 2 s_W^2)}{2 c_W}$\\
  	\hspace{2em}$g_{G_3^0G_3^-W^+}$		& $-\dfrac{1}{2}$ 		& \hspace{2em}$g_{G_3^0H_5^0Z}$		& $\dfrac{1}{\sqrt{3}} \dfrac{s_M}{c_W}$\\
  	\hspace{2em}$g_{G_3^+H_5^{--}W^+}$	& $-\dfrac{1}{\sqrt{2}} s_M$ & \hspace{2em}$g_{G_3^+H_5^-Z}$		& $-\dfrac{1}{2 c_W} s_M$\\
  	\hspace{2em}$g_{G_3^+H_5^{--}W^+}$	& $-\dfrac{1}{\sqrt{2}} s_M$ & \hspace{2em}$g_{H_1^0G_3^0Z}$		& $\dfrac{s_2}{c_W}$\\
  	\hspace{2em}$g_{G_3^0H_5^-W^+}$		& $-\dfrac{1}{2} s_M$ 		& \hspace{2em}$g_{H_{1M}^0G_3^0Z}$		& $\dfrac{s_{2M}}{c_W}$\\
  	\hspace{2em}$g_{H_5^0G_3^-W^+}$		& $\dfrac{1}{2\sqrt{3}} s_M$ & \hspace{2em}$g_{H_1^{0\prime}G_3^0Z}$	& $\sqrt{\dfrac{2}{3}} \dfrac{s_M}{c_W}$\\
  	\hspace{2em}$g_{H_5^0G_3^-W^+}$		& $\dfrac{1}{2\sqrt{3}} s_M$ & \hspace{2em}$g_{H_1^0H_3^0Z}$		& $-\dfrac{s_2 s_M}{2 c_M c_W}$\\
  	\hspace{2em}$g_{H_1^0G_3^-W^+}$		& $\dfrac{1}{2} s_2$ 		& \hspace{2em}$g_{H_{1M}^0H_3^0Z}$		& $-\dfrac{s_{2M} s_M}{2 c_M c_W}$\\
  	\hspace{2em}$g_{H_{1M}^0G_3^-W^+}$		& $\dfrac{1}{2} s_{2M}$ & \hspace{2em}$g_{H_1^{0\prime}H_3^0Z}$	& $\sqrt{\dfrac{2}{3}} \dfrac{c_M}{c_W}$\\
  	\hspace{2em}$g_{H_1^{0\prime}G_3^-W^+}$		& $\sqrt{\dfrac{2}{3}} s_M$ & \hspace{2em}$g_{H_5^+H_5^-\gamma}$	& $s_W$\\
  	\hspace{2em}$g_{H_1^0H_3^-W^+}$		& $-\dfrac{s_2 s_M}{2 c_M}$ 		& \hspace{2em}$g_{H_5^{++}H_5^{--}\gamma}$	& $-2 s_W$\\
  	\hspace{2em}$g_{H_{1M}^0H_3^-W^+}$		& $-\dfrac{s_{2M} s_M}{2 c_M}$ 		& \hspace{2em}$g_{H_3^+H_3^-\gamma}$		& $s_W$\\
  	\hspace{2em}$g_{H_1^{0\prime}H_3^-W^+}$		& $\sqrt{\dfrac{2}{3}} c_M$ & \hspace{2em}$g_{H_{3M}^+H_{3M}^-\gamma}$		& $s_W$\\
  	\hspace{2em}$g_{H_1^0H_{3M}^-W^+}$		& $-\dfrac{s_{2M}}{2 c_M}$ 		& \hspace{2em}$g_{G_3^+G_3^-\gamma}$		& $s_W$\\
  	\hspace{2em}$g_{H_{1M}^0H_{3M}^-W^+}$		& $\dfrac{s_2}{2 c_M}$		 & \hspace{2em}$g_{H_1^0H_{3M}^0Z}$		& $\dfrac{s_{2M}}{2 c_M}$\\
  	~		& ~	& \hspace{2em}$g_{H_{1M}^0H_{3M}^0Z}$		& $-\dfrac{s_{2}}{2 c_M}$ \\
  	\hline
  \end{tabular}
  \end{table*}
  \begin{table*}[!htbp]
  \renewcommand{\arraystretch}{2}
  \centering
  \caption{\label{table:feyn12}$SV_1V_2$ type couplings: $V_1$ and $V_2$ are vector gauge bosons and $S$ is a Higgs boson. Common factor: $\imath g M_W \askGmu$.}
  \bigskip
  \begin{tabular}{|C{0.15\textwidth}C{0.33\textwidth}||C{0.15\textwidth}C{0.33\textwidth}|}
  	\hline
  	\hspace{2em}$g_{H_5^0W^+W^-}$		& $\dfrac{s_M}{\sqrt{3}}$	& \hspace{2em}$g_{H_5^0ZZ}$		& $-\dfrac{2}{\sqrt{3}} \dfrac{s_M}{c_W^2}$\\
  	\hspace{2em}$g_{H_5^{++}W^-W^-}$		& $\sqrt{2} s_M$		& \hspace{2em}$g_{H_5^{+}W^-Z}$	& $-\dfrac{s_M}{c_W}$\\
  	\hspace{2em}$g_{H_1^0W^+W^-}$		& $s_2$				& \hspace{2em}$g_{H_1^0ZZ}$		& $\dfrac{s_2}{c_W^2}$\\
  	\hspace{2em}$g_{H_{1M}^0W^+W^-}$		& $s_{2M}$				& \hspace{2em}$g_{H_{1M}^0ZZ}$		& $\dfrac{s_{2M}}{c_W^2}$\\
  	\hspace{2em}$g_{H_1^{0\prime}W^+W^-}$ & $\dfrac{2 \sqrt{2}}{\sqrt{3}} s_M$	& \hspace{2em}$g_{H_1^{0\prime}ZZ}$	& $\dfrac{2 \sqrt{2}}{\sqrt{3}} \dfrac{s_M}{c_W^2}$\\
  	\hline
  \end{tabular}
  \end{table*}
  \begin{table*}[!htbp]
  \renewcommand{\arraystretch}{2}
  \centering
  \caption{\label{table:feyn22}$S_1S_1^\dagger V_1V_2$ type couplings. Common factor: $\imath g^2 \askGmu$.}
  \bigskip
  \begin{tabular}{|C{0.15\textwidth}C{0.33\textwidth}||C{0.15\textwidth}C{0.33\textwidth}|}
  	\hline
  	$g_{H_5^0H_5^0W^+W^-}$		& $\dfrac{5}{3}$				& $g_{H_5^0H_5^0ZZ}$		& $\dfrac{2}{3 c_W^2}$\\
  	$g_{H_5^+H_5^-W^+W^-}$		& $-\dfrac{3}{2}$			& $g_{H_5^+H_5^-ZZ}$		& $-\dfrac{(c_W^4 + s_W^4)}{c_W^2}$\\
  	$g_{H_5^{++}H_5^{--}W^+W^-}$	& $1$					& $g_{H_5^{++}H_5^{--}ZZ}$		& $2 \dfrac{(1 - 2 s_W^2)^2}{c_W^2}$\\
  	$g_{H_3^0H_3^0W^+W^-}$		& $-\dfrac{(1 + c_M^2)}{2}$		& $g_{H_3^0H_3^0ZZ}$	& $-\dfrac{1}{2 c_W^2} (1 + 3 c_M^2)$\\
  	$g_{H_3^+H_3^-W^+W^-}$		& $-(\dfrac{1}{2} + c_M^2)$		& $g_{H_3^+H_3^-ZZ}$	& $-\left[ \dfrac{s_M^2}{2} \dfrac{(1 - s_W^2)^2}{c_W^2} + c_M^2\dfrac{(c_W^4 + s_W^4)}{c_W^2}\right]$\\
  	$g_{H_{3M}^0H_{3M}^0W^+W^-}$		& $-\dfrac{1}{2}$		& $g_{H_{3M}^0H_{3M}^0ZZ}$	& $\dfrac{1}{2 c_W^2}$ \\
  	$g_{H_{3M}^+H_{3M}^-W^+W^-}$		& $-\dfrac{1}{2}$		& $g_{H_{3M}^+H_{3M}^-ZZ}$	& $-\dfrac{(1 - 2 s_W^2)^2}{2 c_W^2}$ \\
   	$g_{G_3^0G_3^0W^+W^-}$		& $-\dfrac{(1 + s_M^2)}{2}$		& $g_{G_3^0G_3^0ZZ}$		& $-\dfrac{1}{2 c_W^2} (1 + 3 s_M^2)$\\
  	$g_{G_3^+G_3^-W^+W^-}$		& $-(\dfrac{1}{2} + s_M^2)$	& $g_{G_3^+G_3^-ZZ}$		& $-\left[ \dfrac{c_M^2}{2} \dfrac{(1 - s_W^2)^2}{c_W^2} + s_M^2\dfrac{(c_W^4 + s_W^4)}{c_W^2}\right]$\\
  	$g_{H_1^0H_1^0W^+W^-}$		& $\dfrac{1}{2}$				& $g_{H_1^0H_1^0ZZ}$		& $\dfrac{1}{2 c_W^2}$\\
  	$g_{H_{1M}^0H_{1M}^0W^+W^-}$		& $\dfrac{1}{2}$				& $g_{H_{1M}^0H_{1M}^0ZZ}$		& $\dfrac{1}{2 c_W^2}$\\
  	$g_{H_1^{0\prime}H_1^{0\prime}W^+W^-}$	& $\dfrac{4}{3}$		& $g_{H_1^{0\prime}H_1^{0\prime}ZZ}$	& $\dfrac{4}{3 c_W^2}$\\
  	$g_{H_5^+H_5^-\gamma\gamma}$		& $-2 s_W^2$			& $g_{H_5^+H_5^-Z\gamma}$		& $-\dfrac{s_W}{c_W} (1 - 2 s_W^2)$\\
  	$g_{H_5^{++}H_5^{--}\gamma\gamma}$		& $8 s_W^2$		& $g_{H_5^{++}H_5^{--}Z\gamma}$	& $4 \dfrac{s_W}{c_W} (1 - 2 s_W^2)$\\
  	$g_{H_3^+H_3^-\gamma\gamma}$		& $-2 s_W^2$			& $g_{H_3^+H_3^-Z\gamma}$		& $-\dfrac{s_W}{c_W} (1 - 2 s_W^2)$\\
  	$g_{H_{3M}^+H_{3M}^-\gamma\gamma}$		& $-2 s_W^2$		& $g_{H_{3M}^+H_{3M}^-Z\gamma}$		& $-\dfrac{s_W}{c_W} (1 - 2 s_W^2)$ \\
  	$g_{G_3^+G_3^-\gamma\gamma}$		& $-2 s_W^2$			& $g_{G_3^+G_3^-Z\gamma}$		& $-\dfrac{s_W}{c_W} (1 - 2 s_W^2)$\\
  	\hline
  \end{tabular}
  \end{table*}
  \begin{table*}[!htbp]
  \renewcommand{\arraystretch}{2}
  \centering
  \caption{\label{table:feyn22n}More $S_1S_2V_1V_2$ type couplings. Common factor: $\imath g^2 \askGmu$.}
  \bigskip
  \begin{tabular}{|C{0.15\textwidth}C{0.33\textwidth}||C{0.15\textwidth}C{0.33\textwidth}|}
  	\hline
  	$g_{H_1^{0\prime}H_5^0W^+W^-}$		& $\dfrac{\sqrt{2}}{3}$		& $g_{H_1^{0\prime}H_5^0ZZ}$		& $-\dfrac{2 \sqrt{2}}{3 c_W^2}$\\
  	$g_{H_3^+H_5^-W^+W^-}$		& $-\dfrac{c_M}{2}$			& $g_{H_3^+H_5^-ZZ}$			& $c_M \dfrac{(1 - 2 s_W^2)}{c_W^2}$\\
  	$g_{H_3^0G_3^0W^+W^-}$		& $-\dfrac{c_M s_M}{2}$		& $g_{H_3^0G_3^0ZZ}$			& $-\dfrac{3}{2} \dfrac{c_M s_M}{c_W^2}$\\
  	$g_{H_3^+G_3^-W^+W^-}$		& $-c_M s_M$				& $g_{H_3^+G_3^-ZZ}$			& $-\dfrac{c_M s_M}{2 c_W^2}$\\
  	$g_{H_5^+G_3^-W^+W^-}$		& $-\dfrac{s_M}{2}$			& $g_{H_5^+G_3^-ZZ}$			& $s_M \dfrac{(1 - 2 s_W^2)}{c_W^2}$\\
  	~ & ~ & $g_{H_3^+H_5^-Z\gamma}$		& $c_M \dfrac{s_W}{c_W}$\\
  	\hline
  \end{tabular}
  \end{table*}
%
%
%

\section{Partial decay widths of neutral Higgs}\label{sec:hdecay}

In this section we will discuss various production and decay channels relevant for studying properties of $\Htil$, $\Htill$ and $\Htilll$ \cite{HXreport3}. Out of these $H^0 ~\rightarrow ~\gamma\gamma, ~g g$ - type decays (and also the Higgs boson production through $g g ~\rightarrow ~H$) have only one loop contributions at the leading order (LO) and decays like $H^0 \rightarrow WW,~ZZ,~f\bar{f}$ can take place through tree level interactions. We show calculation of the decay width $\Gamma(H\rightarrow\gamma\gamma)$ up to LO in QCD. We will show how all the other relevant decay widths can be calculated easily from the corresponding SM values modified by a multiplicative factor. We calculate these widths in \ewnur model from the SM values given in \cite{HXreport3}. 


\subsection{$\mathbf{H \rightarrow gg}$}

A custodial singlet Higgs boson decays to two gluons through triangle loops of fermions at LO. Unlike $H^0 \rightarrow \gamma\gamma$ channel this channel does not have a `clean' signal at a hadron collider like the LHC due to the large QCD background. However, gluon-gluon fusion channel ($gg\rightarrow H$) is the most dominant production channel for a neutral Higgs and hence, $Hgg$ coupling becomes important while studying the signal strengths of Higgs bosons for various decay channels. 

The production cross section of $gg\rightarrow H^0$ is related to the width of $H^0\rightarrow gg$ by
\be
	\sigma(gg\rightarrow H^0) \propto \Gamma(H^0 \rightarrow gg)\,,
\ee
where the constant of proportionality includes phase space integrals and the mass of $H^0$ (refer to Eq. (2.30) in \cite{higgshunter}). Therefore, for a given mass of Higgs,
\be
	\dfrac{\sigma_{\text{\ewnur}}(gg\rightarrow H^0)}{\sigma_{SM}(gg\rightarrow H^0)} = \dfrac{\Gamma_{\text{\ewnur}}(H^0 \rightarrow gg)}{\Gamma_{SM}(H^0 \rightarrow gg)}\,.
\ee
Hence, to calculate the signal strengths $\mu(H\text{-decay})$, we use $\Gamma(H^0 \rightarrow gg)$ instead of $\Gamma(H^0 \rightarrow gg)$, since we are only interested in the ratios of the production cross-sections in the SM and the \ewnur model.

Consider a general scalar mass-eigenstate $H^0$ that is also a CP-even state in some model of BSM Physics. The relevant part of the interaction Lagrangian is \cite{higgshunter}
\begin{flalign}\label{eq:Lphi2gam}
	\mathcal{L}_{int} = &\dfrac{-m_f}{v_{H^0}}\; \bar{\psi} \psi H^0\; +\; g\: M_W\; \lambda_W ~W_\mu^+ W^{\mu-} H^0&\nonumber\\[.5em]
	&+\; \dfrac{g\; \lambda_S}{M_W} ~S^+ S^- H^0\:,&
\end{flalign}
where $v_{H^0}$ is the vacuum expectation value of $H^0$, $v\; =\; 2 M_W / g \sim \sqrt{\sum_{\text{all $H^0$'s}} v_{H^0}^2}$, $\psi$ is a fermion of mass $m_f$, $S^\pm$ is a charged BSM scalar. For SM $\lambda_W = 1/\sqrt{2}, ~\lambda_S = 0$. For a general (CP-even) Higgs boson $H^0$ that couples to the SM quarks with Yukawa coupling in the equation above, the decay width of $H^0 \rightarrow gg$ is given by
\be\label{eq:widthhgg}
	\Gamma(H^0 \rightarrow gg) = \dfrac{\alpha_S^2\; g^2 m_{H^0}^3}{128 \pi^3 M_W^2} ~\left| \sum_i ~\dfrac{1}{2\; v_{H^0}/v} F_{1/2}(\tau_i) \right|^2\,,
\ee
where, for a loop of quark having mass $m_i$, $\tau_i ~= ~ 4m_i^2 / ~m_{H^0}$ \cite{higgshunter}, and $F_{1/2}(\tau)$ is given by
\be
	F_{1/2}(\tau) = -2\; \tau [1 + (1 - \tau) f(\tau)]\,.
\ee
\bea\label{eq:ftau}
	f(\tau) &=& \left[\sin^{-1}\left(1/\sqrt{\tau}\right)\right]^2, \;~~~~~~~~~~~~\text{if } \tau \geq 1, \nonumber\\[0.5em]
			&=& \frac{1}{4}\; \left[\text{Log}\left(\eta_+/\eta_-\right)\; -\; \imath \pi \right]^2, ~~~\text{if } \tau < 1;
\eea
where
\be
	\eta_\pm\; \equiv\; (1 \pm \sqrt{1 - \tau})\;.
\ee
%
%

In \cite{HXreport3} the partial width of $H^0 \rightarrow gg$ in SM is calculated up to the NLO in QCD. We calculate $\Gamma(H^0 \rightarrow gg)$ in the \ewnur model using these SM values. Using Eq. (\ref{eq:Hmasseigen}) and Table~\ref{table:h_ferm} this decay width can be given by
\begin{flalign}\label{eq:H1gg}
	&\Gamma^{EW\nu_R}(\Htilde\;\rightarrow\; gg) = \Gamma^{SM}(H^0_{SM}\;\rightarrow\; gg)\; \times \dfrac{1}{\left|F_{1/2}(\tau_{top})\right|^2} &\nonumber\\[0.5em]
	&\times\;\Bigg| \dfrac{a_{1,1}}{s_2}F_{1/2}(\tau_{top})+\; \dfrac{a_{1,1M}}{s_{2M}}\; \sum_i F_{1/2}(\tau_{MF_i})\Bigg|^2\; &
\end{flalign}
where $H^0$ denotes $\Htil$, $\Htill$ and $\Htilll$; $\sum_i$ is over all the mirror quarks; 
 $\tau_{MF_i} = 4\; m_{MFi}^2/m_{H^0}^2$. $a_{1,1}$ and $a_{1,1M}$ are elements of the square matrix in Eq~(\ref{eq:Hmasseigen}) - they are coefficients of $H^0_1$ and $H^0_{1M}$ in $\Htilde$ respectively.


\subsection{$\mathbf{H^0 ~\rightarrow ~\gamma\gamma}$}\label{sec:diphoton}

For a custodial singlet Higgs boson, the decay to two photons also proceeds through one loops at LO. It is a `clean' channel due to the absence of a large QCD background. Therefore, in the study of 125-\gev Higgs boson, decay to diphoton is an important channel at CMS and ATLAS \cite{ATLAS_diphoton_july2014, CMS_125_diphoton_july2014}.

For a general Higgs mass eigenstate $H^0$ having couplings as given in Eq~(\ref{eq:Lphi2gam}) the decay width of $H^0 ~\rightarrow ~\gamma\gamma$ is given by \cite{higgshunter}:
\be\label{eq:phi2gam}
\Gamma\;(H^0 \rightarrow \gamma\gamma) = \dfrac{\alpha^2\; g^2}{1024\; \pi^3} ~\dfrac{m_{H^0}^3}{M_W^2} ~\left|\sum_i ~N_{ci}\; Q_{i}^2\; F_s(\tau_i)\; \right|^2\,.
\ee
Here $\sum_i$ is performed over all the particles of spin-$s$ which contribute to $H^0\rightarrow \gamma\gamma$, $s =$ spin-0, spin-1/2, and spin-1 is the spin of $i^{th}$ particle, $Q_i$ is the electric charge in units of $e$, and
\bea\label{eq:Fphi2gam}
	F_1(\tau) &=& \lambda_W\, \tau\; \left[3 \; +\; (4 - 3\,\tau) f(\tau)\right]\,,\nonumber\\
	F_{1/2}(\tau) &=& -2\; \tau [1 + (1 - \tau) f(\tau)]\,,\nonumber\\
	F_0(\tau) &=& 2\; \lambda_S\; [1 - \tau f(\tau)]\,,
\eea
with $\tau\; =\; 4\; m_i^2/ m_{H^0}^2$ and $f(\tau)$ is given by Eq~(\ref{eq:ftau}).

Setting $v_{H^0} = v$ gives the $H^0_{SM} \rightarrow \gamma\gamma$ decay width in SM. Here, the contributions from the charged leptons and quarks (except the top quark loop) are negligible as compared to the contribution from $W^\pm$. Note that $F_1(\tau)$ includes contributions only from the transverse polarization of the $W$-boson; the contribution from the Nambu-Goldstone bosons must be added separately using $F_0(\tau_W)$ \footnote{The formulae given above in Eq~(\ref{eq:Lphi2gam}), Eq~(\ref{eq:Fphi2gam}) are a bit different from Eqs~(2.15), (2.17) in \cite{higgshunter}. We try to give formulas for a general BSM model (e.g. using a general $v_{H^0}$, $\lambda_W$ and $\lambda_S$)}.

Based on Eq~(\ref{eq:phi2gam}) we define the amplitude for $H^0 \rightarrow\gamma\gamma$ as
\be
	\mathcal{A}(H^0 \rightarrow \gamma\gamma) = \sqrt{\dfrac{\alpha^2\; g^2}{1024\; \pi^3} ~\dfrac{m_{H^0}^3}{M_W^2}} ~\left(\sum_i ~N_{ci}\; Q_i^2\; F_s(\tau_i)\; \right)\,.
\ee
Then, in the \ewnur model, we see from Eq~(\ref{eq:Hmasseigen}) that
\begin{flalign}
	\Gamma^{EW\nu_R}&(\Htilde\;\rightarrow\; \gamma\gamma) = \Big| a_{1,1}\; \mathcal{A}^{EW\nu_R}(H_1^0\;\rightarrow\; \gamma\gamma) &\nonumber\\[.5em]
	&+\; a_{1,1M}\; \mathcal{A}^{EW\nu_R}(H_{1M}^0\;\rightarrow\; \gamma\gamma)&\nonumber\\[0.5em]
	&+\; a_{1,1'}\; \mathcal{A}^{EW\nu_R}(H_{1}^{0\prime}\;\rightarrow\; \gamma\gamma)\Big|^2\;,&
\end{flalign}
where $a_{1,i}$ with $(i = 1, 1M, 1^\prime)$ are the coefficients of $H^0_1$, $H^0_{1M}$ and $H^{0\prime}_1$ in $\Htilde$ mass eigenstate, respectively; these are the elements in the first row of the mixing matrix $\{a_{i,j}\}$ in Eq~(\ref{eq:Hmasseigen}).

The $W^\pm$, $G_3^\pm$ loops and the scalar loops with $H^\pm_{3}$, $H^\pm_{3M}$, $H^\pm_5$ and $H^{\pm\pm}_5$ contribute to $\mathcal{A}^{EW\nu_R}(H_1^0\rightarrow \gamma\gamma)$, $\mathcal{A}^{EW\nu_R}(H_{1M}^0\rightarrow \gamma\gamma)$ as well as $\mathcal{A}^{EW\nu_R}(H_1^{0\prime}\rightarrow \gamma\gamma)$. The SM fermion loops contribute only to $\mathcal{A}^{EW\nu_R}(H_1^0\rightarrow \gamma\gamma)$, while the charged mirror fermion loops contribute only to $\mathcal{A}^{EW\nu_R}(H_{1M}^0\rightarrow \gamma\gamma)$. No fermion loops contribute to $\mathcal{A}^{EW\nu_R}(H_1^{0\prime}\rightarrow \gamma\gamma)$.

Various Feynman rules necessary for these calculations can be read from Tables~\ref{table:h_ferm}
$-$\ref{table:feyn22n} and the three-point scalar Feynman rules can be obtained from Eq.~(\ref{eq:pot}).

Before the custodial symmetry is broken all the members of a scalar custodial multiplet are degenerate (refer to Eq.~(\ref{eq:multiplet})), e.g. $H_3^0$ and $H_3^+$ have the same mass and so on. But once the custodial symmetry is broken at the loop level, different custodial multiplet members can have different masses. This mass splitting can also be due to some terms  in the Lagrangian that break the custodial symmetry explicitly (not given explicitly in this paper). 

Assuming such non-degenerate scalars, the partial width of $\Htil \rightarrow \gamma\gamma$ depends on the following variable parameters in \ewnur model:
%
%
\begin{itemize}
	\item
	Masses of $H_3^\pm$, $H_{3M}^\pm$, $H_5^\pm$ and $H_5^{\pm\pm}$,
	\item
	$s_2, ~s_{2M}, ~s_M$,
	\item
	Masses of charged mirror leptons and mirror quarks,
	\item
	Scalar self-couplings: $\lambda_1, ~\lambda_2, ~\lambda_3, ~\lambda_4, ~\lambda_5, ~\lambda_8$,
	\item
	Elements of the $3 \times 3$ $\{a_{i,j}\}$ mixing matrix in Eq.~(\ref{eq:Hmasseigen}).
\end{itemize}
Note that not all of these parameters are independent, e.g. once we fix $s_2$, $s_{2M}$, it automatically fixes $s_M$. Another example is: scalar self-couplings and mixing matrix elements must vary so as to give the mass of the lightest scalar mass eigenstate about 125 \gev.


\subsection{Tree level decays of $\Htil$}\label{sec:htreedecay}

Tree level decay channels of a neutral (CP-even) scalar include decays to two fermions and to $W^+W^-, ~ZZ$. In this subsection, first we show how the decay widths of these decays in the \ewnur model are related to the widths in SM. 

Although at the LO these decays have only the tree level contributions, the NLO QCD+EW corrections become significant at about 5\% accuracy \cite{HXreport3}. Because the decay widths of these channels at tree level in the \ewnur model and in SM are related by a multiplicative factor (as explained below), by using SM decay widths to calculate the \ewnur decay widths, these NLO contributions are automatically included in our results. 

For vertices involving mirror fermions, the QCD+EW corrections are different from the corrections for SM quarks (in SM, non-negligible QCD corrections only come from the top quark). Because masses of the mirror quarks, that we have considered in this paper, are of the same order as the top quark, the NLO corrections due to the mirror quarks can be assumed to have the same magnitude as those due to the top quark, up to a $\sim 5\%$ accuracy. The different tree level couplings in the \ewnur model can be found in Tables~\ref{table:h_ferm} and \ref{table:feyn12}. 

Note that, in general, the predictions for the signal strengths of various decay channels in the \ewnur model are stated up to $\sim 5\%$ accuracy. When the Yukawa couplings of $H_{1M}^0$ with the mirror fermions become large in {\em Dr.~Jekyll} scenario (section \ref{sec:jekyll}), extra QCD corrections can be dominant so as to reduce this accuracy.

\subsubsection{$\mathbf{\Htilde~\rightarrow ~WW, ~ZZ}$}

The $H_1^0 VV$, $H_{1M}^0 VV$ and $H_1^{0\prime} VV$ couplings ($V = W^\pm, ~Z$) in the \ewnur model are suppressed by $s_2 = v_2/v$, $s_{2M} = v_{2M}/v$ and $s_{M} = 2\sqrt{2}v_{M}/v$ respectively, as compared to $H^0_{SM}VV$ couplings in SM. Hence, using Eq.~(\ref{eq:Hmasseigen}) the decay widths for the custodial singlet Higgs mass eigenstates $\Htil,~\Htill,~\Htilll$ are given by
\begin{flalign}\label{eq:WWZZwid}
	\Gamma^{EW\nu_R}&(\Htilde~\rightarrow WW, ~ZZ) = \Gamma^{SM}(H^0_{SM}\rightarrow WW, ~ZZ) \times\nonumber&\\[0.5em]
	&|a_{1,1} ~s_2 + a_{1,1M} ~s_{2M} + a_{1,1'} ~\frac{2\sqrt{2}}{\sqrt{3}} ~s_M|^2 \,.&
\end{flalign}
%

\subsubsection{$\mathbf{\Htilde ~\rightarrow ~f\, \bar{f}}$}

The decays of $\Htil$, $\Htill$, $\Htilll$ to two fermions take place through the tree level Yukawa couplings at the LO, when the decaying scalar is at least twice as much massive as the fermions. 

It can be seen from Table~\ref{table:h_ferm} that the Yukawa couplings of the charged SM fermions with $H_1^0$ and $H_{1M}^0$ are enhanced by factors $1/s_2$ and $1/s_{2M}$, respectively, as compared to the corresponding couplings with $H^0_{SM}$ in SM. Also, $H^{0\prime}_1$ does not couple to particle-antiparticle pairs of charged fermions. Hence, the decay widths to SM fermions can be calculated from the corresponding SM decay widths given in \cite{HXreport3} and using Eq~(\ref{eq:Hmasseigen}). Decay widths calculated in this way also include NLO QCD corrections that are taken into account in \cite{HXreport3}.

The partial widths of decays to SM fermions are given, in terms of the corresponding widths in SM, by
%
\begin{flalign}\label{eq:ffwid}
	\Gamma^{EW\nu_R}&(\Htilde\rightarrow f ~\bar{f}) = \Gamma^{SM}(H^0\rightarrow f ~\bar{f}) \times \left|\dfrac{a_{1,1}}{s_2}\right|^2\,,&
\end{flalign}
%
%
On the other hand, the partial widths of decays to two charged mirror fermions need to be calculated explicitly. We calculate these up to LO, i.e. up to $\sim 5\%$ accuracy, since the NLO-QCD corrections become important for further accuracy . These partial widths are given by
\begin{flalign}
	\Gamma^{EW\nu_R}&(\Htilde\rightarrow f^M ~\bar{f}^M) = \frac{g^2}{32 \pi} ~\frac{m_{f^M}^2}{M_W^2} ~\frac{a_{1,1M}^2}{s_{2M}^2} ~m_{\Htilde}\nonumber&\\[0.5em]
	& \times \Big(1 - \frac{4 ~m_{f^M}^2}{m_{\Htilde}^2}\Big)^{3/2}\,.&
\end{flalign}
%

\subsection{Total width of $\mathbf{\Htilde}$}
%
The total widths of a mass eigenstates $\Htil$, $\Htill$, $\Htilll$ can be calculated by adding individual partial widths of various decay channels. We consider all the kinematically allowed channels among $W^+W^-$, $ZZ$, $\gamma\gamma$, $gg$, $b\bar{b}$, $t\bar{t}$, $\tau\bar{\tau}$, $\mu\bar{\mu}$, $c\bar{c}$, $l^M\bar{l}^M$ and $q^M\bar{q}^M$.

For $\Htil$, Eq.~(\ref{eq:Hwidth}):
\begin{eqnarray*}
\Gamma_{\Htil} = &&~\Gamma_{\Htil\rightarrow b\bar{b}} ~+ ~\Gamma_{\Htil\rightarrow \tau\bar{\tau}} ~+ ~\Gamma_{\Htil\rightarrow c\bar{c}} ~+ ~\Gamma_{\Htil\rightarrow W^+ W^-}\nonumber\\[0.5em]
&& + ~\Gamma_{\Htil\rightarrow ZZ} ~+ ~\Gamma_{\Htil\rightarrow gg} ~+ ~\Gamma_{\Htil\rightarrow \gamma \gamma}\,.
\end{eqnarray*}
For $\Htill$, Eq.~(\ref{eq:widthH1M}):
\begin{eqnarray*}
\Gamma_{\Htill} = &&~\sum_{i=1}^3\Gamma_{\Htill\rightarrow q_i^M\bar{q}_i^M} ~+ \sum^3_{j=1} ~\times ~\Gamma_{\Htill\rightarrow l_j^M\bar{l}_j^M}\nonumber\\[0.5em]
 &&+ ~\Gamma_{\Htill\rightarrow t\bar{t}} ~+ ~\Gamma_{\Htill\rightarrow b\bar{b}} \nonumber\\[0.5em]
&& + ~\Gamma_{\Htil\rightarrow \tau\bar{\tau}} ~+ ~\Gamma_{\Htil\rightarrow c\bar{c}} ~+\Gamma_{\Htill\rightarrow W^+ W^-}\nonumber\\[0.5em]
&& + ~\Gamma_{\Htill\rightarrow ZZ} ~+ ~\Gamma_{\Htill\rightarrow gg} ~+ ~\Gamma_{\Htill\rightarrow \gamma \gamma}\,.
\end{eqnarray*}
%

\section{Amplitude of $H_{3}^0\rightarrow WW/ZZ$}\label{sec:AH3}
The processes such as $H_3^0\rightarrow WW/ZZ$ in this model only take place at loop level. At 1 loop, the Feynman diagrams are:
 \begin{figure}[H]
  \includegraphics[scale=0.4]{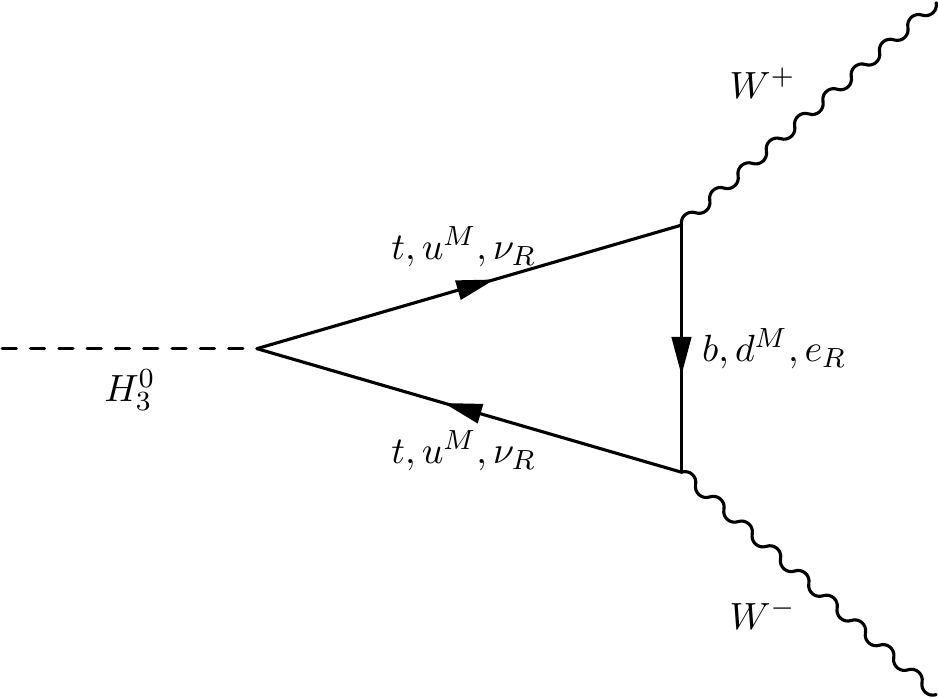}
   \includegraphics[scale=0.4]{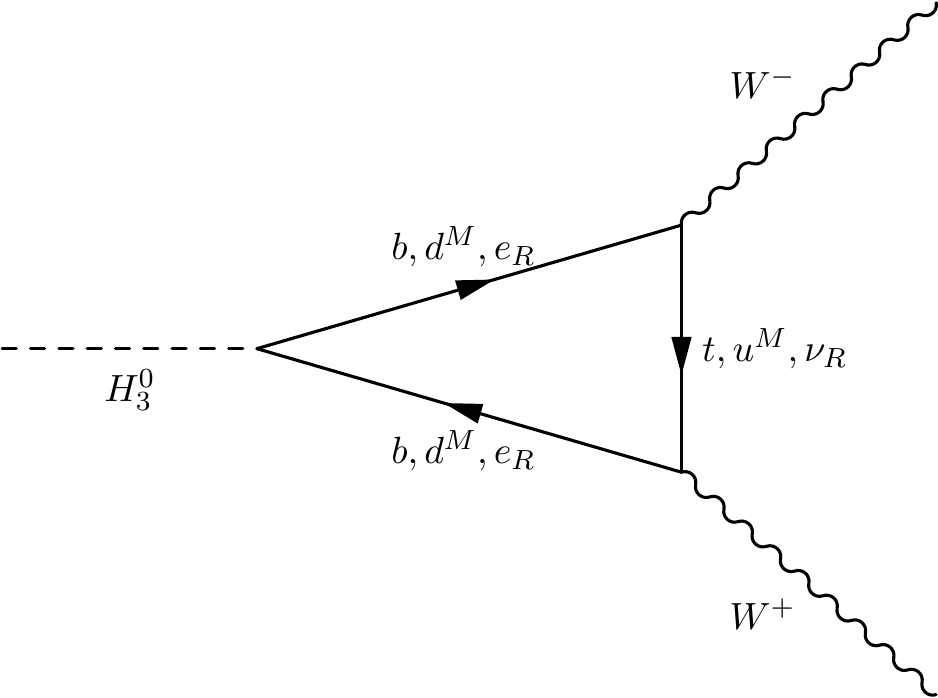}
     \caption{\label{fig:H3oWWdiag}{\small Feynman diagram of $H_3^0\rightarrow W^+W^-$. We have three generations of mirror quarks and three generations of mirror leptons }}
\end{figure}
\begin{figure}[H]
\centering
    \includegraphics[scale=0.4]{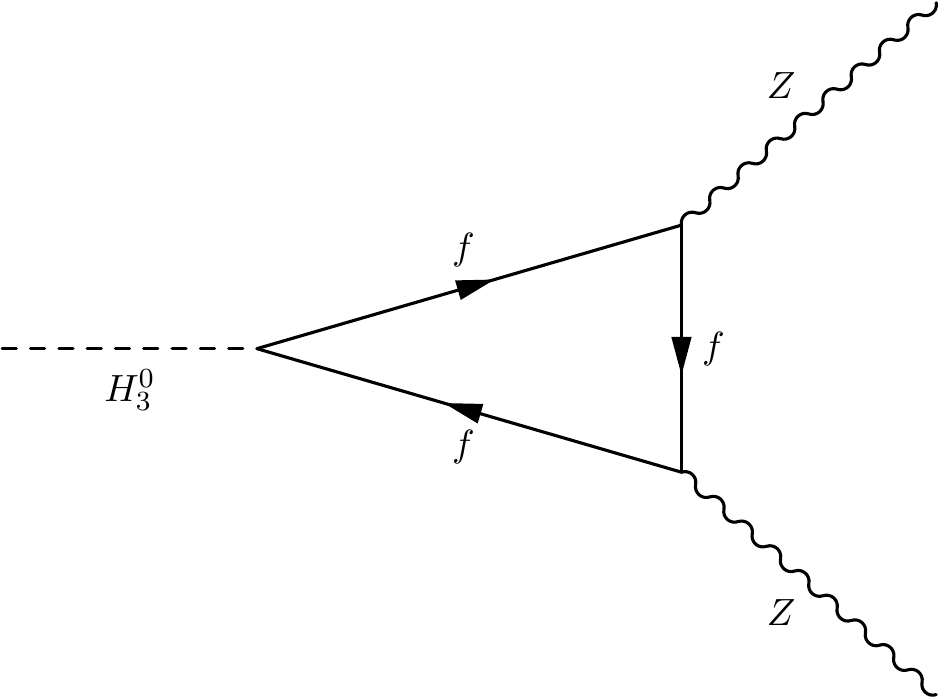}
     \caption{\label{fig:H3oZZdiag}{\small Feynman diagram of $H_3^0\rightarrow ZZ$. Here, $f = u_1^M, d_1^M, u_2^M, d_2^M, u_3^M, d_3^M$, and three charged mirror leptons $l^M$}}  
  \end{figure}
 With the couplings in Table \ref{table:h_ferm}, the amplitude can be expressed as \cite{GunionKao}
 \be\label{eq:A_H3oVV}
  A(H_3^0\rightarrow VV) = m_u^2t_MA_u^V - m_d^2t_MA_d^V
 \ee
Here, $u, ~d$ represent $t, ~b; ~u_M^i, ~d_M^i; ~\nu_R^j, ~l_M^j$. Two intermediate functions $A_u^V, A_d^V$ are expressed in terms of loop functions, $C,\; F$ \cite{tHooft_Veltman}:
 \begin{itemize}
  \item $H_3^0\rightarrow W^+W^-$
   \begin{eqnarray}\label{eq:A_W_ud}
    A_u^W = \frac{1}{2}[C(m_W^2;m_u^2,m_d^2)+F(m_W^2;m_u^2,m_d^2)] \nonumber\\
    A_d^W = \frac{1}{2}[C(m_W^2;m_d^2,m_u^2)+F(m_W^2;m_d^2,m_u^2)]\nonumber\\
    \end{eqnarray}
   \item $H_3^0\rightarrow ZZ$
    \begin{eqnarray}\label{eq:A_Z_f}
     A_f^Z =[ \dfrac{(T_3-Qsin^2\theta_W)^2}{cos^2\theta_W} + (\dfrac{Qsin^2\theta_W}{cos\theta_W})^2]C(m_Z^2;m_f^2)\nonumber\\
                 +[ \dfrac{(T_3-Qsin^2\theta_W)^2}{cos^2\theta_W} - (\dfrac{Qsin^2\theta_W}{cos\theta_W})^2]F(m_Z^2;m_f^2) \nonumber\\     
    \end{eqnarray}
 \end{itemize}
  $C, F$ are generally defined in terms of the 't Hooft-Veltman scalar loop integrals \cite{tHooft_Veltman}. However, in this case we have top quark and heavy mirror fermions, which allows us to use asymptotic forms in the high-mass limit:
  \begin{eqnarray}\label{eq:tHooft_Veltman}
    C(m_V^2;m_u^2,m_d^2) = \int\limits_{0}^{1}dx\int\limits_{0}^{x}dy\dfrac{1}{D}\,,\nonumber\\
    F(m_V^2;m_u^2,m_d^2) =- \int\limits_{0}^{1}dx\int\limits_{0}^{x}dy\dfrac{1}{D}\,,
  \end{eqnarray}
where 
   \begin{eqnarray}\label{eq:D}
     D = m_{H_3^0}^2(1-x)(1-y)+m_V^2(1-x)\nonumber\\
          +m_V^2y(x-y)-m_u^2(1-y)-m_d^2y\,.
   \end{eqnarray}

%
%
\FloatBarrier

\end{document}